\documentclass[apj]{emulateapj}
\usepackage{multirow}

\def\gtorder{\mathrel{\raise.3ex\hbox{$>$}\mkern-14mu
             \lower0.6ex\hbox{$\sim$}}}
\def\ltorder{\mathrel{\raise.3ex\hbox{$<$}\mkern-14mu
             \lower0.6ex\hbox{$\sim$}}}

\slugcomment{Manuscript version \today. Accepted in ApJ.}
\shorttitle{Radio transients}

\begin{document}

\title{Sensitive Search For Radio Variables And Transients In The Extended
  Chandra Deep-Field South}


\author{K.~P.~Mooley\altaffilmark{1},
D.~A.~Frail\altaffilmark{2},
E.~O.~Ofek\altaffilmark{3},
N.~A.~Miller\altaffilmark{4},
S.~R.~Kulkarni\altaffilmark{1} 
\&\
A. Horesh\altaffilmark{1}
}

\altaffiltext{1}{Cahill Center for Astronomy, MC 249-17,
 California Institute of Technology, Pasadena, CA 91125, USA}
\altaffiltext{2}{National Radio Astronomy Observatory, P.O. Box O,
  Socorro, NM 87801}
\altaffiltext{3}{Benoziyo Center for Astrophysics, Faculty of Physics,
  The Weizmann Institute for Science, Rehovot 76100, Israel}
\altaffiltext{4}{Department of Astronomy, University of Maryland,
  College Park, MD, 20742-2421, USA}
\begin{abstract}
%
  We report on an analysis of the Extended Chandra Deep Field South 
  (E-CDFS) region using archival data from the Very Large Array, with 
  the goal of studying radio variability and transients at the sub-mJy
  level. The 49 epochs of E-CDFS observations at 1.4 GHz sample
  timescales from one day to 3 months. We find that only a fraction
  (1\%) of unresolved radio sources above 40 $\mu$Jy are variable at
  the 4$\sigma$ level. There is no evidence that the fractional
  variability changes along with the known transition of radio
  source populations below one milliJansky. Optical identifications 
  of the sources show that the variable radio emission 
  is associated with the central regions of an active galactic nucleus or  
  a star-forming galaxy. After
  a detailed comparison of the efficacy of various source-finding
  algorithms, we use the best to carry out a transient search. No
  transients were found. This implies that the areal density of transients
  with peak flux density greater than 0.21\,mJy is less than 0.37 deg$^{-2}$ 
  (at a confidence level of 95\%). This result is 
  approximately an order of magnitude below the transient rate
  measured at 5 GHz by Bower {\it et al.} (2007) but it is consistent
  with more recent upper limits from Frail {\it et al.} (2012).
  Our findings suggest that the radio sky at 1.4 GHz is relatively
  quiet.  For multi-wavelength transient searches, such as the
  electromagnetic counterparts to gravitational waves, this frequency
  may be optimal for reducing the high background of false positives.
\end{abstract}

\keywords{catalogs --- radio continuum: galaxies --- surveys}

\section{Introduction}\label{sec:Introduction}

For more than four decades the largest science yield of variable and transient radio emission has come from
single-dish radio telescopes, which have surveyed the sky for pulsed
and periodic emission from pulsars and related phenomena on typical
timescales of seconds to milliseconds. In contrast, interferometric
imaging surveys, which are best-suited to probe timescales ranging
from seconds, hours and days, have only just begun. There are a
limited number of surveys of these so-called {\it long duration}
transients and variables at frequencies of 1--10 GHz, each with a
different sensitivity, cadence and field-of-view \citep{ofb+11,bfs+11}.
Fully exploring this phase space is one of the main science drivers
for a new generation of synoptic radio imaging facilities, such as ASKAP 
\citep{jtb+08} and Apertif/WSRT \citep{ovv10}, that are 
being built in the coming years \citep[see also][]{lbb+09}.

Our knowledge of the variable GHz sky is especially lacking at
sub-milliJansky flux density levels. For {\it persistent} sources
there is a well-known flattening of the Euclidean-normalized radio
source counts below about 1 mJy, corresponding to a change in the
radio source populations. This flattening is likely due to the
emerging importance of star forming galaxies and low-luminosity active 
galactic nuclei (AGNs) at redshifts of order unity 
\citep{condon2007,sdm+08,sss+08,pad11,condon2012}. Thus, while the variability
studies above milliJansky levels are dominated by radio-loud AGNs with
compact, flat-spectrum components \citep{sre+06,lov08}, variability at
these deeper flux density levels may probe new source populations.

There are indications that the mJy transient sky is exciting.  Over
the last decade astronomers have detected transient decimetric
emission from a variety of sources: transient, bursting and pulsed
radio emission from magnetars \citep{ccr+05,gkg+05,crh+06}, short-lived 
radio afterglows of short-duration gamma-ray bursts
\citep{ffp+05}, emission from (a transient) jet in a dwarf nova
\citep{krk+08}, a new population of sub-relativistic supernovae
\citep{scp+10}, a mysterious population of bursting radio sources
\citep{hlk+05,hwl+09}, and relativistic outflows from tidal disruption
events \citep{zbs+11,ckh+12}. However, with few exceptions, most
of what we know about the transient radio sky has come via radio
follow-up of objects identified by synoptic telescopes at optical,
X-ray or gamma-ray wavelengths.  Clearly, this titillating trove 
of (serendipitous) discoveries call for systematic exploration
of the decimetric sky on timescales of minutes and longer.

A ready source of archival data for searching for sub-milliJansky
transients and variables comes from deep continuum imaging surveys of 
the GHz radio sky undertaken to study the extragalactic radio source populations.
Many such surveys have been carried out, reaching
noise levels of 4--10 $\mu$Jy and with arcsecond resolution
\citep[e.g.][]{ssc+07,zmnw10}. 
In order to reach these deep flux density limits, it is
standard to observe these fields for many epochs with a cadence that
samples timescales of days, weeks and months. An added benefit is that
these deep surveys are accompanied by rich multi-wavelength continuum
datasets and optical/infrared spectroscopic measurements. Thus the
counterpart of any unusual variable or transient source can be readily
identified and its redshift determined.

In this paper we present a search for transients and variables at
sub-milliJansky flux density levels using data taken as part of a deep
radio continuum survey toward a region known as the Extended Chandra
Deep Field-South \citep[E-CDFS;][]{mfk+08}. In \S\ref{SurveyObs} we describe the
original survey and our re-reduction of the data. In
\S\ref{sec:variables} we describe how the variability light curves for
599 point sources were extracted. The transient search is described in
\S\ref{Sec:Method}. The interpretation and implication of these
results for radio source variability and transients is discussed in
\S\ref{Discussion}.

\section{Observations and Data Reduction}
\label{SurveyObs}

\begin{figure}[htp]
\centering
\includegraphics[width=3.4in,viewport=50 215 490 590,clip]{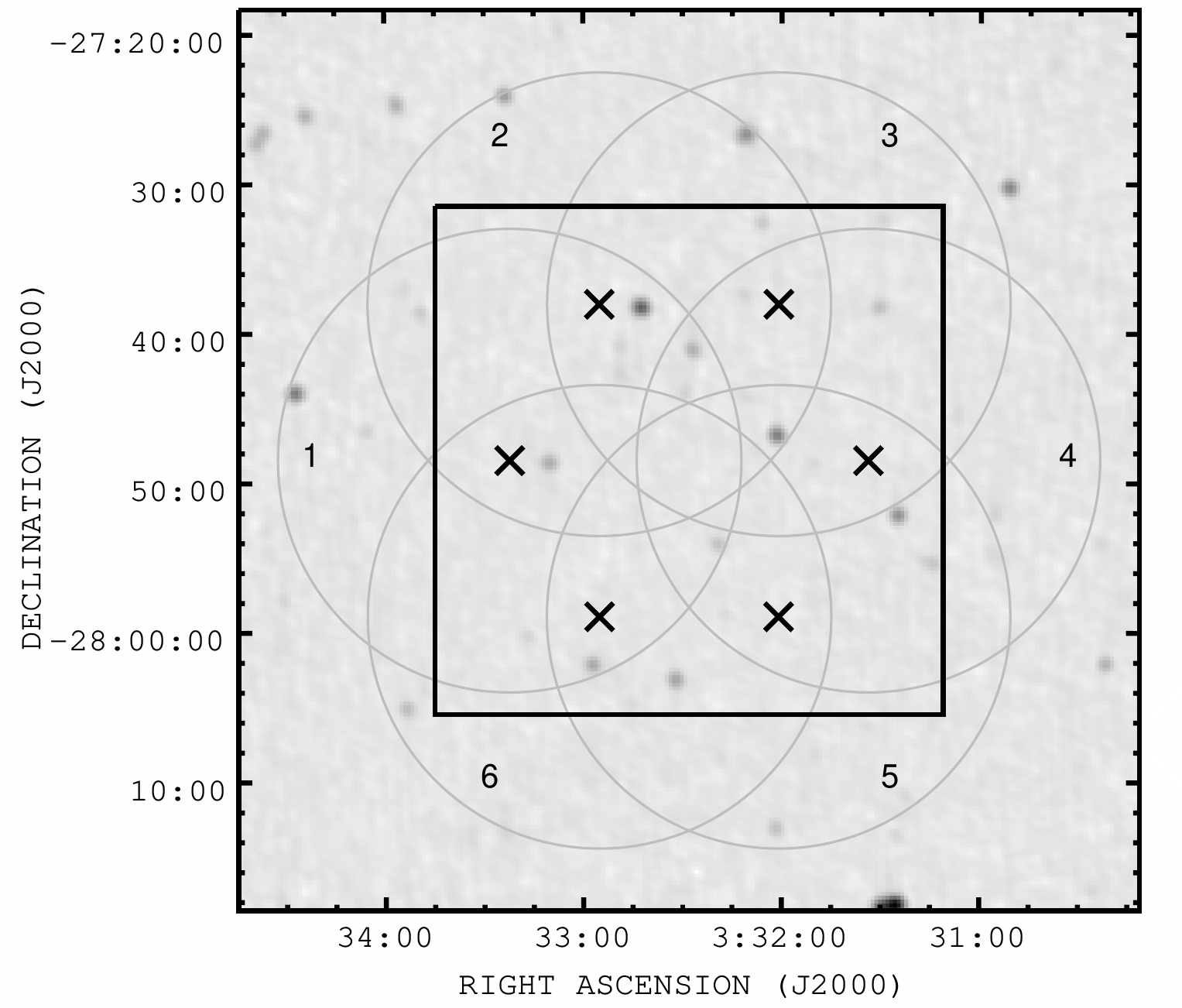}
\caption{Hexagonal grid of six pointings (crosses), each pointing separated 12$^\prime$ from its nearest neighbor.
The $34\arcmin \times 34\arcmin$ (black square) region represents the extent of the final image from the \cite{mfk+08} data release.
$15\arcmin$-circles (grey) corresponding to the 50\% beam attenuation in different pointings are also shown. 
The $60\arcmin \times 60\arcmin$ background image is from NVSS.}
\label{Fig:obsCoverage}
\end{figure}

The E-CDFS is an intensely studied region having a plethora of multi-wavelength 
data available \citep[viz. X-ray, ultraviolet, optical, infrared and radio; see][and references therein]{mfk+08}. 
Here, we use data from the radio survey undertaken by \cite{mfk+08}\footnote{In 
the past, radio observations of this field have also been carried out by 
\cite{kfm+08}, \cite{amk+06}, and \cite{naa+06}.}.
The observations were made at a frequency of 1.4 GHz, using the
National Radio Astronomy Observatory (NRAO) VLA in its A configuration
(Project code AM\,889). A hexagonal grid of six pointings was made,
with each pointing separated 12$^\prime$ from its nearest neighbor
(see Table~\ref{Tab:ListOfFields} and Figure~\ref{Fig:obsCoverage}). 
All the observations were carried out in 2007 
between June 15 and September 23. Only a single pointing was observed
for each epoch. Each epoch was a 5-hr track centered on 03:30 LST\footnote{This is of great benefit to variability studies, which are otherwise plagued by changes in the observing setup.}.
There were a total of 49 epochs, with a combined allocation of 245 hrs
(see Table~\ref{Tab:ListOfEpochs}). The total number of epochs for
each pointing $N_{{\rm ep}}$ is given in Table~\ref{Tab:ListOfFields}.
For more details about the specifics of the observational setup see
\cite{mfk+08}.

The original purpose of these data was to average together all
pointings and epochs in order to create a deep ($\sigma_{rms}=$ 5--8 
$\mu$Jy) continuum image of the E-CDFS. In order to explore variability and to search for
transients, we needed to work with the single epoch, single pointing
images instead. Despite this, many of the data reduction steps that we
followed were similar to \cite{mfk+08}. We summarize the
process here, pointing out slight differences. All calibration and
imaging was carried out in the Astronomical Image Processing System
({\it AIPS}) package\footnote{http://www.aips.nrao.edu/}.

The calibrated {\it uv} data for each epoch were imaged and deconvolved
separately. In order to image the entire field-of-view at full
resolution, we created 37 different facets in a flys-eye pattern.
Each of the 37 facets was an image of 1024$^2$ pixels with a
0.5$^{\prime\prime}$ pixel size offset from one another. There were
another 23-25 outlier fields 128$^2$ pixels each made of cataloged
bright radio sources outside the primary beam of each pointing but
within a 2$^\circ$ radius.

The {\it AIPS} task {\tt IMAGR} was used to deconvolve each image to the rms
noise level. In order to have a consistent set of images, we applied a
Gaussian taper to the (30\% level) of 100 k$\lambda$ and 70 k$\lambda$
in the {\it u} and {\it v} directions to the visibility data, and we
restored the final images to a synthesized beam of
2.8$^{\prime\prime}\times 1.6^{\prime\prime}$ (position angle$\simeq
0^\circ$). After deconvolution, the 37 facets for each epoch were
combined using the {\it AIPS} task {\tt FLATN} to form a single 5120$^2$
pixel image 42.7$^\prime$ across. A correction for the attenuation
from the primary beam was not applied at this stage in order that the
images used in analysis 
had uniform noise statistics
across the entire image.  The rms noise $\sigma_{rms}$ for each epoch
is given in Table~\ref{Tab:ListOfEpochs}.

%
\begin{deluxetable}{llll}
\tablecolumns{4}
\tabletypesize{\scriptsize}
\tablewidth{0pt}
\tablecaption{List of survey pointings}
\tablehead{
\colhead{Pointing ID}   &
\colhead{R.A. (J2000.0)}   &
\colhead{Dec. (J2000.0)}  &
\colhead{$N_{{\rm ep}}$}
}
\startdata
ECDFS1 & 03~33~22.25 & $-$27~48~30.0 &7\\
ECDFS2 & 03~32~55.12 & $-$27~38~03.0 &9\\
ECDFS3 & 03~32~00.88 & $-$27~38~03.0 &8\\
ECDFS4 & 03~31~33.75 & $-$27~48~30.0 &8\\
ECDFS5 & 03~32~00.88 & $-$27~58~57.0 &9\\
ECDFS6 & 03~32~55.12 & $-$27~58~57.0 &8
\enddata
\tablecomments{$N_{{\rm ep}}$ is the number of epochs per pointing.}\label{Tab:ListOfFields}
\end{deluxetable}

\begin{deluxetable}{lllc}
\tablecolumns{4}
\tablecaption{Observing Epochs}
\tablehead{
\colhead{Epoch} &
\colhead{Date} & 
\colhead{Pointing}&
\colhead{$\sigma_{rms}$}\\
\colhead{}   &
\colhead{UT} & 
\colhead{}   &
\colhead{$\mu$Jy $bm^{-1}$}
}
\startdata
1 & 2007 Jun. 15  & ECDFS2 & 26.3 \\
2 & 2007 Jun. 24  & ECDFS3 & 28.0 \\
3 & 2007 Jun. 25  & ECDFS4 & 28.6 \\
4 & 2007 Jul. 01  & ECDFS6 & 25.9 \\
5 & 2007 Jul. 06  & ECDFS5 & 29.2 \\
6 & 2007 Jul. 12  & ECDFS1 & 26.6 \\ 
7 & 2007 Jul. 13  & ECDFS2 & 26.6 \\
8 & 2007 Jul. 14  & ECDFS3 & 25.5 \\
9 & 2007 Jul. 15  & ECDFS4 & 26.4 \\
10 & 2007 Jul. 16  & ECDFS5 & 26.8 \\
11 & 2007 Jul. 17  & ECDFS6 & 26.1 \\
12 & 2007 Jul. 19  & ECDFS1 & 26.0 \\
13 & 2007 Jul. 20  & ECDFS2 & 34.2 \\
14 & 2007 Jul. 21  & ECDFS3 & 26.0 \\
15 & 2007 Jul. 22  & ECDFS4 & 25.3 \\
16 & 2007 Jul. 23  & ECDFS5 & 27.8 \\
17 & 2007 Jul. 24  & ECDFS6 & 27.5 \\
18 & 2007 Jul. 26  & ECDFS5 & 31.8 \\
19 & 2007 Jul. 27  & ECDFS2 & 27.2 \\
20 & 2007 Jul. 28  & ECDFS3 & 30.5 \\
21 & 2007 Jul. 29  & ECDFS4 & 27.1 \\
22 & 2007 Jul. 30  & ECDFS1 & 28.0 \\
23 & 2007 Aug. 02  & ECDFS6 & 27.0 \\
24 & 2007 Aug. 03  & ECDFS1 & 27.4 \\
25 & 2007 Aug. 04  & ECDFS2 & 26.7 \\
26 & 2007 Aug. 05  & ECDFS3 & 25.9 \\
27 & 2007 Aug. 06  & ECDFS4 & 29.1 \\
28 & 2007 Aug. 09  & ECDFS5 & 31.2 \\
29 & 2007 Aug. 10  & ECDFS6 & 30.0 \\
30 & 2007 Aug. 11  & ECDFS1 & 31.9 \\
31 & 2007 Aug. 13  & ECDFS2 & 31.3 \\
32 & 2007 Aug. 14  & ECDFS3 & 31.5 \\
33 & 2007 Aug. 16  & ECDFS4 & 30.8 \\
34 & 2007 Aug. 17  & ECDFS5 & 45.7 \\
35 & 2007 Aug. 18  & ECDFS6 & 28.6 \\
36 & 2007 Aug. 21  & ECDFS1 & 29.1 \\
37 & 2007 Aug. 23  & ECDFS2 & 30.9 \\
38 & 2007 Aug. 25  & ECDFS3 & 30.8 \\
39 & 2007 Aug. 26  & ECDFS4 & 29.8 \\
40 & 2007 Aug. 28  & ECDFS5 & 33.9 \\
41 & 2007 Aug. 31  & ECDFS6 & 29.3 \\
42 & 2007 Sept. 06 & ECDFS2 & 30.9 \\
43 & 2007 Sept. 07 & ECDFS3 & 29.3 \\
44 & 2007 Sept. 08 & ECDFS4 & 29.3 \\
45 & 2007 Sept. 09 & ECDFS5 & 28.9 \\
46 & 2007 Sept. 10 & ECDFS6 & 30.2 \\
47 & 2007 Sept. 11 & ECDFS1 & 30.4 \\
48 & 2007 Sept. 12 & ECDFS5 & 37.9 \\
49 & 2007 Sept. 23 & ECDFS2 & 31.6 
\enddata
\tablecomments{List of the 49 epochs. Each epoch consisted of a 5-hr
  track centered on 03:30 LST.}
\label{Tab:ListOfEpochs}
\end{deluxetable}

\section{Variability Analysis}\label{sec:variables}

The source catalog we used to investigate variability was taken from
the second data release \citep[DR2;][]{mfk+13} of \cite{mfk+08}. This catalog was
generated by combining all the data from Tables \ref{Tab:ListOfFields}
and \ref{Tab:ListOfEpochs} to make a single deep 34$^\prime\times
34^\prime$ image with a typical sensitivity of 7.4 $\mu$Jy.
\cite{mfk+13} identified sources using the {\it AIPS} task {\tt SAD}
down to 4$\sigma$, and then inspected the residual map to identify
missed sources as well as accepted sources which were poorly fit by
{\tt SAD}. These missing sources were then added to the preliminary
source list.  Further flagging and follow-up 
was done in order to produce a modified source list in
which all sources with peak flux density greater than 5 times the
local rms noise (i.e. 5$\sigma$), were fit using the {\it AIPS} task
{\tt JMFIT}.  Also, the effect of bandwidth smearing on sources within
the six individual pointings was assessed using {\tt JMFIT}, and the
resolution information was thus preserved in the output catalog.
Lastly, the sources in the DR2 catalog were compared with the catalog of 
\cite{kfm+08}. 

The DR2 catalog contains almost twice as many sources (883 vs. 464)
compared to the first data release \citep[see][]{mfk+13,mfk+08} owing to a
more careful data reduction. Of the 883 sources in the DR2 catalog we
created a point-source-only catalog of 736 objects used for exploring
variability. With this careful approach outlined above, we expect the
\citeauthor{mfk+08} DR2 catalog to contain all real sources above 
5$\sigma$ (however, see \S\ref{Sec:Algorithms}).

\begin{figure}[htp]
\centering
\includegraphics[width=3.2in,viewport=20 10 540 440,clip]{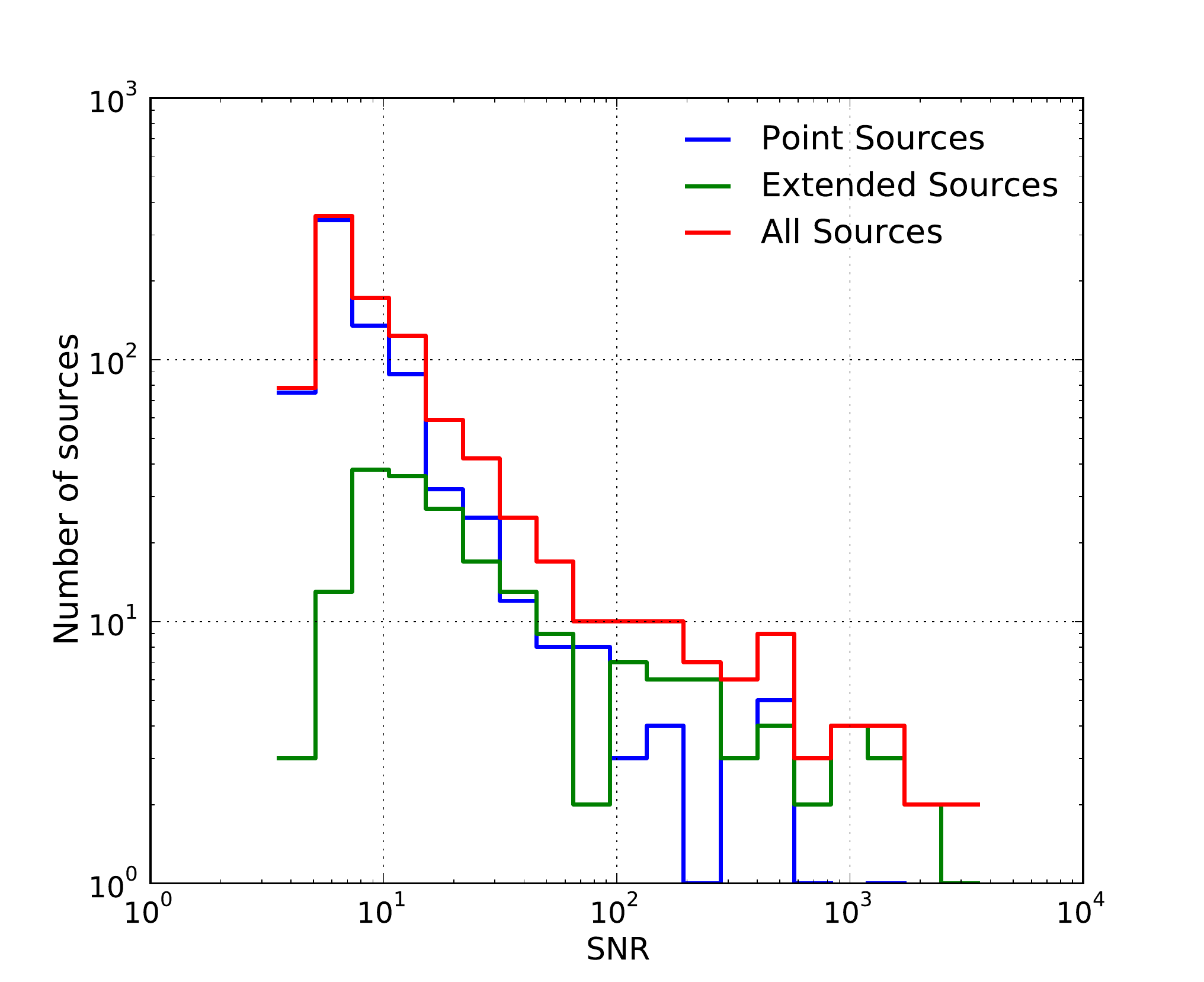}
\caption{A histogram of the signal-to-noise ratio (SNR) of the sources in the DR2 catalog.}
\label{Fig:DR2_SNR}
\end{figure}

We thus justify our use of the 
DR2 catalog for investigating the variability of the sub-milliJansky
population.  In \S\ref{Sec:Method} we use both the DR2 image and its
source catalog as a testbed for different source-finding algorithms.
The signal-to-noise ratio (SNR) of sources in the DR2 catalog is shown in
Figure~\ref{Fig:DR2_SNR}.

It is important to quantify the uncertainty in the peak flux density
when analyzing variability of sources in different epochs. Following
\cite{thwb+11}, we attribute this uncertainty primarily to a
combination of four\footnote{Far out in the primary beam, orthogonally polarized beams can be far offset on 
the sky, making amplitude calibration difficult. For short observations, the effects of this ``beam squint'' can be 
much larger than all other uncertainties combined. However, these offsets tend to average out when observations are 
made over sufficiently long times. Since the VLA's beam squint is oriented almost exactly East-West \citep{cp10}, 
our observational setup is optimal for averaging out the effects of beam squint.} causes: (i) local image rms noise, (ii)
uncertainty in the primary beam, (iii) flux density calibration
amplitude, (iv) pointing errors. Let us denote the measured peak flux
density of a source in epoch $i$ by $f_i$, such that the primary beam
correction is $b = b(\theta)$, where $\theta$ is the angular distance
of the source from the phase center. We wish to calculate the
uncertainty in the quantity $(f_i/b)$. The image local noise ($\Delta
f_i$) scales as $\Delta f_i/b$.  If we denote the fractional
uncertainty in the beam as $\epsilon_b$ and that in the flux
density calibration amplitude by $\epsilon_c$, then the corresponding
errors scales as $(f_i/b)\epsilon$.  We adopt a value of 4\% for 
$\epsilon_c$, intermediate between the conservative estimate of \citeauthor{thwb+11}, 5\%, and the one quoted by \cite{ofb+11}, 3\%.  The typical pointing error ($\Delta \theta$) of a VLA
antenna is between 10\arcsec--20\arcsec. The resultant uncertainty scales as 
$(f_i/b^2) (-db/d\theta) \Delta\theta$.  All these four
error terms, added in quadrature, would give the total uncertainty,
$\sigma_i$, in the peak flux density corrected for the primary beam
attenuation. However, as shown below, the pointing-related error term is much
smaller than the rest, and hence can be neglected.  Thus,

\begin{equation}
\sigma_i = \frac{1}{b} \sqrt{\Delta f_i^2 + f_i^2 (\epsilon_b^2 + \epsilon_c^2)}
\label{Eq:ferr}
\end{equation}


Polynomial coefficients (and the associated error) that express the
average angular dependence of $b$ can be found in the AIPS task
\texttt{PBCOR}, while measurements of the VLA beam power response to
beyond the first null are given in \citet{cp10}.  Using the beam
response profile from \citeauthor{cp10}, we can estimate the error
terms (i)--(iv) above for a typical source in the DR2 catalog, having
measured flux density of $300\pm 30~\mu$Jy.  If the source lies at the
half-power radius ($\theta = 15\arcmin$), then these correction
factors are about 10\%, 4\%, 4\% and 1\% of the primary-beam-corrected
flux density ($570~\mu$Jy) respectively. The pointing-related
uncertainty is thus negligible.

In light of \cite{ofb+11}, we use two measures of variability \citep[see
also][]{scheers11}, the modulation index defined as the standard
deviation divided by the mean,

\begin{equation}
m= \frac{1}{\bar{f}}\sqrt{\frac{1}{{\rm N}-1} \sum_{i=1}^{N}{(f_{i}-\bar{f})^{2}}}, 
\label{Eq:modi}
\end{equation}

\noindent and the $\chi^2$,
\begin{equation}
\chi^{2}= \sum_{i=1}^{N}\frac{(f_{i}-\bar{f})^{2}}{\sigma_{i}^{2}},
\label{Eq:chi2}
\end{equation}

\noindent where $N$ is the number of epochs, 
and $\bar{f}$ is the mean flux density of the source over 
all the epochs considered for the variability analysis. 
In these two equations, the primary beam correction is implicit in $f_{i}$ and $\bar{f}$.
$\chi^{2}$ gives a measure of the deviation from stochastic epoch-to-epoch fluctuations 
in the peak flux density, and we define ``significant variability'' beyond a level\footnote{For 
Gaussian noise, 4$\sigma$ corresponds to a probability of about 1/16,000, 
while the number of measurements in our variability analysis (several to tens of epochs 
multiplied by a few hundred sources) ranges from about 1,500 to 15,000.}
of 4$\sigma$.
The modulation index indicates the strength of variability, i.e. the fractional variation of the 
peak flux density.

\subsection{Single-Pointing Variability}\label{sec:single}

A variability analysis was carried out on each pointing in Table 
\ref{Tab:ListOfFields} separately. Peak flux densities were measured 
for all point sources in the DR2 catalog brighter than 40 $\mu$Jy and 
within a 15$\arcmin$ radius of the pointing centers (i.e. the 50\%
response radius of the primary beam of the VLA antennas; 
see also Figure~\ref{Fig:obsCoverage}). This
approach has the merit of being simple and robust. Since the angular
distance of a source from its pointing center is constant, the
accuracy of the correction for the primary beam attenuation $b(\theta)$ is
unimportant. The modulation index and $\chi^2$ measures of variability 
are insensitive to a constant $b(\theta)$.

The limitation of this approach is that the resulting light curves are
constructed for only 7--9 epochs. Many of our investigated point sources 
are found in multiple pointings and therefore light 
curves can be constructed with many more epochs, resulting in higher cadence
over the full 100 days of observing. A full variability analysis of
this kind is carried out in \S\ref{sec:full}.

\begin{figure*}[htp]
  \centering
\includegraphics[width=7.1in,viewport=5 0 890 555,clip]{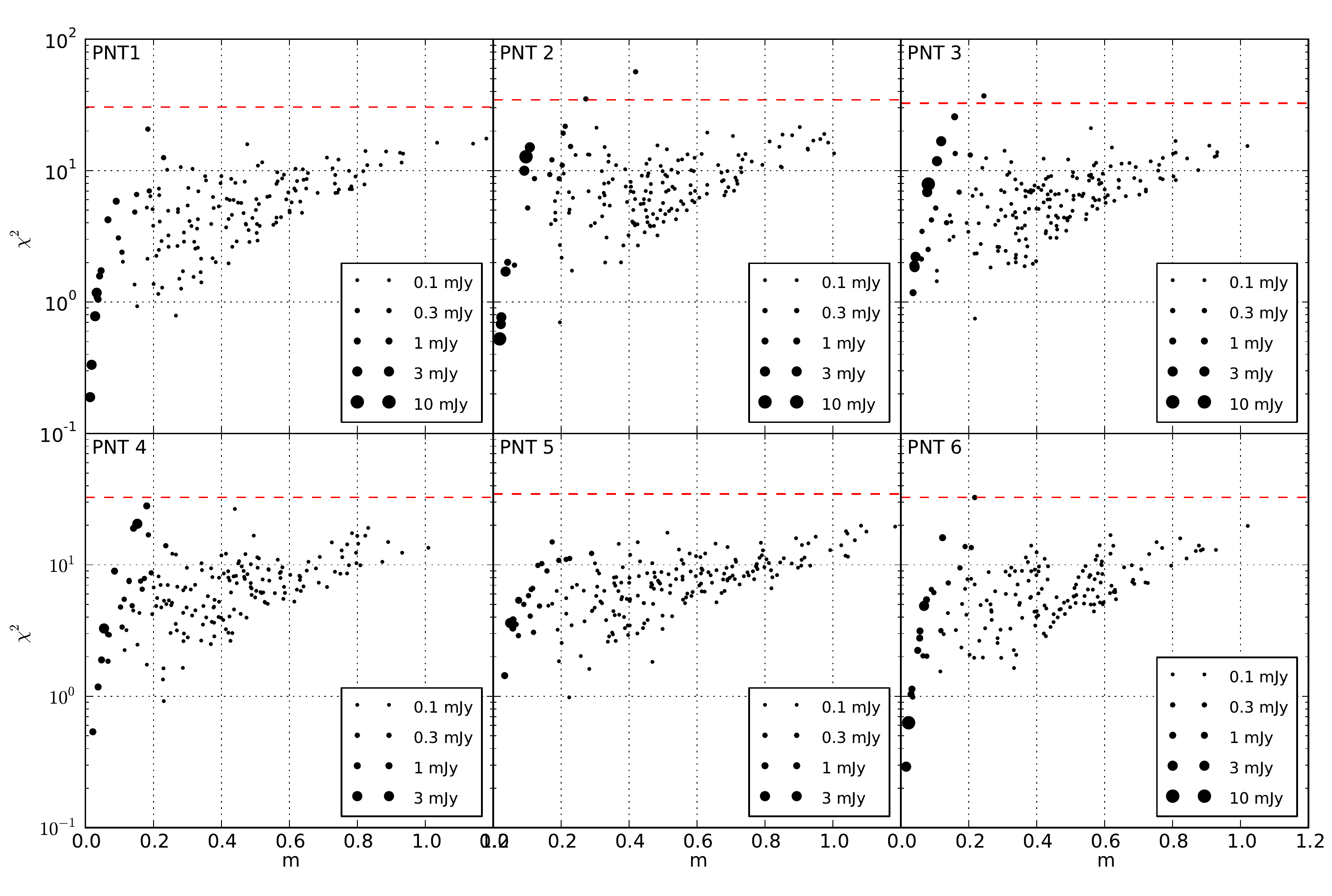}
\caption{Variability plot for the point sources in the \citeauthor{mfk+08} DR2
 catalog, shown separately for all pointings. The peak flux density is denoted by the 
symbol size. The red dashed line  represents the 4$\sigma$ level for the appropriate 
number of degrees of freedom (one less the number of epochs in each pointing)
 for each pointing. The number of epochs in each pointing lies between 7 and 9.}
\label{Fig:chi2VSm_singlePontings}
\end{figure*}

In Figure~\ref{Fig:chi2VSm_singlePontings} we show variability plots
(i.e. $\chi^2$ vs. $m$) for the DR2 sources in each of the six
pointings.  There are approximately 175 sources per pointing.  Some of
the bright ($>3$ mJy) sources show significant variability but with only
low modulation indices ($\sim$10\%).
We define strong variables as sources having $m>0.5$ (i.e. higher than
50\% fractional variability).  Only two genuine variables were
found in this single-pointing analysis; no strong variables were found
(Table~\ref{Tab:Variables}, upper panel).


\subsection{Full Variability Analysis}\label{sec:full}

In order to undertake variability analysis using data from all the
epochs, the single-epoch peak flux densities were required to be
corrected for (i) bandwidth-smearing, and (ii) the primary beam
response, as any given source will lie at different angular distance
from the phase center in different pointings.  We applied bandwidth
smearing correction from the approximation given in \cite{bs89}
(Eqn. 13-19 therein).

There were two choices for the primary beam profile --- one derived as
the empirical beam profile for VLA-FIRST \citep{thwb+11}, and the
other found in the AIPS task \texttt{PBCOR}. To test which of
these profiles best represented our data, we adopted the
following approach.  We first normalized the peak flux densities of
$>300~\mu$Jy sources from all epochs using their peak flux density from
the DR2 catalog and plotted them as a function of distance from the
pointing center.
The resultant beam profile matched with the VLA-FIRST profile better
than the one from {\tt PBCOR} (to within 1\%, but only for
$\theta<12\arcmin$; scatter of 6.5\%). Hence we used the former beam profile for our
primary beam correction, $b(\theta)$; the associated error
($\epsilon_b$), was also taken from \cite{thwb+11}.  
Thus, for a reliable all-epoch variability analysis, we restricted our 
search to the point sources in the DR2 catalog which were located
within $12\arcmin$ from the pointing centers of their respective 
epochs. This also appears to be the radius beyond which our bandwidth-
smearing approximation starts to break down. 
Thus, for example, a source located at $\alpha = 03^{\rm h}33^{\rm m}00^{\rm s}$ and $\delta = -28\arcdeg00\arcmin00\arcsec$ would be present in 
pointings 6, 5, and 1, but not in 2,3, and 4 because the separation between the source 
and the centers of pointings 2,3, and 4 is larger than 12 arcmin.
Further, as we did with
the single-pointing variability (\S\ref{sec:single}), we restricted
our analysis to sources whose mean flux density was brighter than 40 
$\mu$Jy. This full variability analysis was carried out on 599 point
sources.

\begin{figure}[htp]
\centering
\includegraphics[width=3.5in]{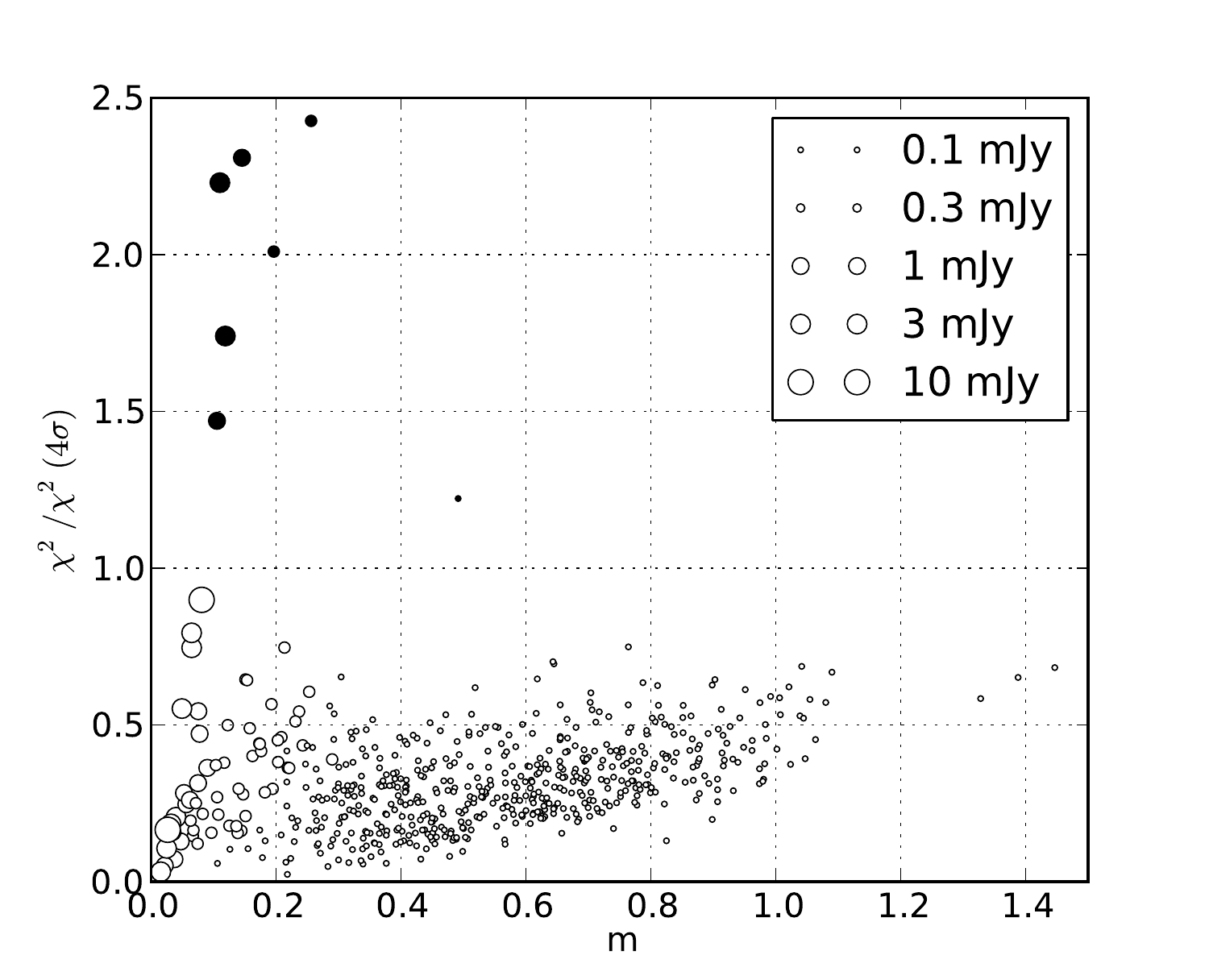}
\caption{$\chi^2$ normalized by its value at the 4$\sigma$ level, plotted against 
the modulation index, $m$, for sources in the \citeauthor{mfk+08} DR2 catalog using 
peak fluxes from all pointings, and corrected with empirically-derived beam attenuation profile.
The 4$\sigma$ level is different for different sources, depending on 
the number of epochs in which they are present. The mean peak flux density is denoted by the 
symbol size. Filled circles indicate significant variables (lying above a normalized $\chi^2$ of unity).
}
\label{Fig:chi2.imval_allepoch}
\end{figure}

Depending on the number of pointings in which a source is present,
this analysis allowed us to exploit the higher cadence over the entire
duration of the observing program.  The resulting light curves are now
more densely sampled with 15--26 epochs, rather than the 7--9 epochs for
the single-pointing variability case. In
Figure~\ref{Fig:chi2.imval_allepoch} we show variability plots for the
DR2 sources, taking into account all the epochs.

\begin{figure}[htp]
\centering
\includegraphics[width=3.5in]{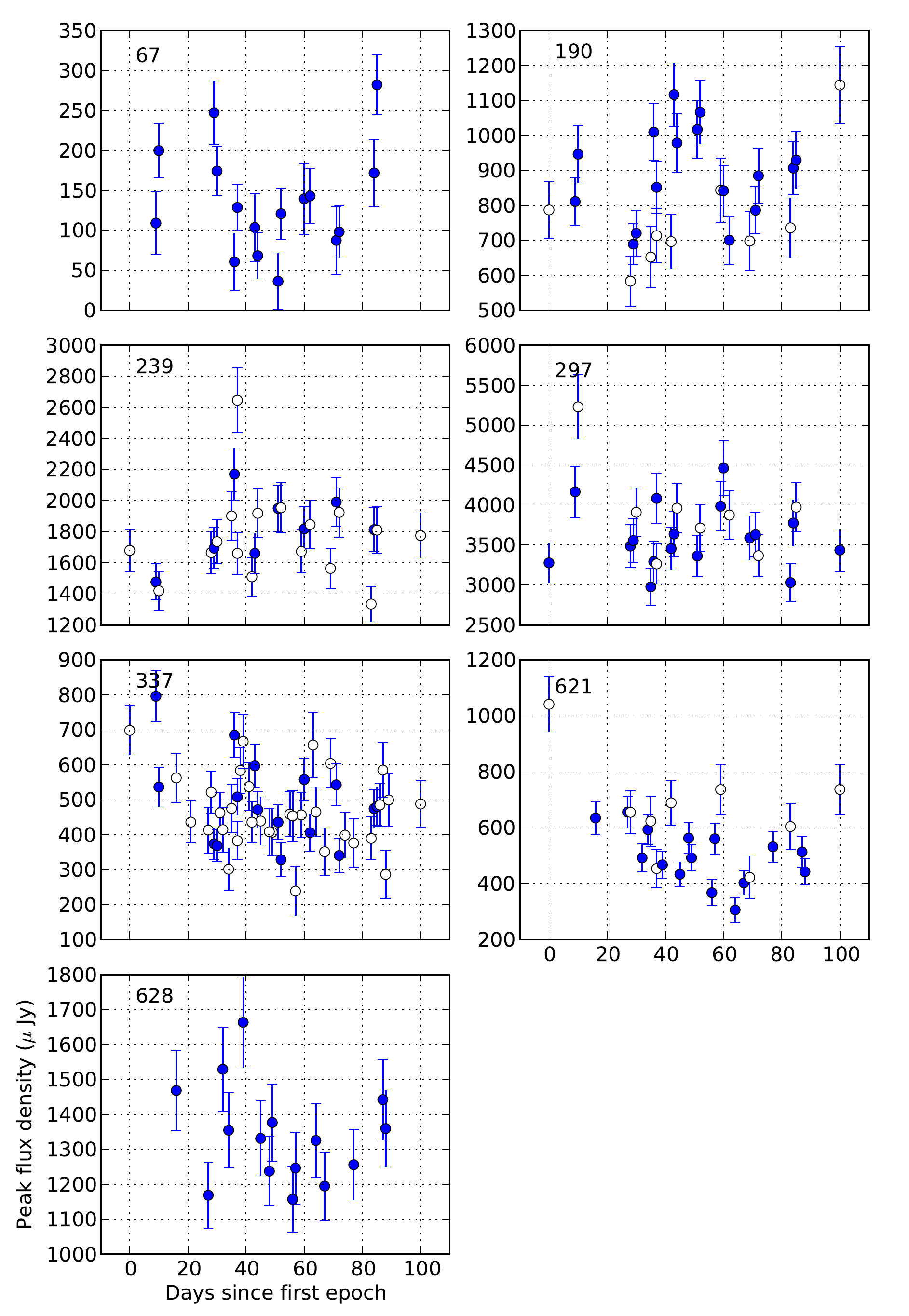}
\caption{Light curves of the variable sources from \citeauthor{mfk+08} DR2 catalog listed in Table~\ref{Tab:Variables}. 
Filled symbols represent the flux densities considered for variability analysis (i.e. where $\theta < 12\arcmin$). 
Flux densities in epochs where a source lies within the 50\% power circle of the beam (i.e. where $\theta \lesssim 15\arcmin$) are plotted for reference as open symbols.
The error bars take into account the background rms, primary beam correction, and bandwidth smearing (no taper).}
\label{Fig:lightCurves}
\end{figure}

Seven significant variables were found via this procedure, but no
strong variables ($m>0.5$). Both of the variables identified in the
single-pointing analysis in \S\ref{sec:single} are also seen here. The
results of the variability study are given in 
Table~\ref{Tab:Variables} and the light-curves for the significant
variables are shown in Figure~\ref{Fig:lightCurves}.  We can compare
our variability criteria with the \cite{cif03} measure for
variability, i.e.  V$_c=(S_1-S_2)/S$, where $S_1$ and $S_2$ are the
maximum and minimum flux densities respectively observed, and $S$ is
their mean. This measure of variability for the seven variable sources 
found in this work are listed in Table~\ref{Tab:Variables}.

\begin{table*}[htp]
\centering
\caption{Variables among \citeauthor{mfk+08} DR2 sources}
\label{Tab:Variables}
\begin{tabular}{r cc c ccr c ccc cr} 
\hline \hline
ID   & $\alpha_{J2000}$ & $\delta_{J2000}$            & $\overline{f}$ & $m$       & V$_c$ & $N_{\rm ep}$ & $z$  & log $\overline{L}_R$ & $\alpha_{\rm R}$ & $\alpha_{\rm IR}$ & $M_R$ & Energy \\
     & (h, m, s)        & (\arcdeg, \arcmin, \arcsec) & ($\mu$Jy)      &           & (\%)  &              &      & (cgs)                &                  &                   & (mag) & Source \\
\hline 
\multicolumn{13}{c}{Single Pointing Analysis}\\
\hline
337 & 03 32 18.03  & $-$27 47 18.8 & \,\,\,558 $\pm$ 14   & 0.26   & 76  & 8   \\
621 & 03 33 15.00  & $-$27 51 51.3 & \,\,\,663 $\pm$ 22   & 0.23   & 93  & 9   \\
\hline
\multicolumn{13}{c}{Full Variability Analysis}\\
\hline
67  & 03 31 27.07  & $-$27 44 09.9 & 136  $\pm$ 9         & 0.50    & 181 & 16 & 1.005 & 30.6 & $+0.23$ & $> -3.8$ & $-21.04$ & SF+AGN\\ 
190 & 03 31 52.13  & $-$27 39 26.6 & \,\,\,891  $\pm$ 12  & 0.15    & 48  & 16 & 2.296 & 32.1 & $-0.35$ & $> -1.3$ & $-22.07$ & AGN   \\ 
239 & 03 32 00.85  & $-$27 35 57.1 & 1822 $\pm$ 27        & 0.12    & 38  & 8  & 0.266 & 30.5 & $-0.30$ & $> -2.3$ & $-17.26$ & SF+AGN\\ 
297 & 03 32 11.66  & $-$27 37 26.3 & 3600 $\pm$ 36        & 0.11    & 41  & 17 & 0.605 & 31.5 & $+0.89$ &~~~$-0.9$ & $-23.98$ & AGN   \\ 
337 & 03 32 18.03  & $-$27 47 18.8 & \,\,\,494  $\pm$ 11  & 0.26    & 95  & 16 & 0.734 & 30.8 & $-0.08$ & $> -3.4$ & $-22.69$ & SF+AGN\\ 
621 & 03 33 15.00  & $-$27 51 51.3 & \,\,\,497  $\pm$ 10  & 0.20    & 70  & 15 & 1.107 & 31.2 & $+0.24$ & $> -3.4$ & $-20.71$ & SF+AGN\\ 
628 & 03 33 16.74  & $-$27 56 30.4 & 1341 $\pm$ 16        & 0.11    & 38  & 15 & 0.685 & 31.2 & $-0.40$ &~~~$-2.2$ & $-21.21$ & SF+AGN\\ 
\hline
\multicolumn{13}{p{6.5in}}{Notes$-$ (1) ID is the source-ID as given in the \citeauthor{mfk+08} DR2 catalog. 
(2) $\bar{f}$ refers to the mean flux density corrected for the primary beam and bandwidth smearing. 
(3) V$_c$ is the \cite{cif03} variability criteria as described in \S~\ref{sec:full}. 
(4) Redshift $z$ is according to \cite{bon12} or \cite{tvg+09}. The redshifts of ID 67, ID 239 and 
ID 297 as per the COMBO-17 survey catalog \citep{wolf2004} are 0.548, 0.947 and 1.574 respectively.
(4) The 1.4 GHz spectral luminosity (erg s$^{-1}$ Hz$^{-1}$) is $\overline{L}_R = 4\pi d_l^2\bar{f}/(1 + z)$, 
where the luminosity distance $d_l$ assumes cosmological parameters from \cite{komatsu2011}.
(5) The spectral indices ($S \propto \nu^\alpha$) between 1.4 GHz and 5.5 GHz and between 24~$\mu$m and 70~$\mu$m 
are tabulated as $\alpha_{\rm R}$ and $\alpha_{\rm IR}$ respectively. 
In the absence of a 70~$\mu$m counterpart, a $3\sigma$ upper limit to the flux density at this wavelength is considered.
Note that the 1.4 GHz and 5.5 GHz measurements are non-simultaneous.
(6) The absolute R-band magnitude $M_R$ has been calculated using the redshift $z$ and the 
apparent magnitude from the NASA/IPAC Extragalactic Database.
(7) The energy sources within the galaxy as indicated by the radio and mid-to-far-infrared properties are listed 
in the last column. SF: star-formation, AGN: active galactic nucleus (see \S\ref{sec:VariablesNotes}).
}

\end{tabular}
\end{table*}

In order to undertake multiwavelength identifications we had to align the reference 
frames of all the data sets. 
The radio and optical source positions were brought to the
same reference frame by calculating the radio-source position offsets with respect to
{\it Hubble} Space Telescope source positions from GOODS-S
\citep{giavalisco2004} and the GEMS \citep{rix2004}. Optical
counterparts (from these two HST catalogs) were searched toward 
radio sources within 1\arcsec. Only point-like sources having a single counterpart
were chosen.  A histogram of the offsets of these counterparts in
right ascension ($\Delta \alpha = \alpha_{\rm radio} - \alpha_{\rm
  optical}$) and declination ($\Delta \delta = \delta_{\rm radio} -
\delta_{\rm optical}$) was then computed to find the most-likely
offset (peak of the histogram). The associated error was taken to be 
the standard deviation about this most-likely offset added in 
quadrature with the cataloged mean positional error of the radio source.  
In addition to the \cite{mfk+13} DR2 catalog,
we repeated this procedure for other radio-source catalogs in the
E-CDFS region, viz. \cite{kfm+08}, \cite{naa+06}, and \cite{amk+06}
using $\geqslant$5$\sigma$ sources. The computed radio vs. optical positional
offsets along with the mean positional errors listed in the respective
radio catalogs are shown in Table~\ref{Tab:Offsets}.

\begin{table}[htp]
\centering
\tiny{
\caption{Radio Positional Offsets wrt. Optical HST Catalogs}
\label{Tab:Offsets}
\begin{tabular}{p{0.6cm}c p{1.3cm}p{1.5cm}p{1.3cm}p{1.5cm}} 
\hline \hline
                       &  arcsec                 & (1)               & (2)                &  (3)             & (4) \\
\hline
\multirow{2}{*}{}      & $\sigma_{\alpha}$       & $\sim0.1$         & 0.37               & 0.31             & $^b$ \\
                       & $\sigma_{\delta}$       & $\sim0.1$         & 0.57               & 0.58             & $^b$ \\
\hline
\multirow{2}{*}{GEMS}  & $ \Delta \alpha $       & $~~~0.18\pm0.31$  & $~~0.15\pm0.45$    & $-0.09\pm0.41$   & $-0.17\pm0.40^a$\\
                       & $ \Delta \delta $       & $-0.32\pm0.32$    & $-0.34\pm0.53$     & $-0.20\pm0.44$   & $~~0.11\pm0.30^a$\\
\hline
\multirow{2}{*}{GOODS} & $ \Delta \alpha $       & $-0.20\pm0.23$    & $~~0.15\pm0.42^a$  & $0.03\pm0.44^a$  & $-0.10\pm0.39^a$\\
                       & $ \Delta \delta $       & $~~~0.22\pm0.28$  & $-0.18\pm0.37^a$   & $0.03\pm0.44^a$  & $-0.18\pm0.33^a$\\
\hline
\multicolumn{6}{l}{Column headers$-$ (1) \cite{mfk+13}, (2) \cite{kfm+08}}\\
\multicolumn{6}{l}{\qquad\qquad\qquad\qquad\, (3) \cite{naa+06}, (4) \cite{amk+06}}\\
\multicolumn{6}{l}{Notes$-$ (a) Few ($\leqslant 15$) sources available to calculate the offsets.}\\
\multicolumn{6}{l}{\qquad\quad\,\, (b) Positional uncertainty not mentioned in catalog; assumed to be 0.1\arcsec.}\\
\multicolumn{6}{l}{\qquad\quad\,\, (c) All offsets are in arcseconds.}\\
\end{tabular}
}
\end{table}


\begin{figure*}[htp]
\centering
\includegraphics[width=1.5in,height=1.9in,viewport=55 90 575 675,clip]{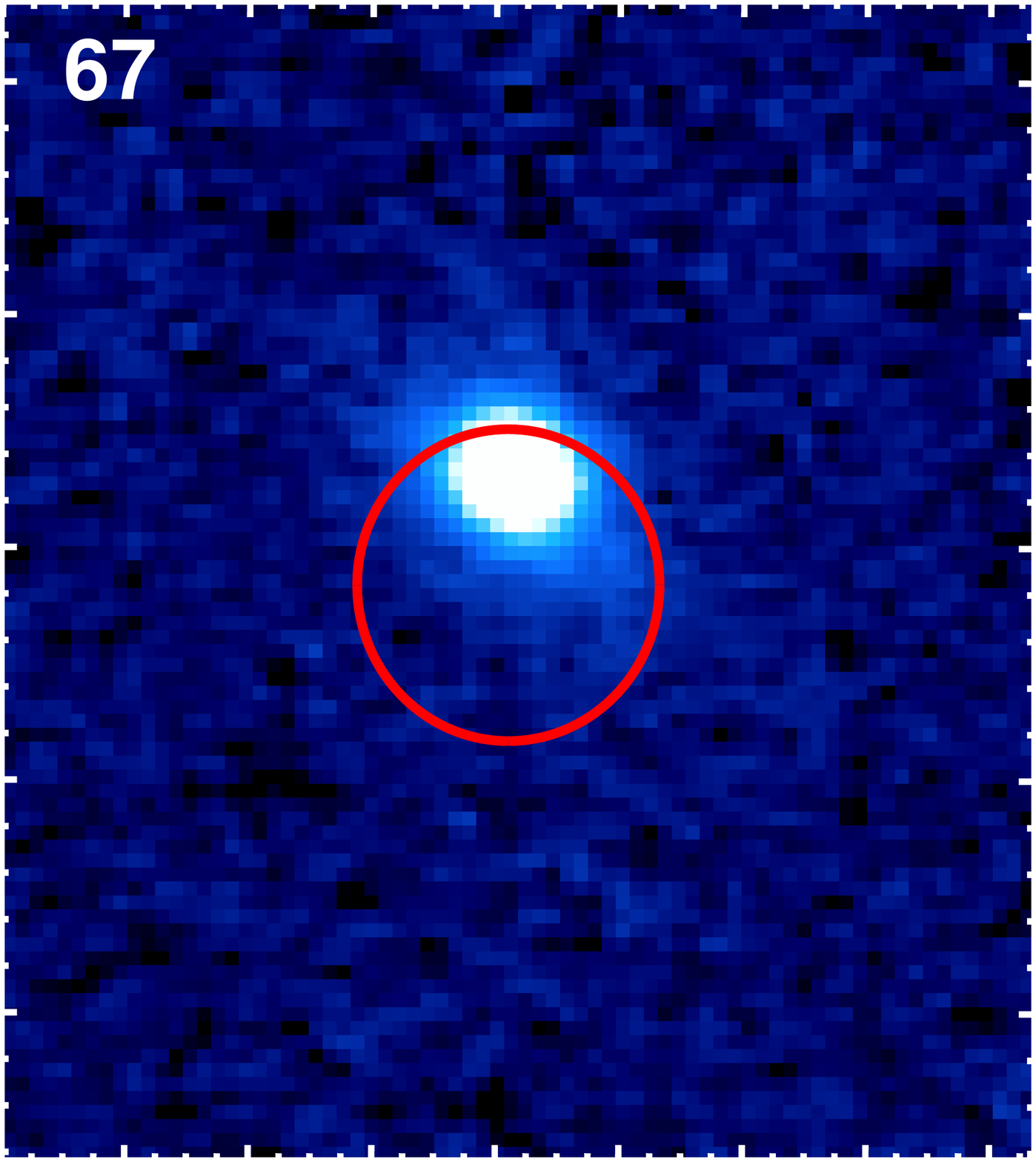}  
\includegraphics[width=1.5in,height=1.9in,viewport=55 85 575 675,clip]{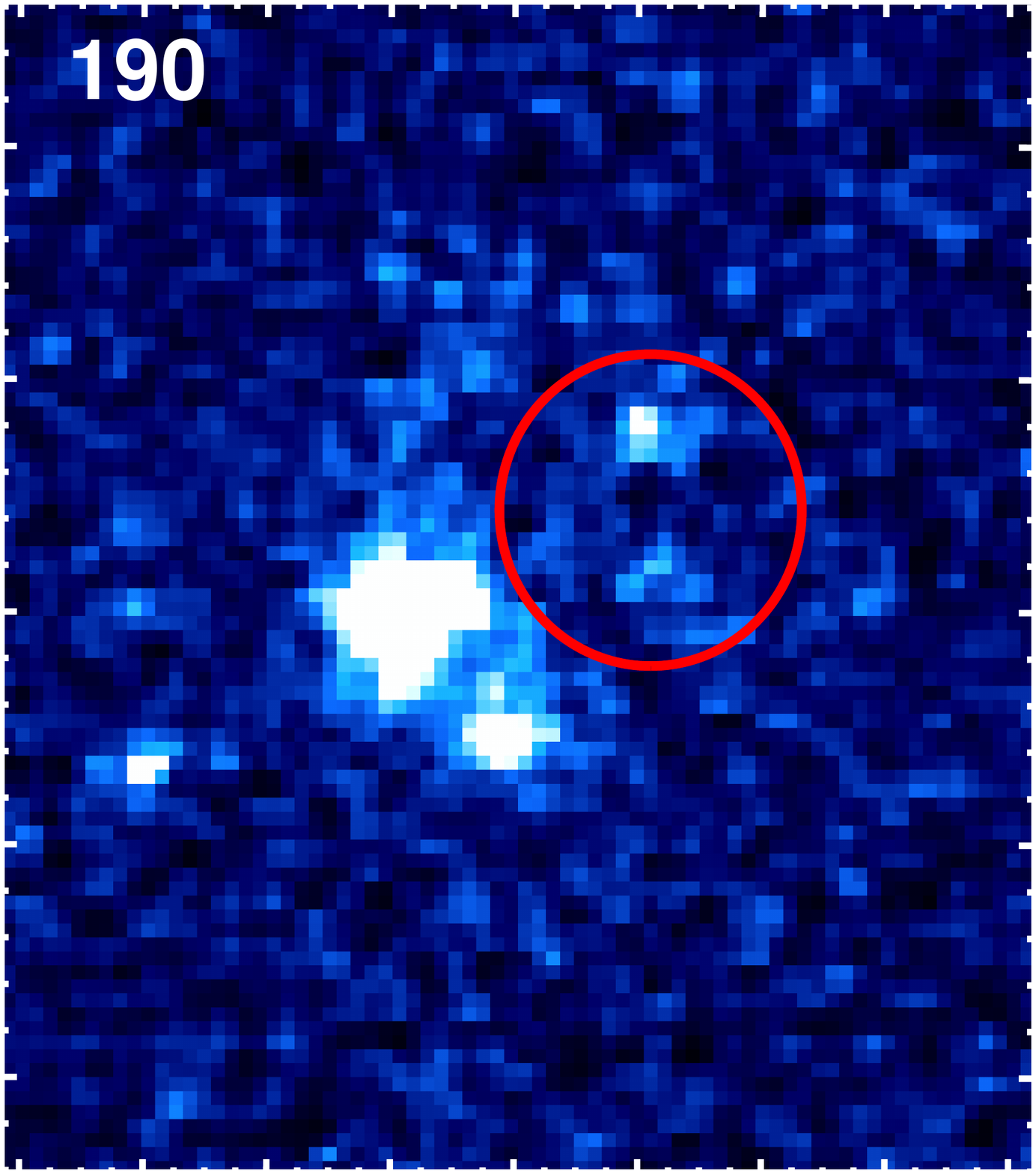} 
\includegraphics[width=1.5in,height=1.9in,viewport=55 85 575 675,clip]{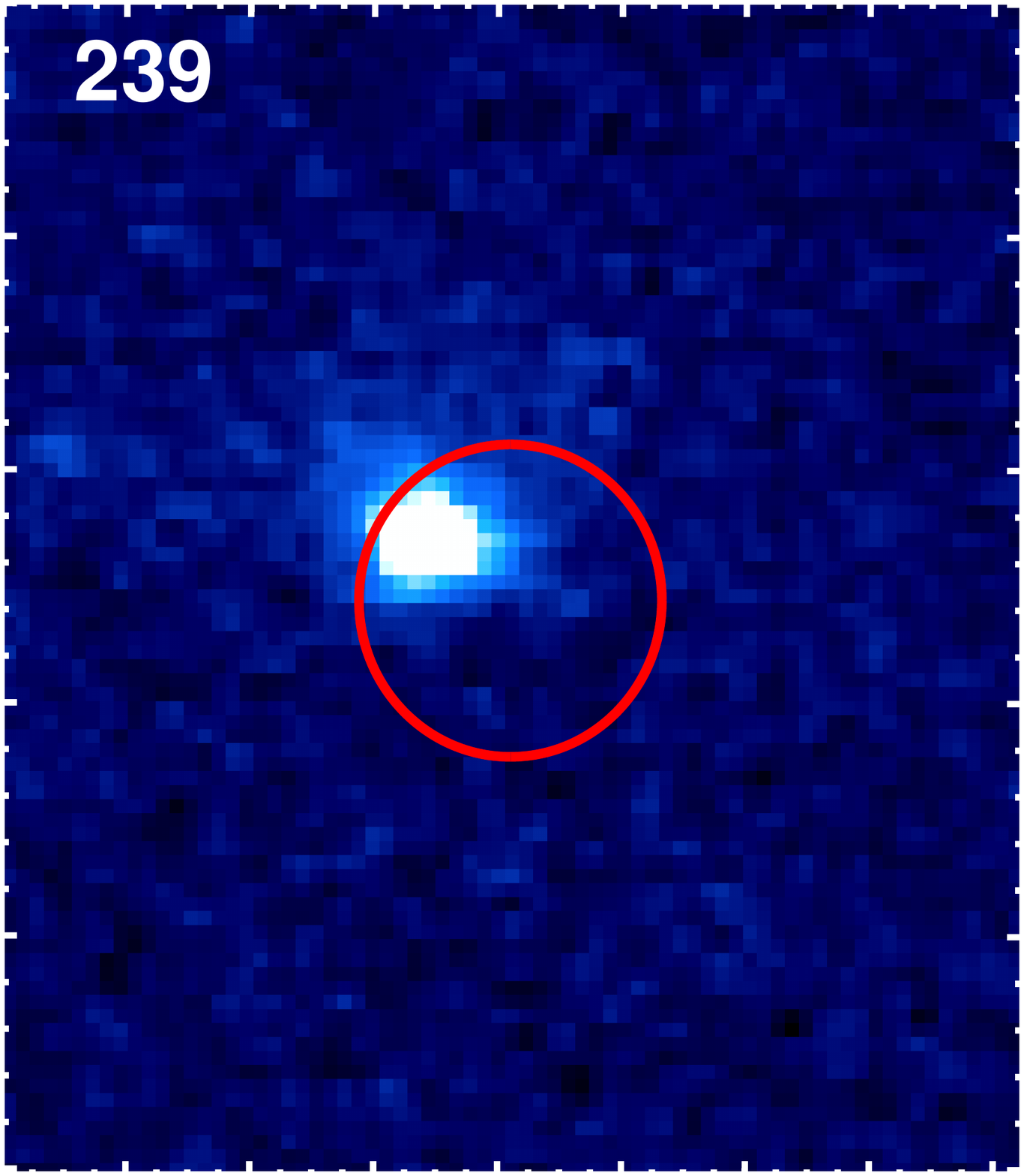} 
\includegraphics[width=1.5in,height=1.9in,viewport=55 85 575 675,clip]{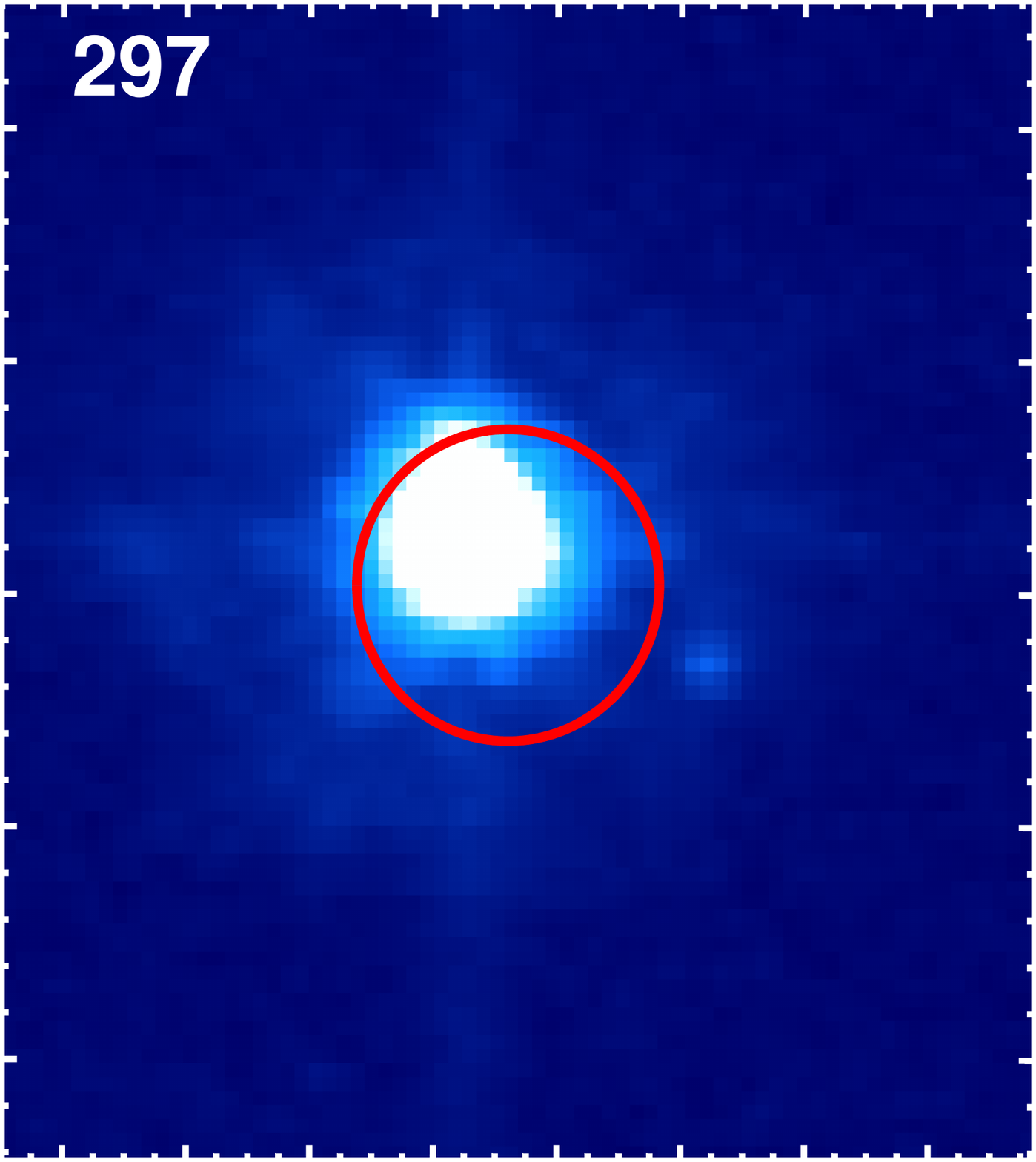} 
\includegraphics[width=1.5in,height=1.9in,viewport=50 85 575 680,clip]{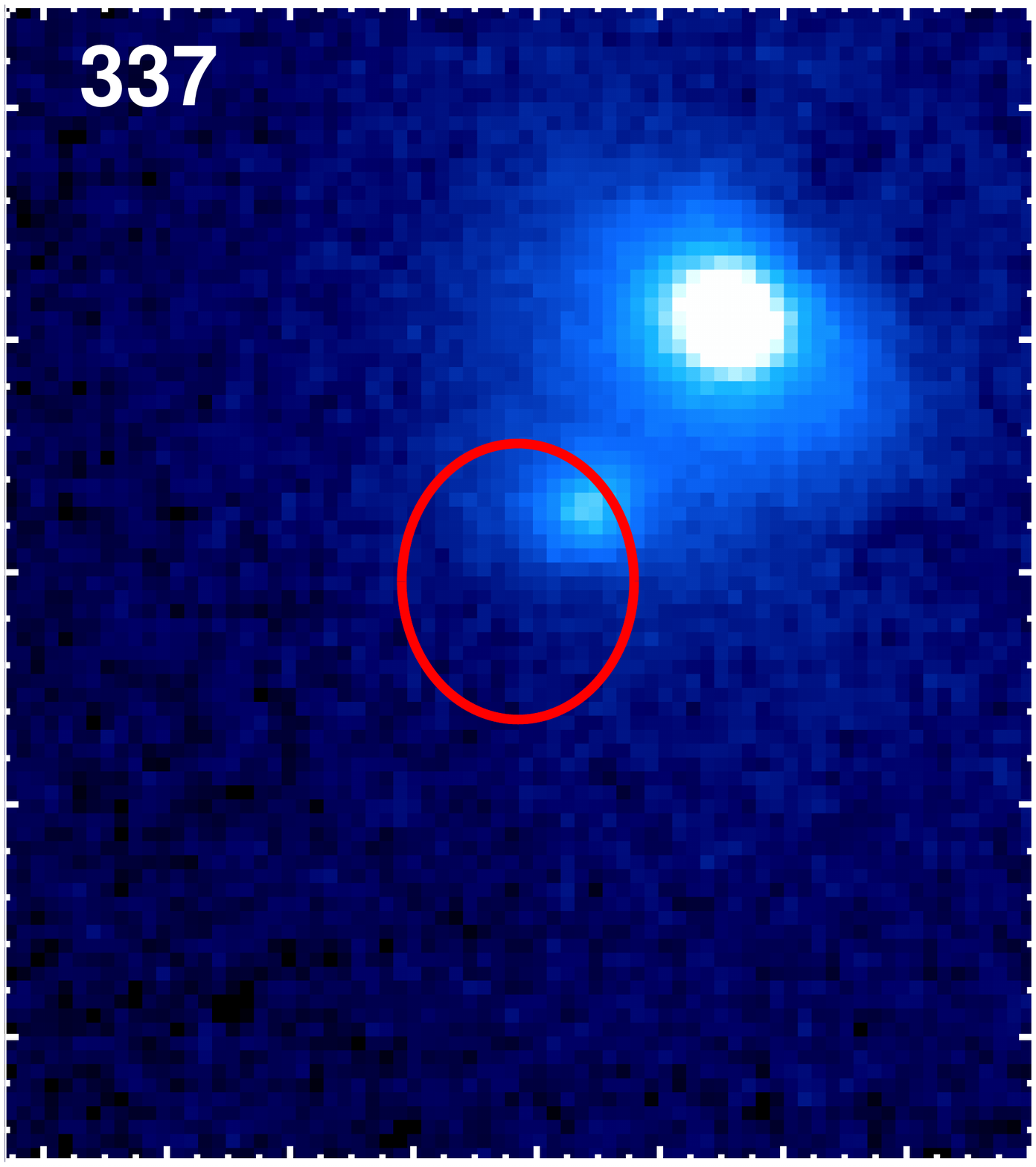} 
\includegraphics[width=1.5in,height=1.9in,viewport=50 85 575 675,clip]{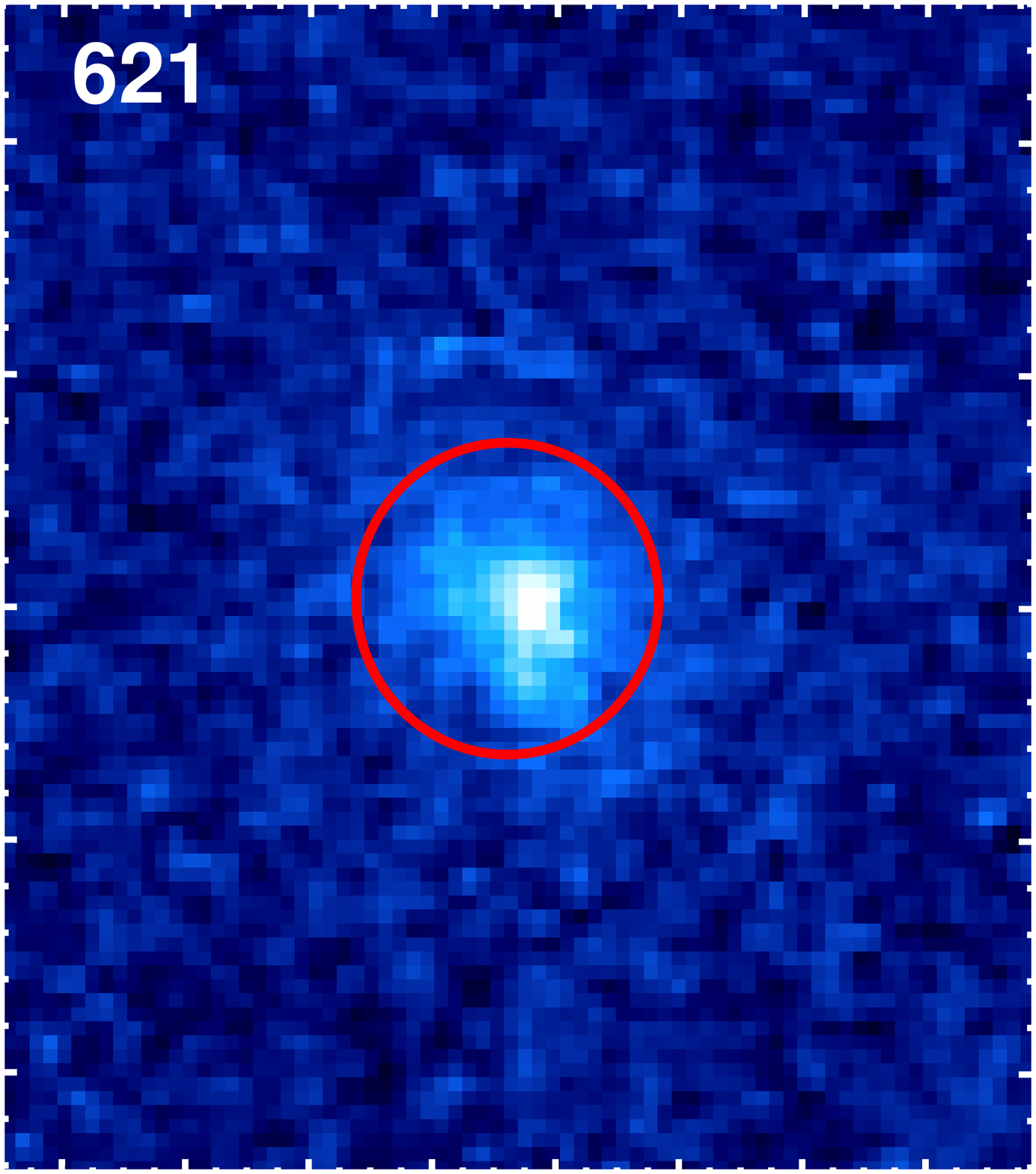} 
\includegraphics[width=1.5in,height=1.9in,viewport=55 85 575 675,clip]{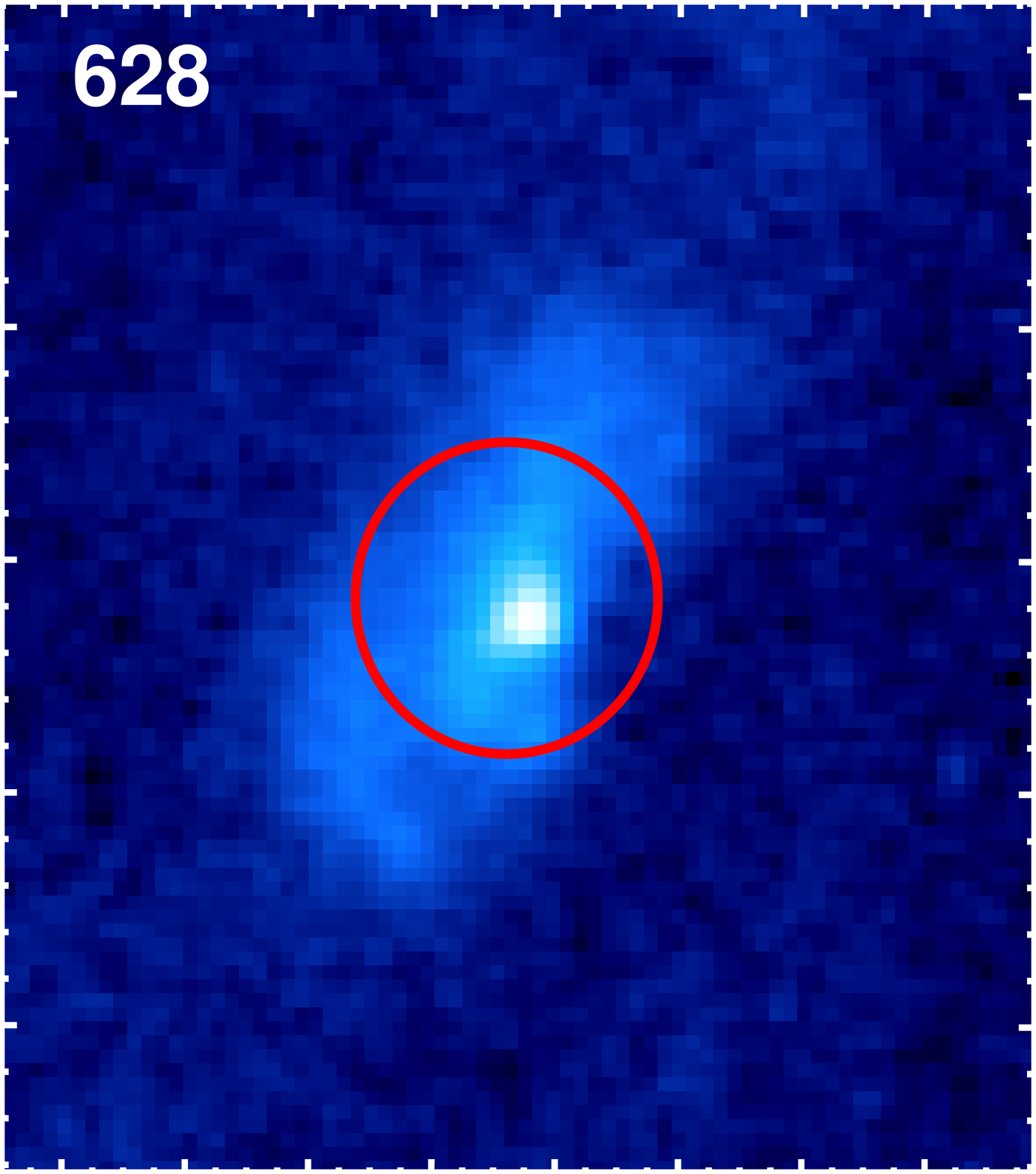} 
\caption{$2.2\arcsec \times 2.5\arcsec$ GEMS (all sources except ID 337) and GOODS-S (ID 337) F606W ACS-WFC image cutouts of variable sources found in this work.
The red error ellipses denote the radio source positions from \cite{mfk+13}, shift-corrected to the HST source positions.
All the positional uncertainties are 1$\sigma$ (see \S~\ref{sec:full} and Table~\ref{Tab:Offsets}).}
\label{Fig:cutouts}
\end{figure*}

\quad \newline 
\subsection{Notes on Variables Found}\label{sec:VariablesNotes}

An important question that we can address is whether
the variability at sub-milliJansky levels is dominated by
normal star-forming galaxies or by AGN. For AGN-dominated
samples above 1 mJy, variability at frequencies of a few GHz or below is thought to be
dominated by propagation effects (i.e. refractive interstellar
scintillation) and not by intrinsic changes in the source \citep{gh00,of11}. 
For a disk galaxy we expect there to be steady emission
from diffuse synchrotron emission and the sum of all supernovae.
Intrinsic variability can be induced by (i) a stellar explosion
(supernova, low-luminosity gamma-ray burst), and (ii) nuclear radio emission
(AGN).  Mapping the radio emission to the center of the galaxy would
favor (ii) whereas if the radio emission is mapped to the disk then
(i) is favored.

\begin{figure}[htp]
\centering
\includegraphics[width=3.3in,viewport=10 10 550 420,clip]{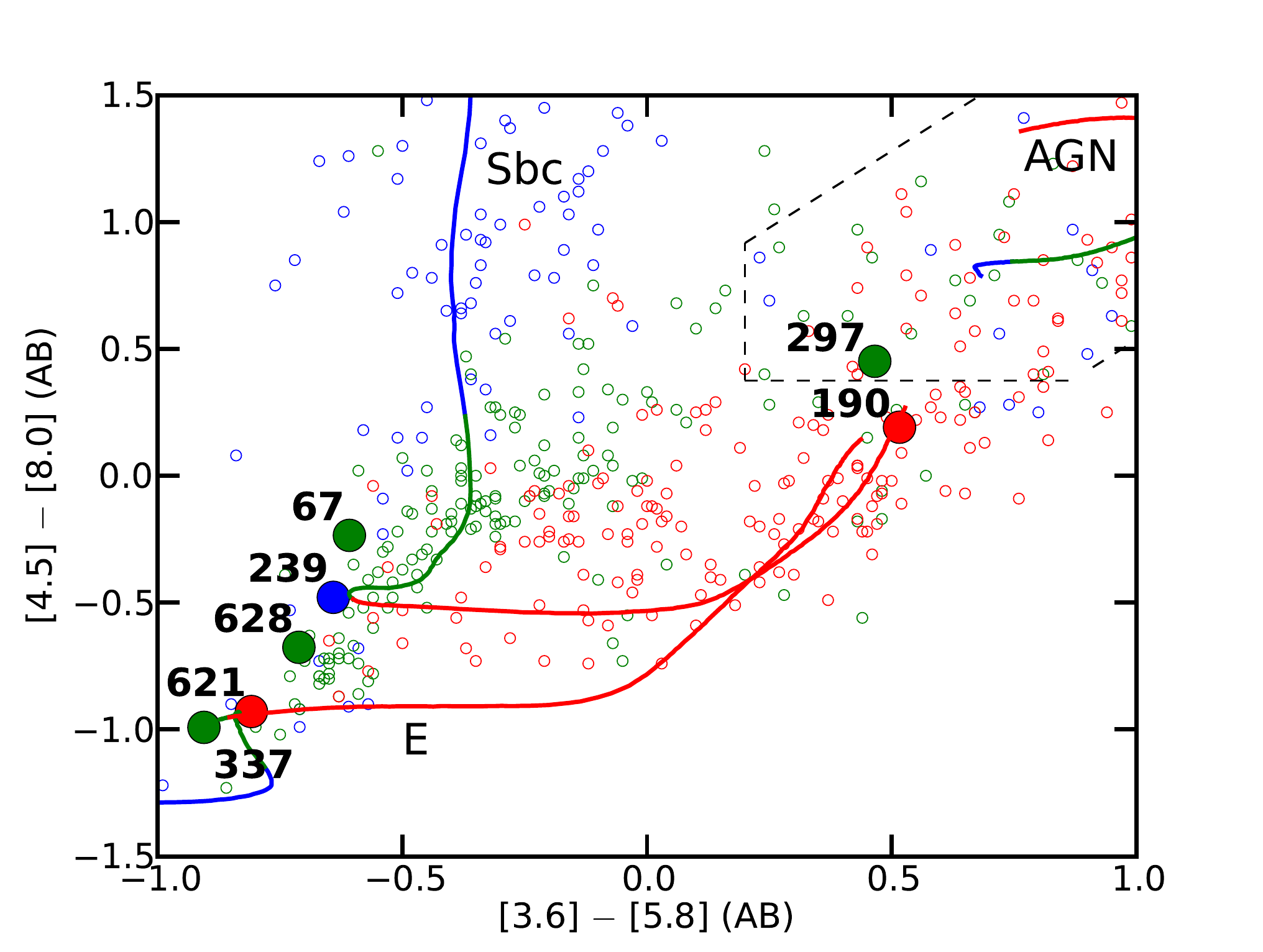}
\caption{{\it Spitzer}/IRAC color-color diagram for the mid-infrared counterparts of the variable sources (filled circles).
For comparison, the radio sources from AEGIS20 \citep{willner2012} are also shown (unfilled circles).
The AGN selection region in the upper-right corner, bounded by the dashed lines, is from \cite{donley2012}.
Curves represent the colors of three template spectral energy distributions (E: elliptical galaxy; Sbc: spiral galaxy; AGN) 
from \cite{assef2010} as redshift increases from 0 to 3.
The redshift-dependent color-coding is --- blue: $z \leqslant 0.5$, green: $0.5 < z \leqslant 1.1$ and red: $z > 1.1$.
}
\label{Fig:IRACColors}
\end{figure}

To this end, we overplotted radio source positions of the seven
variables given in Table~\ref{Tab:Variables} on HST 
image cutouts from GOODS-S and the GEMS projects (Figure~\ref{Fig:cutouts}).  All
of the radio sources have an optical counterpart on these HST images.
The redshifts of these objects vary from 0.3 to 2.3. Thus, the post-offset 
radio source-position uncertainties lie between 1.4 and 2.7 kpc.
Within most error ellipses, there is a centrally compact source, suggesting that this is the
source of the variable emission.
Light-curves of the variable sources are given in Figure~\ref{Fig:lightCurves}.  Detailed
notes on each of these objects are given below, and key physical parameters 
are listed in Table~\ref{Tab:Variables}. 
All of the variable sources have luminosities in excess of 10$^{30}$ erg s$^{-1}$ Hz$^{-1}$, 
where the luminosity functions of AGNs and star-forming galaxies intersect \citep{condon2002}.
Most of the variable sources have a roughly flat spectrum between 1.4 GHz and 5.5 GHz suggesting
the presence of AGN.
Note that the spectral indices between 1.4 GHz and 5.5 GHz are based 
on non-simultaneous measurements having different resolutions.
Additionally, mid-infrared colors, far-infrared spectral indices, and mid-to-far-infrared  
luminosities indicate that most of the variable sources reside in star-forming galaxies.
In Figure~\ref{Fig:IRACColors} we show the mid-infrared color-color diagram for the variable 
sources using {\it Spitzer}/IRAC photometry from the SIMPLE survey catalog \citep{damen2011}.
The \cite{donley2012} AGN selection region is overplotted.
For comparison, the colors of the radio sources from the AEGIS20 sample \citep{willner2012} are shown 
along with those of three template spectral energy distributions (SEDs) --- an elliptical galaxy, 
an Sbc galaxy, and an AGN --- from \cite{assef2010}.
This figure along with the far-infrared flux densities identify two variables with AGNs 
and the rest as star-forming galaxies.

Summarizing, the high-resolution optical images together with photometric information 
from radio and mid-to-far-infrared suggest that variability arises from the central regions 
of an AGN or star-forming galaxy.

{\bf ID 67:}
This source has a mean 1.4 GHz flux density of $136 \pm 9~\mu$Jy and shows
the strongest variation among all seven variables. Two flaring bursts are
evident from the light curve (Figure~\ref{Fig:lightCurves}), one of
them lasting for about 25 days, and the flux density of the other
increasing by at least a factor of 2.5 in 12 days. This source also
appears in the \citeauthor{kfm+08} 1.4 GHz catalog, where its flux density is 
$90 \pm 16~\mu$Jy. The photometric redshift of z=1.005
\citep{bon12} implies a mean radio luminosity of log $L_R = 30.6$ erg s$^{-1}$ Hz$^{-1}$. The 
radio spectral index (defined as $\alpha$ where 
S$_\nu\propto\nu^\alpha$) between 1.4 GHz and 5.5 GHz is $+0.23$
\citep{hhl+12}. The morphology of the host galaxy as seen from the HST
image appears to be that of a bright, compact nucleus surrounded by
faint extended structure. The half-light radius of the galaxy according to 
the {\it GALFIT} parameters from the GEMS catalog is $620 \pm 10$ pc.
The K-corrected\footnote{For the $5.8~\mu$m and $24~\mu$m 
luminosities, the K-correction has been applied based on the 8~$\mu$m and 
70~$\mu$m flux densities. In the absence of a 70~$\mu$m detection, the 
$3\sigma$ upper limit has been used.} 
mid-to-far-infrared spectral luminosities derived from the 
FIDEL and GOODS \citep{magnelli2009, magnelli2011}, and SIMPLE {\it Spitzer} 
surveys are $\nu L_\nu(5.8~\mu {\rm m}) = 6.6 \times 10^8 ~\rm L_\sun$ and 
$\nu L_\nu(24~\mu{\rm m}) < 5.4 \times 10^{10} ~\rm L_\sun$. By comparing these 
quantities with the \cite{chary2001} template SEDs (see Figure 4 of that paper) and 
from the {\it Spitzer}/IRAC color-color diagram (Figure~\ref{Fig:IRACColors}), 
we interpret that the host galaxy is star-forming.

Taken together, the radio luminosity, radio spectral index, optical
morphology and mid-to-far-infrared flux densities argue that ID 67 is a 
star-forming galaxy harboring a low-luminosity AGN.

{\bf ID 190:}
This source has a mean 1.4 GHz flux density of $891 \pm 12~\mu$Jy. The light
curve appears to fluctuate between high and low flux density states on
a timescale of order 20 days. This source is also found in the
\citeauthor{kfm+08} and \citeauthor{naa+06} catalogs where the flux
density is $970 \pm 30$ and $810 \pm 19~\mu$Jy respectively,
and a spectral index of $-0.6$ between 1.4 and 4.8 GHz. The 5.5 GHz flux
density from \citeauthor{hhl+12} is $555 \pm 17~\mu$Jy, for which
we derive a spectral index of $-0.35$.  The photometric redshift of
z=2.296 \citep{bon12} implies a mean radio luminosity of log $L_R =
32.1$ (cgs). The HST image of this galaxy as seen in 
Figure~\ref{Fig:cutouts} shows that the radio source is offset from
the brightest emission in the field. On the basis of its departure from 
the radio-FIR correlation, \citeauthor{naa+06} classify this source as an
AGN. The radio and far-infrared spectral indices (Table~\ref{Tab:Variables}) 
and mid-infrared colors (Figure~\ref{Fig:IRACColors}) 
are consistent with this identification. Note that between 24~$\mu$m and 
70~$\mu$m, a spectral index greater than $-1.5$ is representative 
of AGNs \citep[e.g.][]{condon2002}.


{\bf ID 239:}
This source has a mean 1.4 GHz flux density of $1822 \pm 27~\mu$Jy. This 
source is also found in the \citeauthor{kfm+08} and 
\citeauthor{naa+06} catalogs where the flux density is $2030 \pm 43$ 
and $1640 \pm 20~\mu$Jy respectively. Its light-curve shows a
gradual increase in flux density, over a period of about 50 days,
followed by an equally gradual decline.  The radio spectral index
between 1.4 GHz and 5.5 GHz is $-0.30$. This slope is more consistent
with AGNs than star-forming galaxies which are expected to have $\alpha
\simeq -0.8$. The redshift z=0.266 \citep{tvg+09} implies a mean radio
luminosity of log $L_R = 30.5$ (cgs). The HST cutout (Figure~\ref{Fig:cutouts}) reveals just a
bright nucleus surrounded by faint extended structure, the half-light
radius being $3.2 \pm 0.5$ kpc.
The K-corrected mid-to-far-infrared spectral luminosities, 
$\nu L_\nu(5.8~\mu {\rm m}) = 3.6 \times 10^8 ~\rm L_\sun$ and 
$\nu L_\nu(24~\mu{\rm m}) < 1.0 \times 10^{9} ~\rm L_\sun$, and the mid-infrared 
colors (Figure~\ref{Fig:IRACColors}) show that the host is a 
star-forming galaxy.

Taken together, the radio luminosity, radio spectral index, optical 
morphology and mid-to-far-infrared flux densities argue that ID 239 is a 
star-forming galaxy harboring a low-luminosity AGN.

{\bf ID 297:}
This source has a mean 1.4 GHz flux density of $3.60 \pm 0.04$ mJy.
\citet{hhl+12} measure a 5.5 GHz flux density of 12.25 mJy, implying a
spectral index of $+0.89$. One the basis of this steep positive spectral
index, \citet{hhl+12} suggest that this is part of a class of
Gigahertz Peaked Spectrum (GPS) sources, thought to be a young AGN.  The
redshift z=0.605 \citep{tvg+09} implies a mean radio luminosity of log
$L_R = 31.5$ (cgs). ID 297 appears to be a stochastically varying source
with no specific trend in its light-curve
(Figure~\ref{Fig:lightCurves}).  The radio position is consistent with
a bright, unresolved HST source. On the basis of departure from radio-FIR 
correlation \citeauthor{naa+06} classify this source as an AGN. 
The mid-infrared colors (Figure~\ref{Fig:IRACColors}) and far-infrared 
spectral index (Table~\ref{Tab:Variables}) are consistent with this identification.


{\bf ID 337:}
This source has a mean 1.4 GHz flux density of $494 \pm 11~\mu$Jy. From its
radio light-curve, ID 337 appears to have repeated outbursts roughly every 
25d.  The flux density of this source in the 
\citeauthor{mfk+13}, \citeauthor{kfm+08}, \citeauthor{naa+06} and
\citeauthor{amk+06} catalogs is $439 \pm 8$, $524\pm14$,
$380\pm16$, and $404\pm34~\mu$Jy respectively, and a spectral index of $-0.2$
between 1.4 and 4.8 GHz.  The 5.5 GHz flux density from
\citeauthor{hhl+12} is $443 \pm 20~\mu$Jy, for which we derive a
spectral index of $-0.08$.  The photometric redshift of z=0.734
\citep{vcd+08} implies a mean radio luminosity of log $L_R = 30.8$ (cgs).

The HST image (Figure~\ref{Fig:cutouts}) shows that the optical
counterpart to the radio source is actually associated with the
fainter (or more extincted) galaxy among a group of two closely separated 
galaxies. This has lead to some confusion over the correct optical
identification. VLA observations carried out in 1999--2001 and 2007 
(\citeauthor{kfm+08} and \citeauthor{mfk+13} respectively) suggest that the radio 
counterpart is the fainter galaxy. ATCA observations carried out in
2003--2004 (\citeauthor{naa+06} and \citeauthor{amk+06}) suggest the bright
galaxy as the optical counterpart. Our optical-radio frame tie
summarized in Table~\ref{Tab:Offsets} supports the fainter optical
source as the likely radio counterpart. The spectral index and radio
luminosity argue that ID 337 is a low-luminosity AGN.
\citeauthor{amk+06} state that the bright radio source is a 
luminous star-forming galaxy, possibly part of a merging system.
Comparison of the K-corrected spectral luminosities, 
$\nu L_\nu(5.8~\mu {\rm m}) = 4.0 \times 10^9 ~\rm L_\sun$ and 
$\nu L_\nu(24~\mu {\rm m}) < 1.1 \times 10^{10} ~\rm L_\sun$, with 
the \cite{chary2001} SED templates and \cite{desai2007}, together 
with the mid-infrared colors (Figure~\ref{Fig:IRACColors}) advocate the 
star-forming nature of the host galaxy.

{\bf ID 621:}
This source has a mean 1.4 GHz flux density of $497 \pm 10~\mu$Jy. The first
epoch in its light-curve reveals an initial brightening of the source
followed by a decline and subsequent small-amplitude variations.  The
maximum flux density is close to 400\% of the quiescent flux density
of about 300$~\mu$Jy. The \citeauthor{mfk+08}, \citeauthor{kfm+08}, and
\citeauthor{naa+06} catalogs list the flux density of this source as
$494 \pm 10$, $565\pm17$, $450\pm18~\mu$Jy respectively, and a spectral
index of $-0.1$ between 1.4 and 4.8 GHz. The 5.5 GHz flux density from
\citeauthor{hhl+12} is $689 \pm 16~\mu$Jy, for which we derive a
spectral index of $0.24$.  The photometric redshift of z=1.107
\citep{bon12} implies a mean radio luminosity of log $L_R = 31.2$ (cgs).

The HST image reveals a bright nucleus surrounded by diffuse 
emission. The half-light radius is $\sim$3 kpc. 
The radio position is consistent with the nuclear source. 
Taken together, the radio luminosity, spectral index and optical 
morphology argue that ID 621 is an AGN. Additionally, the K-corrected 5.8~$\mu$m 
luminosity of $5.0 \times 10^9 ~\rm L_\sun$, the 24~$\mu$m 
luminosity upper limit of $6.8 \times 10^{10} ~\rm L_\sun$, and the 
mid-infrared colors (Figure~\ref{Fig:IRACColors}) suggest that this 
is also a star-forming galaxy.

{\bf ID 628:} 
This source has a mean 1.4 GHz flux density of $1.34 \pm 0.02$ mJy. The
light-curve indicates a steady increase of the quiescent emission to
maximum flux density, followed by a steady decline. The 
\citeauthor{mfk+08}, \citeauthor{kfm+08}, and \citeauthor{naa+06}
catalogs list the flux density of this source as $1.07 \pm 0.02$,
$1.33\pm0.03$, and $0.90\pm0.02~\mu$Jy. The 5.5 GHz flux density from
\citeauthor{hhl+12} is $0.78 \pm 0.02~\mu$Jy, for which we derive a
spectral index of $-0.40$. The photometric redshift of z=0.685
\citep{naa+06} implies a mean radio luminosity of log $L_R = 31.2$ (cgs).

The HST image reveals a bright nucleus surrounded by disk-like diffuse 
emission, the half-light radius being $\sim$10 kpc.
The radio position is consistent with the nuclear source. 
On the basis of departure from radio-FIR correlation \citeauthor{naa+06} classify this source as
an AGN, which is consistent with our flat radio spectral index and radio luminosity.
Additionally, the K-corrected 5.8~$\mu$m 
luminosity of $3.0 \times 10^9 ~\rm L_\sun$, the 24~$\mu$m 
luminosity of $3.7 \times 10^{10} ~\rm L_\sun$, and the 
mid-infrared colors (Figure~\ref{Fig:IRACColors}) suggest that this 
is a normal star-forming or a starburst galaxy.
Taken together, the radio and mid-to-far-infrared properties along with the optical 
morphology argue that ID 628 is an AGN embedded within a star-forming galaxy.


\section{Transient Search}\label{Sec:Method}

For our transient search, we are interested in identifying those point
sources which show up above the flux density limit for a short amount
of time (corresponding to one or more epochs depending on the
cadence), and remain below the limit in all other epochs. It is
therefore important to reliably distinguish noise from real transients
(which will determine how well we can reject false positives), and to
avoid the rejection of transients as noise (reducing the number of
true negatives). Thus, characterizing the effectiveness of
source-finding algorithms in terms of reliability and completeness is
crucial. Another motivation for characterizing source-finding
algorithms is to find the optimum parameter values for use in transient
searches.

In the following subsections we begin by the testing the efficacy of
existing source-finding algorithms (\S \ref{Sec:Algorithms}).
We then apply some of the better-performing algorithms on the E-CDFS
epochs (\S \ref{Sec:AlgorithmsTransients}).

\subsection{Efficacy of Source-finding Algorithms}\label{Sec:Algorithms}

Recently, quantitative tests have been carried out on the reliability
and completeness of source-finding algorithms \citep{hhn+11,hmg+12}.
Of these many publicly available software packages, {\tt sfind} and
{\tt IMSAD} in {\it MIRIAD}, {\it Aegean}, {\it SExtractor}, and {\it
  Selavy}, these studies found that {\tt sfind} and {\it Aegean}
produce the most reliable catalogs.

The analysis of \citet{hhn+11} and \citet{hmg+12} was carried out on two 
simulated data sets. (i) The ASKAP simulation is a $4\arcdeg \times 4\arcdeg$ image of 
a full continuum observation with critically-sampled beams in the 6-km ASKAP 
configuration. Its pixel-scale is 2.75$^{\prime\prime}$ and the rms noise is about 35 mJy, which varies 
across the field. 16 idealized beams one degree 
apart, spaced in a rectangular grid, mimic the effect of the 
phased-array feed. The image contains $\sim$7.7 million sources having flux 
densities greater than 1 $\mu$Jy from the S$^3$-semi-empirical extragalactic 
simulation \citep{wilman2010}.
(ii) For the Hancock et al. simulation, a sky image was created as a 4801$^2$ pixel image
8$^\circ$ across with a 6$^{\prime\prime}$ pixels sampling a
30$^{\prime\prime}$ beam and an rms noise of 25 $\mu$Jy. Sources were
injected at random positions with angular sizes (with random position
angles) from 0$^{\prime\prime}$ to 52$^{\prime\prime}$ and with source
number counts distributed with peak flux densities as N(S)$\propto
{\rm S}^{-2.3}$ such that 15,000 sources having fluxes densities $>1\sigma$
are present in the image.

With our E-CDFS dataset we are able to carry out a similar analysis
using real data with all its attendant residual 
calibration and imaging errors. A comparison of real and simulated
data could be informative. While simulations are useful in determining
which source-finding algorithm work best in general, they do not
explore the parameter space of the algorithm thoroughly.  Hence they
may not provide optimum parameter values for a transient search on a
specific dataset. The deep field of the E-CDFS is well suited for
this comparison. The 4096$^2$ pixel image is 34$^\prime$ across with a
0.5$^{\prime\prime}$ pixels sampling a synthesized beam of
2.8$^{\prime\prime}\times 1.6^{\prime\prime}$ (position angle$\simeq
0^\circ$) and an rms noise 7.4 $\mu$Jy. As outlined in
\S\ref{sec:variables}, great care was taken in constructing the DR2
source catalog so we can be assured of its completeness and
reliability (see also Figure \ref{Fig:DR2_SNR}).


In what follows, we will use the DR2 catalog and the deep E-CDFS image to 
test various source-finding algorithms for completeness and 
reliability for different input parameters. We use the terms ``real
sources'' and ``false sources'' as being those sources present in the DR2
catalog, and those that are not, respectively\footnote{Even though the DR2 catalog was constructed with great care, 
it is likely that $\sim$1 genuine source was missed and a handful of 
spurious sources added (perhaps not truly spurious, but {\tt SAD} sources 
at 4$\sigma$ bumped up to 5$\sigma$ by the {\tt JMFIT} task in 
 {\it AIPS}). Recall that according to Eddington bias, more sub-5$\sigma$ 
sources get bumped up than 5$\sigma$ get bumped down.}. Whether a 
source detected by a source-finding algorithm has a counterpart in the 
DR2 catalog is determined by searching for DR2 sources within 1\arcsec\ from the source position.
Following \cite{hhn+11}, we define two additional terms,
``completeness'', as the fraction of real sources detected by a
source-finding algorithm, and ``reliability'', as the fraction of
detected sources which are real.
Note that in Figures~\ref{Fig:Completeness_algorithms},~\ref{Fig:Reliability_algorithms} and 
~\ref{Fig:HancockTests}, we plot the completeness and reliability within contiguous SNR bins, 
unlike \cite{hhn+11} and \cite{hmg+12} where, for a given SNR, the plots represent the 
completeness and reliability for sources greater than or equal to that SNR.
For source-finding algorithms employing a probabilistic approach of drawing a pixel from the
background and thus calculating false-detection rate (FDR), FDR + Reliability = 100\%.


\subsubsection{{\tt sfind} ({\it MIRIAD})}\label{Sec:sfind_efficacy}
In its default mode, {\tt sfind}\footnote{www.atnf.csiro.au/computing/software/miriad/doc/sfind.html} incorporates a 
statistically robust method for detecting source pixels, called ``False Discovery Rate'', or FDR.
In the FDR algorithm, detected sources are drawn from a distribution of pixels with a robustly known chance of being falsely drawn from the
background. Contiguous, monotonically decreasing adjacent pixels from the FDR-selected ones, are used for fitting 2-D elliptical Gaussians 
to the sources.            
Thus, the fraction of expected false sources is more reliably determined than in sigma-clipping criteria (see methods below).
Details of the FDR method can be found in \cite{hopkins2001}.
The run-time for {\tt sfind} searching for sources down to 5$\sigma$ in the DR2 image is about 20 seconds.
For all the tests carried out on {\tt sfind}, MIRIAD version 4.2.3 (optimized for CARMA; CVS Revision 1.11, 2011/04/26) was used.

Here, we explore the completeness and reliability of {\tt sfind} by tweaking the two relevant parameters: 
(i) $alpha ~(\alpha)$, the percentage of probable background pixels that can be accepted in the analysis, and 
(ii) $rmsbox$, the size of the smoothing box used for estimating the background and the standard deviation of the image.
Table~\ref{Tab:sfind_completeness} lists the parameter values tested.

In general, completeness of {\tt sfind} increases, and its reliability decreases (Table~\ref{Tab:sfind_completeness}) with increasing $\alpha$, as expected for an FDR algorithm.
Also, both these quantities increase with $rmsbox$.
The rise in completeness and reliability is precipitous (between 5\% and 20\%) as $rmsbox$ 
is increased from 5 to 10 beamwidths, after which it flattens off.
There is a slight decrease in the reliability as $rmsbox$ is increased from 20 to 50 beamwidths.
10 beamwidths can then be interpreted as the minimum box size for determining the background rms 
noise without significant contribution from the sources themselves, 
whereas 50 beamwidths would correspond to the size where the calculated rms starts deviating significantly from the true local rms.
However, in the case of bright sources in the field, these $rmsbox$ limits might be somewhat larger.
Completeness and reliability are $\sim$88\% for $\alpha$=10 and $rmsbox$=20, but 
better reliability (by a few percent) can be obtained at an equal expense of completeness by using $\alpha$=5 or 2.
Further decrease in $\alpha$ substantially decreases the completeness without any significant improvement in the reliability.
Thus, we determine the optimum values for the input parameters to be $5 \leqslant \alpha \leqslant 10$ and 
$10 \leqslant rmsbox \leqslant 50$ beamwidths.
For the FDR algorithm, we expect the reliability to be $1-\alpha$.
However, {\tt sfind} reliability is lesser than this expected value by a few percent, due to the 
acceptance of sidelobes of bright sources and fitting of extended sources with multiple elongated 
and overlapping (unphysical) components.
To some extent, the acceptance of these false sources can be reduced by choosing a lower value of $\alpha$.
Some examples of the components fit by {\tt sfind} to different kinds of sources in different environments in the DR2 image are shown in Figure~\ref{Fig:cutouts}.
In general, compared to other source-finding algorithms excepting {\tt SAD}, {\tt sfind} has a better reliability for a given completeness, 
and vice-versa (Tables~\ref{Tab:sfind_completeness}--\ref{Tab:Aegean_completeness}; see also \S\ref{Sec:SAD_efficacy} for a note on the efficacy of {\tt SAD}).
These findings are consistent with the tests carried out on simulated images \citep{hhn+11}.
In the ASKAP simulation, the reliability of the {\tt sfind} catalog is rather flat with respect to $\alpha$, whereas, 
for the \citeauthor{hmg+12} simulation it decreases by several percent as $\alpha$ increases from 0.1 to 10 \citep{hhn+11}.
In the latter, $rmsbox$=20 is found to give slightly better reliability than $rmsbox$=20.
For the $\sim$3--10$\sigma$ sources in these simulations, the completeness for $\alpha$=5 is greater than 
that for $\alpha$=0.1 by 5--10\%. They do not explore the completeness for $\alpha$=10 and $rmsbox$ other than 10 beamwidths.

Figure~\ref{Fig:Completeness_algorithms} shows the completeness of the {\tt sfind} catalog as a function of the SNR of the detected sources for different values of the input parameter $\alpha$.
The reduction in completeness beyond SNR$\sim$20 is due to missed or badly-fit components of extended sources.
The completeness for optimum values of input parameters is shown in the upper panel of Figure~\ref{Fig:Reliability_algorithms}, which 
shows that {\tt sfind} reaches 100\% completeness at a much lower SNR than other algorithms.
The lower panel of Figure~\ref{Fig:Reliability_algorithms} shows the reliability. Curiously, there is a dip in the reliability between SNRs of 6 and 10.
Point sources strewn across the DR2 image but primarily located near its edges (in regions of increased rms; see the lower panel of Figure~\ref{Fig:source_cutouts} for example), 
which are absent in the DR2 catalog and detected by {\tt sfind}, are responsible for this reduced reliability.
Some of these sources get rejected from the {\tt sfind} catalog when a lower value of $\alpha$ or $rmsbox$ is used, 
indicating that they are either not genuine sources on the sky or are sources at a lower SNR.
Indeed, other algorithms find some of these sources to be at a much reduced SNR between 3 and 5.
The results of \citep{hmg+12} also show a dip in the {\tt sfind} reliability, although centered on SNR$\sim$15.
Figure~\ref{Fig:SNR_offset_algorithms} compares the SNR of sources detected by {\tt sfind}, with those of the counterparts from the DR2 catalog.
The largely increased SNR reported for sources at or below 6$\sigma$ in the DR2 catalog, and largely 
reduced SNR for sources above $\sim$100$\sigma$, is anomalous.
Although for SNR above 100 the discrepancy is likely to be due to overlapping components fit to extended sources, 
all the above observations indicate that the SNR reported by {\tt sfind} is somewhat different from that reported by other algorithms.
Sources with 6$<$SNR$<$10 in the {\tt sfind} catalog clearly have over-estimatated (upto 100\%) peak flux densities and different dimensions 
than what would be expected by inspecting the cutouts of these sources.
Thus, fitting of sources rather than differently-reported rms is the cause of the SNR descrepancy, at least at these low SNRs.
The solution of this issue is possibly in correctly setting the {\it fdrpeak} and {\it psfsize} input parameters which 
allow reasonable measurements of sources close to the threshold.
Since there are not many sources in our data having SNR $\gtrsim$70 (Figure~\ref{Fig:DR2_SNR}), the suggested trend in completeness, 
reliability and measured SNR in this domain should be treated with caution.

\subsubsection{{\it SExtractor}}\label{Sec:SExtractor_efficacy}
{\it SExtractor}\footnote{www.astromatic.net/software/sextractor} is a source-finding program widely used in optical astronomy, and is 
particularly oriented towards the reduction of large-scale galaxy-survey data, as well as 
sparsely- to moderately-crowded stellar fields. It analyzes the image in two passes such that, in the first pass, 
a background map is made, and in the second, background-subtraction, filtering and thresholding is done on-the-fly.
Detected sources are then deblended and CLEANed before performing photometry.
The run-time for {\it SExtractor} searching for sources down to 5$\sigma$ in the DR2 image is about 2 seconds.
For all the tests carried out here, the latest release of {\it SExtractor} (version 2.8.6, 09-Apr-2009) has been used.

We adopted a strategy of searching for sources down to N$\sigma$ 
(N=3,5,6,7,10, as set using the {\small $DETECT\_THRESH$} 
parameter in {\it SExtractor}) and then selecting the ones $\geqslant$5$\sigma$ 
prior to the comparison with the DR2 sources.
Two input parameters (apart from possibly {\small $DEBLEND\_NTHRESH$},
{\small $BACK\_FILTTHRESH$} and {\small $CLEAN\_PARAM$}, which we have not tested) are
expected to affect the completeness and reliability of the {\it SExtractor} catalog.
First, the mesh-size, which determines the size of the box used for background rms estimation, is specified by {\small $BACK\_SIZE$}, and 
second, the size of the median filter ({\small $BACK\_FILTERSIZE$}) applied to the background 
grid used for smoothing large artifacts in the image.

In Table~\ref{Tab:SExtractor_completeness}, we list the completeness and reliability as a function of these parameters.
Note that in this table, in order to reflect the true completeness for the specified detection threshold, the completeness values of the {\it SExtractor} catalog for 
6$\sigma$, 7$\sigma$ and 10$\sigma$ thresholds have been normalized by the percentage of sources in the DR2 catalog which are beyond these thresholds respectively.
Thus, the completeness at any threshold is expected to be 100\% if all the DR2 sources beyond that threshold are detected by {\it SExtractor}.
The general trend observed is that, with increasing detection threshold, the completeness of {\it SExtractor} is fairly steady, 
but its reliability increases. However, for 5$\sigma$ and 6$\sigma$ detection thresholds, the completeness is rather low ($\sim$75\% or lower).
Choosing higher detection thresholds such as 7--10$\sigma$ does not seem to improve the completeness substantially.
This is due several factors through which the source-finding in {\it SExtractor} seems to be different from the conventionally-used algorithms in radio astronomy.
Firstly, for any given source, the reported peak flux density decreases as the detection threshold is raised, although there is not much of a change in the reported rms.
Hence, the SNR of sources close to the detection threshold decreases so as to be rejected by {\it SExtractor}.
Choosing a 3$\sigma$ threshold usually gives the correct peak flux density.
This is also the reason why searching down to 3$\sigma$ and selecting sources at the desired, higher threshold, increases the completeness (although with the side effect of reduced reliability).
The explanation for such an effect is hinted by the fact that the dimensions of the fitted sources decrease (this effect is quite significant for sources with SNR$<$10) with 
increasing detection threshold, implying that the number of source pixels considered in the fitting process depends on the threshold.
Secondly, several DR2 sources having SNR$<$10 and detected by other source-finding algorithms (excepting {\tt IMSAD}), are not detected at all by {\it SExtractor}, not even at a reduced SNR.
This might be due the differences in the fitting process rather than discrepant rms.
Lastly, for some extended sources, positions of the components reported by {\it SExtractor} are at least a few synthesized beams away from the 
positions given in the DR2 catalog; the latter almost always agree with the positions reported by other sources-finding algorithms 
(again, with the exception of {\tt IMSAD}; see \S\ref{Sec:IMSAD_efficacy} for anomalies associated with {\tt IMSAD}).
The completeness peaks at a background mesh-size of 5 beamwidths used with 3x smoothing, whereas reliability 
appears to increase steadily or remain constant with increasing mesh-size.
{\small $BACK\_FILTERSIZE$} is not seen to change either the completeness or reliability much.
$SExtractor$ does not seem to achieve a high completeness and reliability ($>$85\%) simultaneously for any given set of values for the input parameters.
For reliability of $>90$\%, we see that completeness $<80$\%.
Beyond a detection threshold of 6$\sigma$, the reliability is quite good, yet less than 100\% owing primarily to the 
acceptance of sidelobes and the fitting of extended sources handled differently.
Nevertheless, the highest reliability for $SExtractor$ is better than the catalogs of other algorithms (except, maybe {\tt SAD}), and hence it is 
best for cases where reliability is strongly favored over completeness. In such a case, we recommend setting the detection threshold to about 10$\sigma$, along with 
a large enough value for {\small $BACK\_SIZE$}.
Tests on {\it SExtractor} with the ASKAP and \citeauthor{hmg+12} simulations \citep{hhn+11} suggest that the reliability is 
almost constant for mesh-size between 10 and 100 beamwidths, but increases with the detection threshold.
They find that the completeness generally decreases or remains constant as mesh-size increases, but its change with respect to detection threshold is not explored.
Our results are thus broadly consistent with the tests on the simulated images.

\begin{figure*}[htp]
\centering
\includegraphics[width=3.6in,height=2.6in,viewport=35 41 528 405,clip]{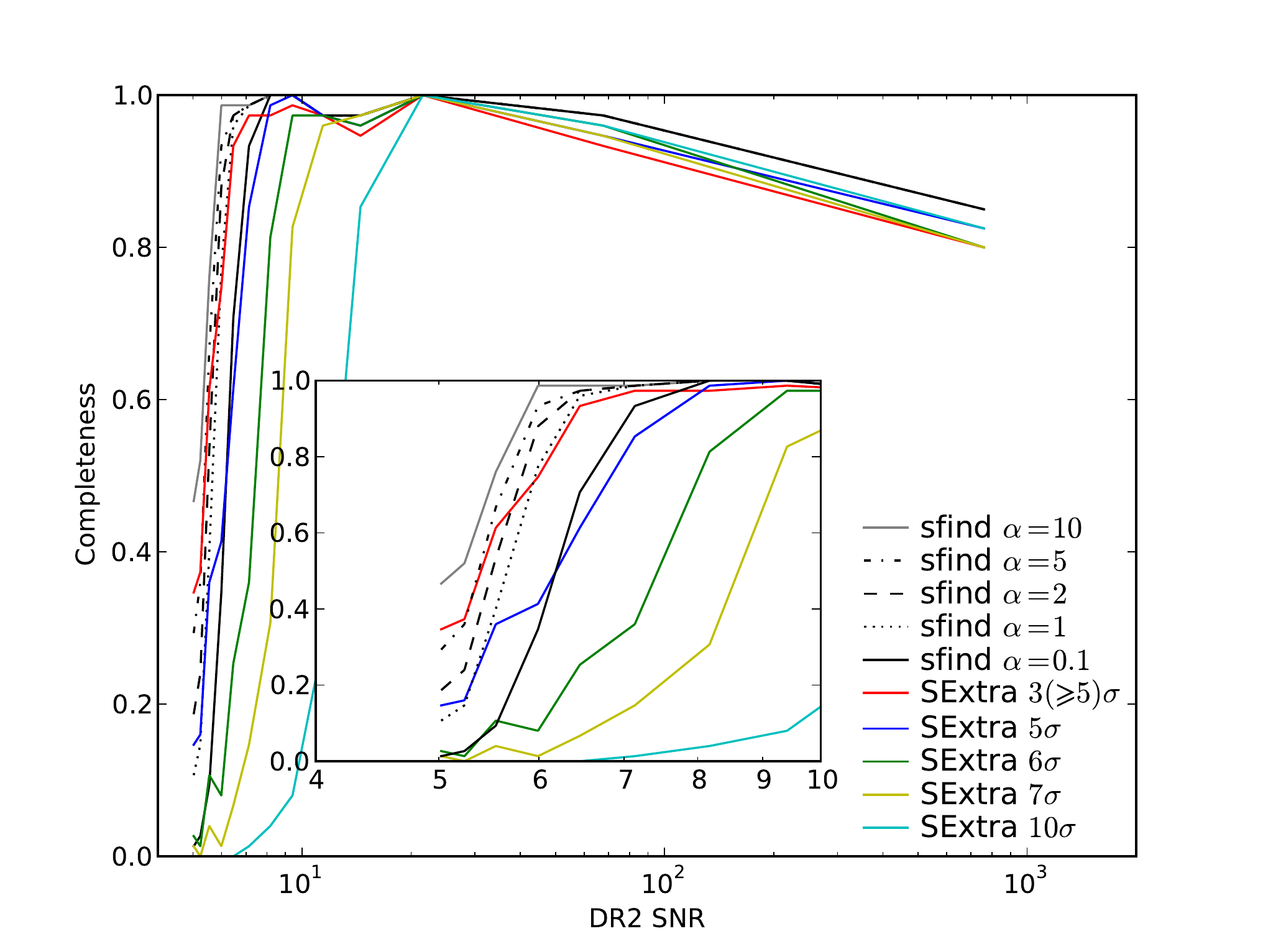}
\includegraphics[width=3.4in,height=2.6in,viewport=73 41 528 405,clip]{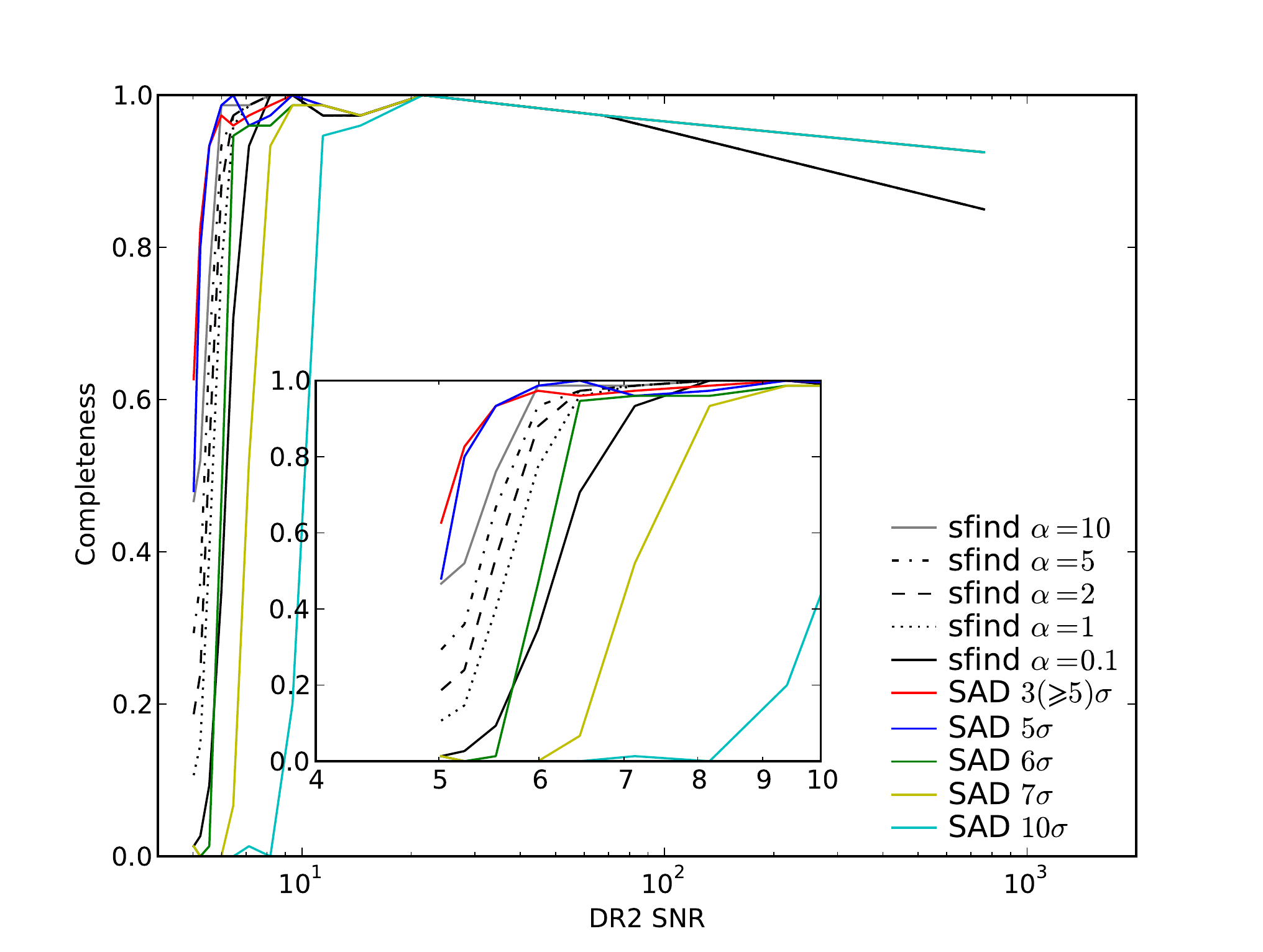}
\includegraphics[width=3.6in,height=2.8in,viewport=35 10 528 405,clip]{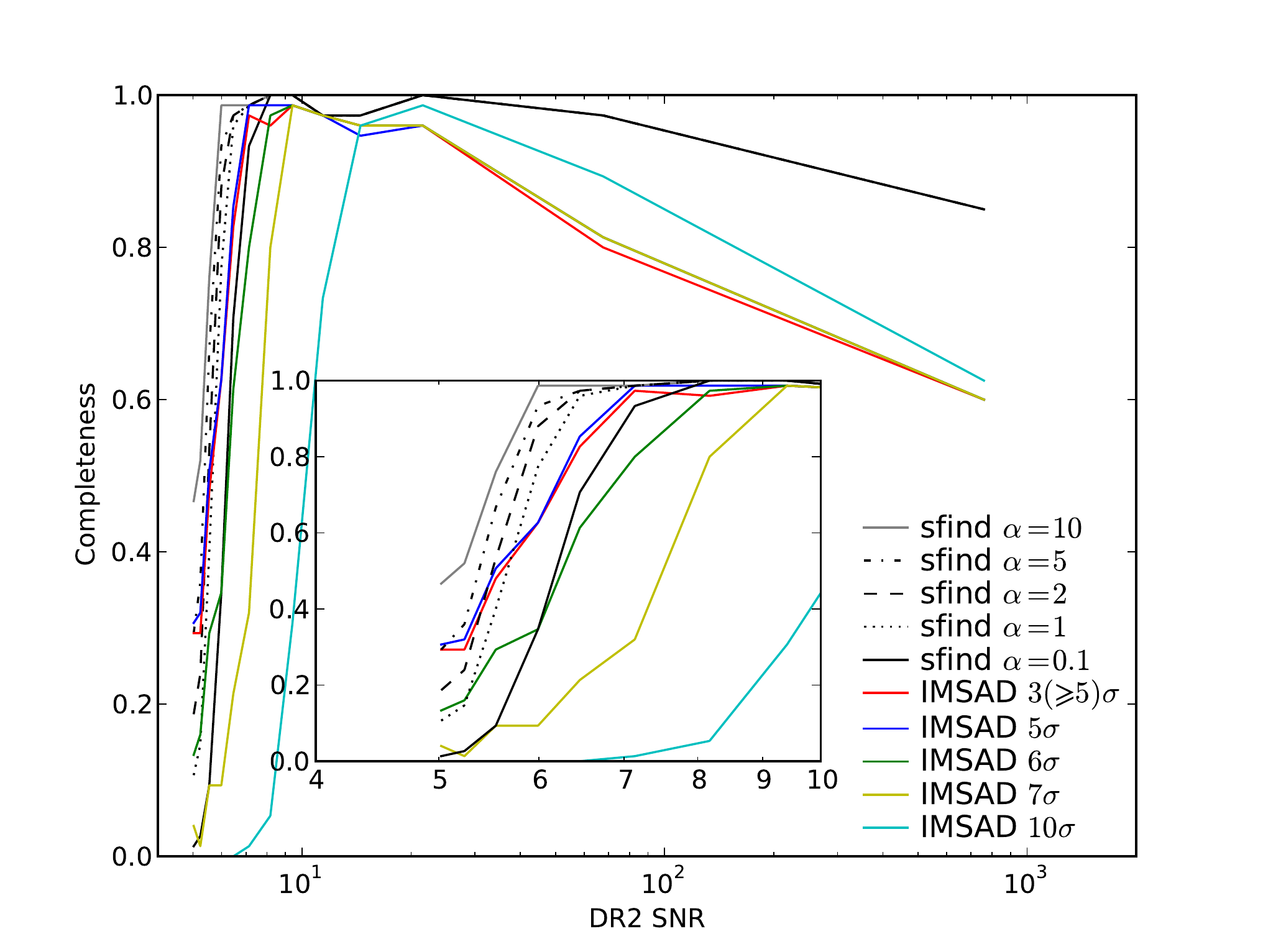}
\includegraphics[width=3.4in,height=2.8in,viewport=73 10 528 405,clip]{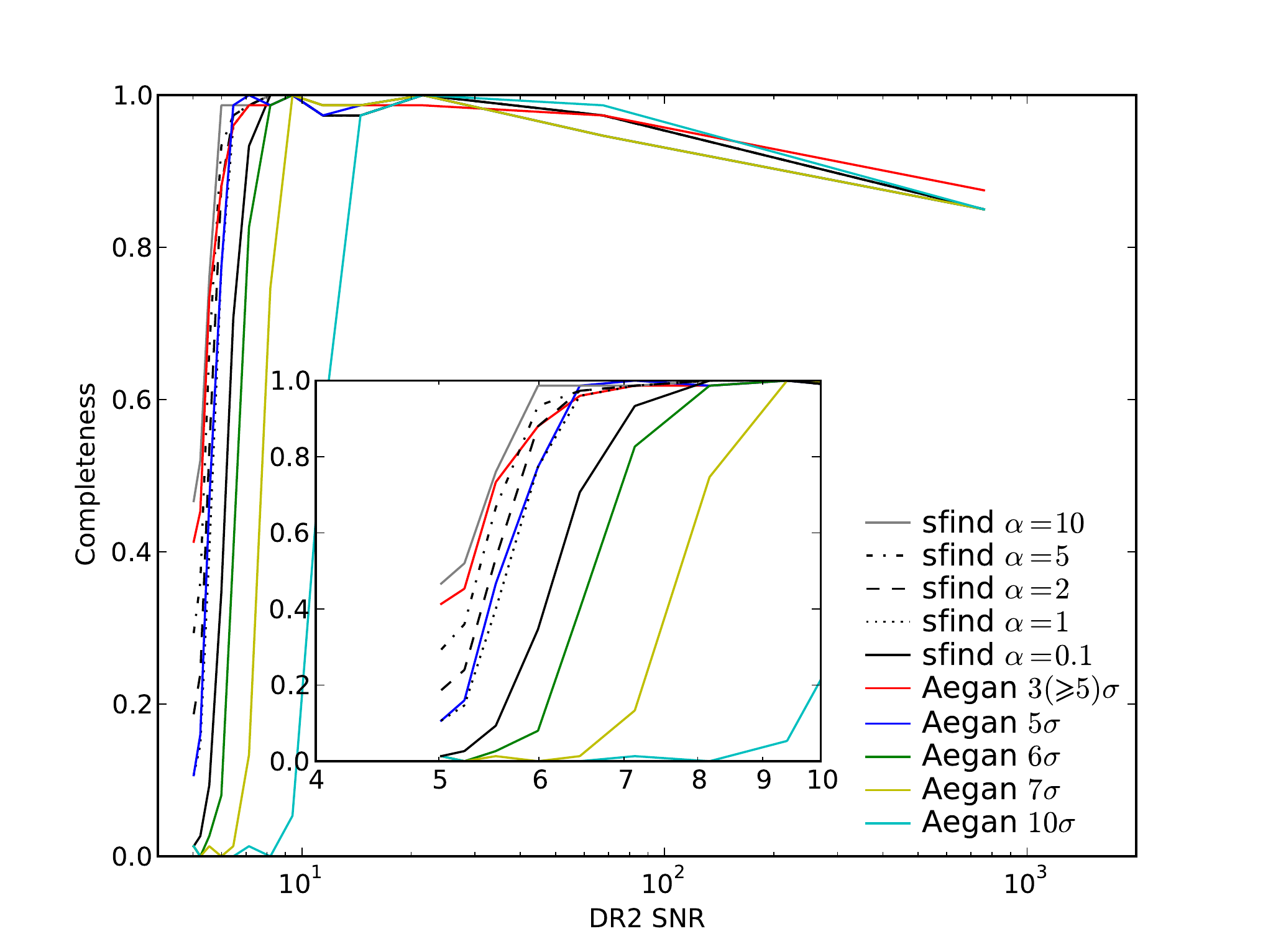}
\caption{Completeness of the catalogs generated by different source-finding algorithms. 
Results for a background mesh-size (rmsbox) of 20 beamwidths, wherever specifiable, are shown.
For {\tt IMSAD} and {\it Aegean}, the results are for the histogram option and csigma=1$\sigma_{\rm cmap}$ respectively.
See \S \ref{Sec:Algorithms} for the definition of completeness used here.
The decreasing completeness beyond $SNR\sim50$ is due to deblending of multiple components of 
extended sources. The inset shows only the region where SNR is between 4 and 10.
Smoothing over every 75 data points has been done before plotting.
Note the low-number statistics for sources with SNR $\gtrsim$70 as implied by Figure~\ref{Fig:DR2_SNR}.}
\label{Fig:Completeness_algorithms}
\end{figure*}

The upper-left panel of Figure~\ref{Fig:Completeness_algorithms} shows the completeness of {\it SExtractor} 
sources as a function of their SNR in the DR2 catalog for mesh-size of 20 beamwidths and different values of the detection threshold.
The completeness for optimum input parameter values is shown in the upper panel of Figure~\ref{Fig:Reliability_algorithms}.
{\it SExtractor} approaches 100\% completeness faster than {\tt IMSAD}, but slower than the other algorithms.
This is possibly related to the approach used for determining source pixels.
The decrease in completeness beyond SNR$\sim$20 can be attributed to the extended-source components being reported differently than other algorithms, as mentioned above.
The lower panel of Figure~\ref{Fig:Reliability_algorithms}, which shows the {\it SExtractor} reliability in relation with other source-finding algorithms, reveals that 
this algorithm gives quite unreliable results for SNR$\lesssim$10.
These low-SNR false sources are all point sources located near the edges of the DR2 image, where the rms is somewhat large (12 $\mu$Jy), but the rms reported 
by {\it SExtractor} is quite small (7--8 $\mu$Jy).
This scenario is presented in the image cutouts in Figure~\ref{Fig:source_cutouts}, which also illustrate the ability of 
{\it SExtractor} to find sources with different morphologies located in different environments.
Figure~\ref{Fig:SNR_offset_algorithms} shows the SNR of sources detected by {\it SExtractor}, in comparison with the corresponding sources in the DR2 catalog.
For SNR$\leqslant$30, the peak flux density of {\it SExtractor} sources sfind,SExtractor is generally less than that of the DR2 catalog sources, suggesting that 
the difference in the calculated peak flux density (and to a smaller extent, the associated uncertainty) is responsible for the observed departure 
of the {\it SExtractor} SNR from the DR2 SNR.

\subsubsection{{\tt SAD} ({\it AIPS})}\label{Sec:SAD_efficacy}
The ``Search and Destroy'' (SAD\footnote{www.aips.nrao.edu/cgi-bin/ZXHLP2.PL?SAD}) algorithm finds 
all the pixels above a specified threshold (typically a multiple of the rms noise, which is assumed to be Gaussian) in the image, 
and merge contiguous pixels above the threshold into islands.
The strength and size of each island is then estimated, followed by least squares Gaussian fitting of 
each island (if rms residual is too high, then multiple Gaussian fits may be applied).
However, note that Gaussian statistics may not be a good model for the 
distribution of values in pixels well above zero flux density on account of thermal noise, 
and calibration and imaging artifacts \citep{cotton2011}.
The run-time for {\tt SAD} searching for sources down to 5$\sigma$ in the DR2 image is about 2 minutes.
Additionally, to prepare the background rms image, the task {\tt RMSD} takes $\sim$15 minutes.
For all the tests carried out on {\tt SAD}, 31DEC11 $AIPS$ was used.

{\tt SAD} has several input parameters which affect the number of sources detected.
Here, we test the effect of the detection SNR ({\small $CPARM$}), rms threshold, flux residual 
threshold ({\small $DPARM, GAIN$}), and the size of rms box ({\small $IMSIZE$}).
The parameter {\small $DPARM(3)$} (along with {\small $GAIN$} added in quadrature) specifies the upper 
limit for the rms in the fitting box, while {\small $DPARM(7)$} and {\small $GAIN$} specify the upper limit on the residual flux in the fitting box.
{\small $GAIN$} thus defines the fraction of the source flux which is acceptable in the residual image.
By default, {\tt SAD} uses the entire image to find the rms.
We used the {\it AIPS} task {\tt RMSD} to prepare rms images using mesh-sizes (specified by the {\small $IMSIZE$} parameter) of 5, 10, 20, and 50 beamwidths.
Although a decremental search in SNR (via {\small $CPARM$}) is recommended in the {\tt SAD} help file, 
we found that such a search results in multiple sources being fit to a single genuine source during each iteration, especially 
when the source is extended. Hence, we rejected this recommendation.
As in the case of {\it SExtractor}, we searched for sources down to N$\sigma$.
Note that in {\tt SAD}, the errors in the flux density are determined by theory from the image rms ({\tt actnoise} keyword in the image header).

In Table~\ref{Tab:SAD_completeness}, we list the completeness and reliability of the {\tt SAD} catalog for different values of the input parameters.
We held the input parameters {\small $DPARM(3)$} and {\small $DPARM(7)$} fixed at 
1000 and 1 (in units of Jy beam$^{-1}$) respectively in order to get optimum completeness (without significant loss of reliability, 
as we found later, but possibly at the expense of correct SNR of the detected sources).
Due to these large input values, we did not find any change in the results with the {\small $GAIN$} parameter. 
However, we found that reliability can be traded for completeness by setting smaller values for {\small $DPARM(7)$} and {\small $GAIN$}.
In general, the completeness and reliability of {\tt SAD} increase with the detection threshold.
Completeness increases with mesh-size. So does reliability, though this quantity decreases significantly as the mesh-size is increased from 
20 to 50 beamwidths, similar to {\tt sfind}.
Our inspection of the rms image for a mesh-size of 50 beamwidths reveals that there are a few pockets where the rms is rather low ($\sim$1$~\mu$Jy).
A profusion of false sources (or very low-SNR sources reported to have an SNR above the threshold) detected in these pockets are responsible for 
the markedly reduced reliability for the case of mesh-size equal to 50 beamwidths.
Search with {\tt actnoise} usually performs at least a few percent worse in terms of completeness and reliability than using mesh-size of 20 beamwidths.
Curiously, the completeness and reliability of {\tt SAD} is $>$90\% for a wide range of input parameters tested.
We determine the optimum values of input parameters to be 10--20 beamwidths mesh-size and 6--7$\sigma$ detection threshold.
Searching down to 3$\sigma$ followed by selection of sources greater than 6--7$\sigma$ may improve completeness to some extent.
\cite{hhn+11} and \cite{hmg+12} have not tested {\tt SAD} on simulated images.

\begin{figure}[htp]
\centering
\includegraphics[width=3.2in,height=2.6in,viewport=35 5 530 410,clip]{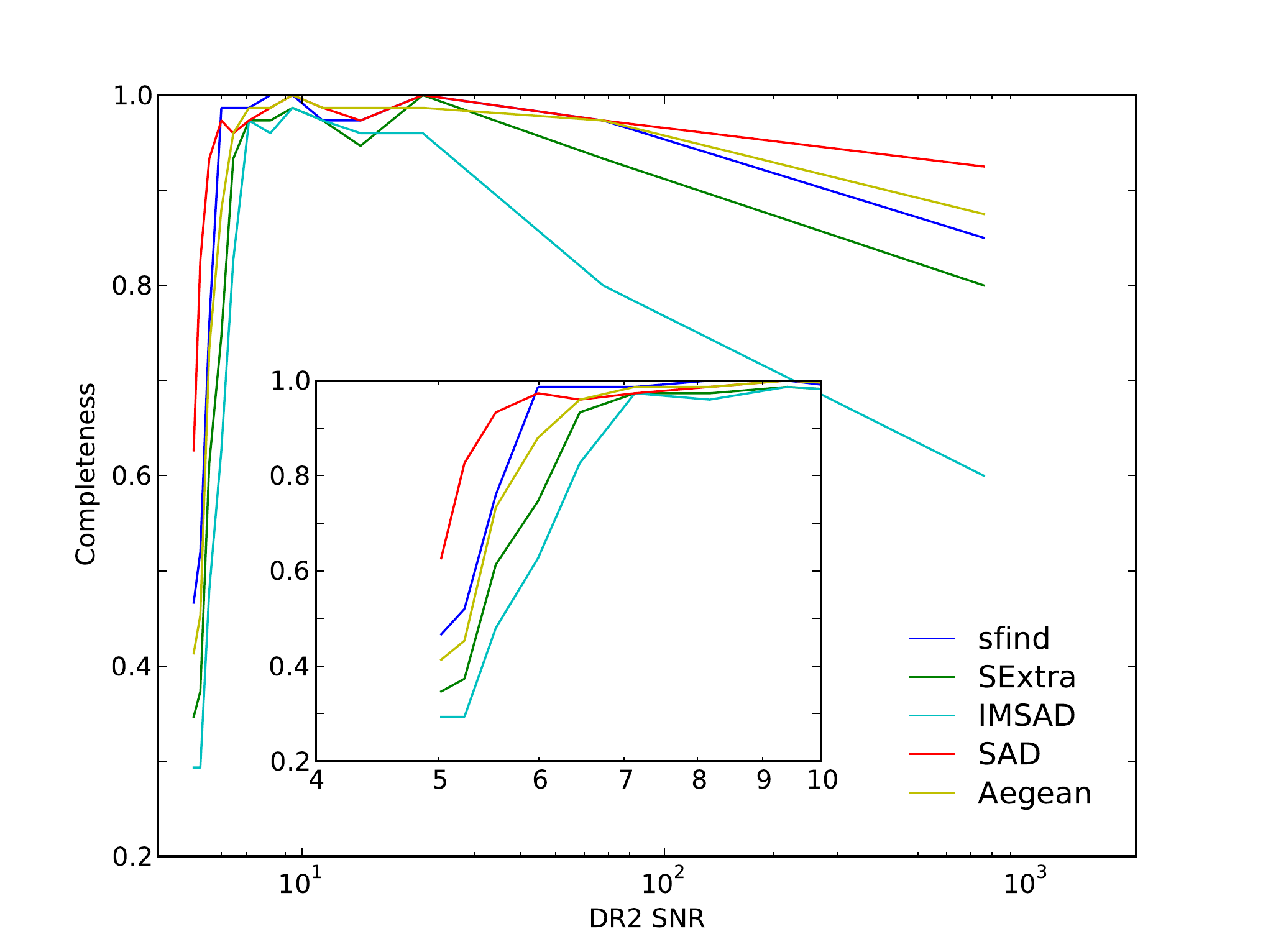}\\
\includegraphics[width=3.2in,height=2.6in,viewport=35 5 530 410,clip]{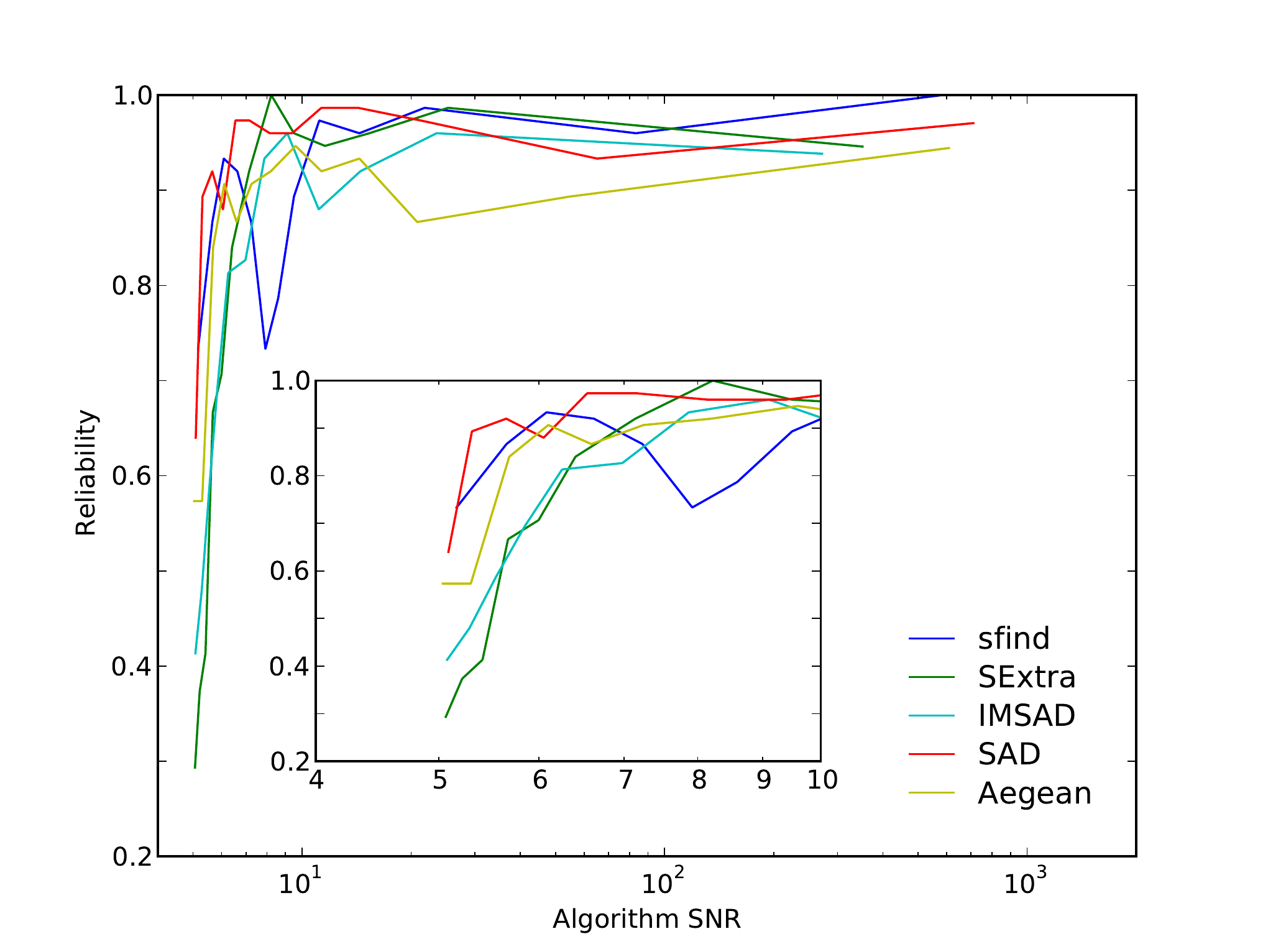}
\caption{The completeness (top) and reliability (bottom) of the catalogs generated by different source-finding algorithms.
See \S \ref{Sec:Algorithms} for the definitions of completeness and reliability used in this work.
Results for a background mesh-size (rmsbox) of 20 beamwidths are shown.
For {\tt IMSAD}, these results are for the histogram option.
For sfind, we have used $\alpha$=10, whereas for {\it SExtractor}, {\tt SAD}, {\tt IMSAD} and {\it Aegean}, 
detection threshold of 3$\sigma$. Only those sources which are $\geqslant$5$\sigma$ have then been 
selected from the respective catalogs prior to comparison with the DR2 catalog.
The inset shows only the region where SNR is between 4 and 10.
Smoothing over every 75 data points has been done before plotting. 
Note the low-number statistics for sources with SNR $\gtrsim$70 as implied by Figure~\ref{Fig:DR2_SNR}.}
\label{Fig:Reliability_algorithms}
\end{figure}

\begin{figure*}[htp]
$~$ \qquad \qquad \qquad \qquad \qquad 
DR2 \qquad \qquad \quad 
{\tt sfind} \qquad \qquad 
{\it SExtractor} \qquad \qquad 
{\tt IMSAD} \qquad \qquad \quad 
{\tt SAD} \qquad \qquad \qquad 
{\it Aegean}\\
\centering
\includegraphics[width=7.5in]{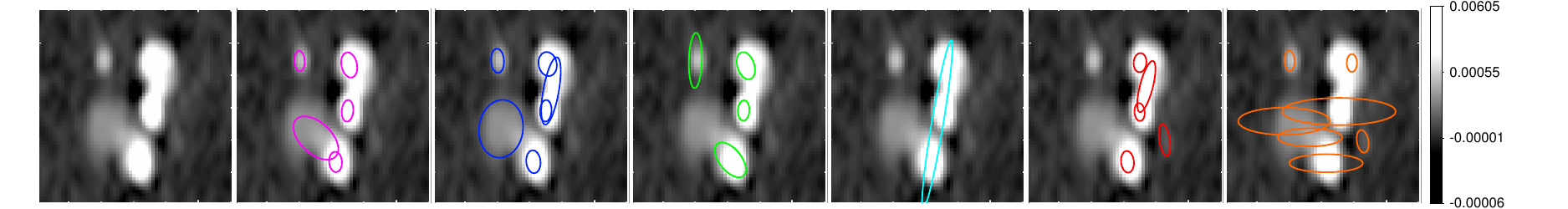}
\includegraphics[width=7.5in]{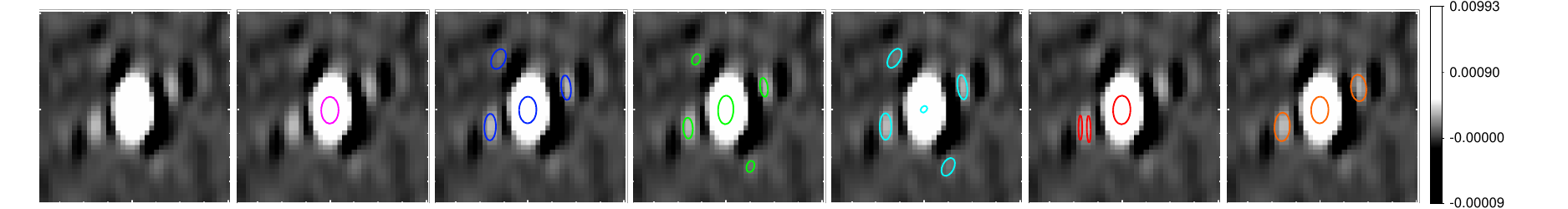}
\includegraphics[width=7.5in]{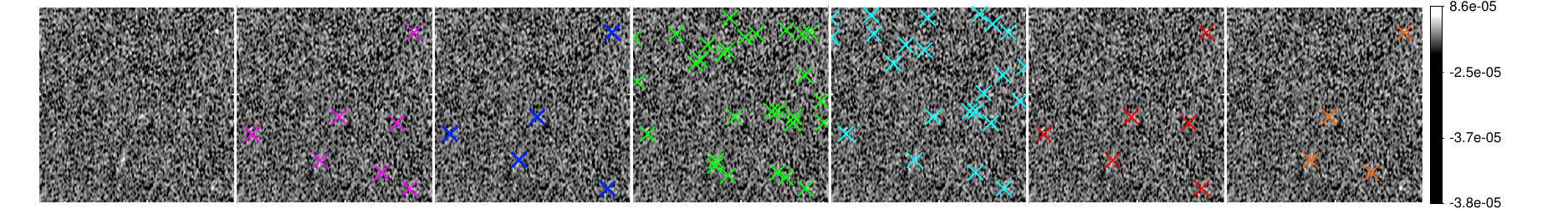}
\caption{Examples of $5\sigma$ and brighter sources detected by various source-finding algorithms in the DR2 image.
Top panel: blended components (30\arcsec ~cutouts centered on 03$^{\rm h}$32$^{\rm m}$32.2$^{\rm s}$,$-28$\arcdeg 03\arcmin 09.4\arcsec), 
middle panel: source with sidelobes (20\arcsec ~cutouts centered on 03$^{\rm h}$32$^{\rm m}$06.1$^{\rm s}$,$-27$\arcdeg 32\arcmin 35.8\arcsec), 
bottom panel: region with a relatively large rms (12 $\mu$Jy) at the corner of the image (3\arcmin ~cutouts centered on 03$^{\rm h}$31$^{\rm m}$19.4$^{\rm s}$,$-27$\arcdeg 32\arcmin 55.6\arcsec).
The logarithmic flux density scale shown in each panel has units of Jy.
The ellipses have major and minor axes and position angles according to the parameters reported by the respective algorithms.}
\label{Fig:source_cutouts}
\end{figure*}

\begin{figure}[htp]
\centering
\includegraphics[width=3.2in,viewport=25 5 535 400,clip]{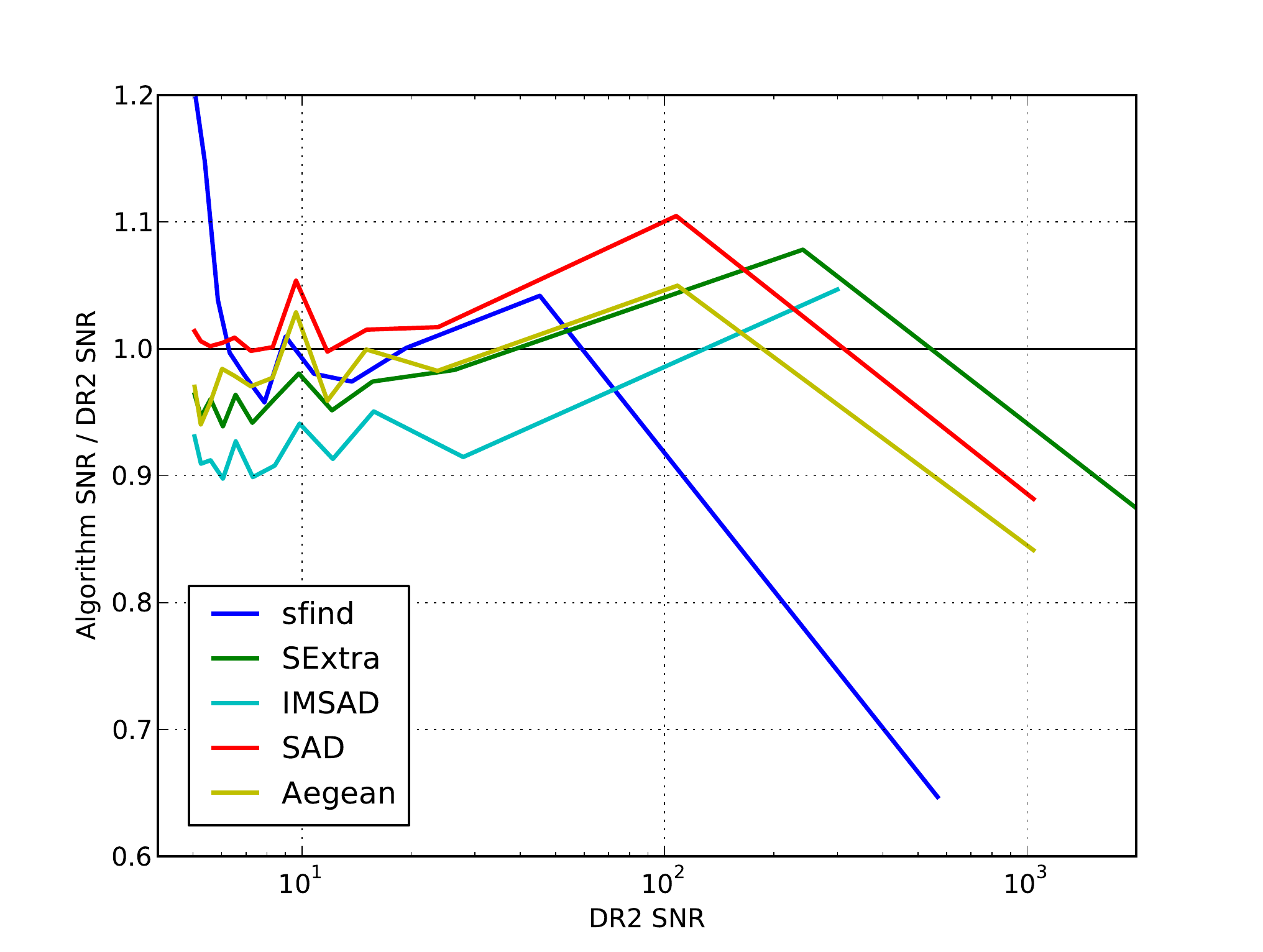}
\caption{The SNR of sources found by different source-finding algorithms relative to their SNR in the DR2 catalog.
Smoothing over every 75 data points has been done before plotting. 
Note the low-number statistics for sources with SNR $\gtrsim$70 as implied by Figure~\ref{Fig:DR2_SNR}.}
\label{Fig:SNR_offset_algorithms}
\end{figure}

The completeness of {\tt SAD} with respect to detection SNR and for different detection thresholds is shown in the upper-right panel of 
Figure~\ref{Fig:Completeness_algorithms}, which depicts the high level of completeness close to the threshold and rapid increase with 
SNR compared to the other algorithms.
The upper panel of Figure~\ref{Fig:Reliability_algorithms} plots the {\tt SAD} completeness for optimum input parameters.
Due to missed sources throughout the SNR spanned, the completeness is seen to hover close to unity, but not quite getting to 100\%.
The source-rejection criteria based on 2-D Gaussian fitting in {\tt SAD}, defined by the several elements of the {\small $DPARM$} input array, is the  
likely cause for such such missed sources.
The lower panel of Figure~\ref{Fig:Reliability_algorithms} shows the reliability for the optimum values input parameters.
As with the case of completeness, the reliability seems to hover close to unity. {\tt SAD} provides the best reliability for sources 
SNR$\lesssim$20, beyond which, {\tt sfind} gives better results.
Departure of the reliability from unity for a wide range of SNR values is due to the fitting of different components to extended sources.
Some examples of how {\tt SAD} fits different kinds of sources in the DR2 image, in comparison with other algorithms, is shown in Figure~\ref{Fig:source_cutouts}.
Figure~\ref{Fig:SNR_offset_algorithms} shows the SNR of sources detected by {\tt SAD}, in comparison with that of their counterparts in the DR2 catalog.
Usually the SNRs agree with each other, but any disagreement between the two is due to the reported uncertainty in the peak.
The peak flux densities reported by {\tt SAD} match those in the DR2 catalog quite well. 

\begin{figure}[htp]
\centering
\includegraphics[width=3.3in,viewport=30 20 535 510,clip]{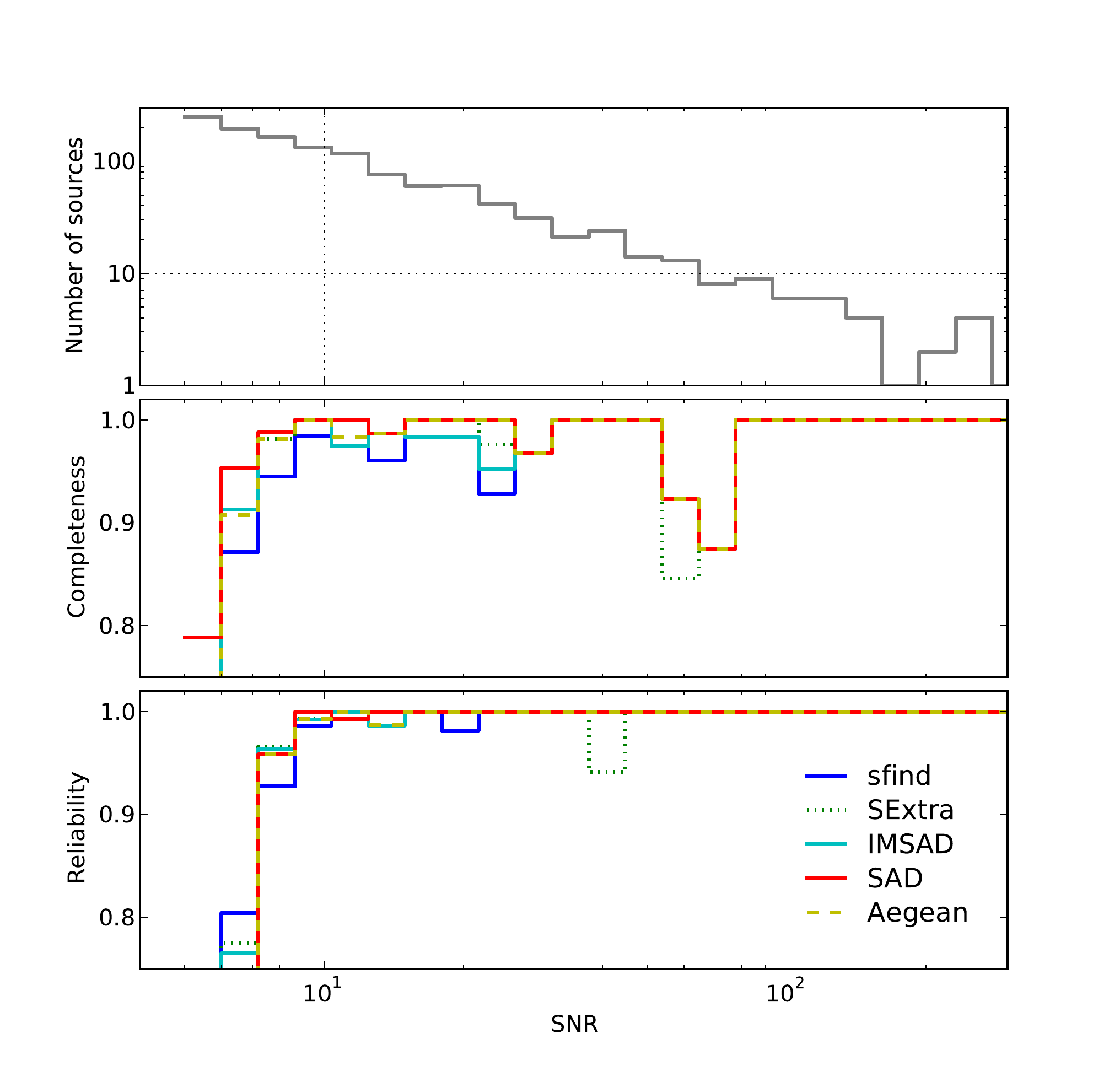}
\caption{Results of the completeness (middle) and reliability (bottom) tests on the \citeauthor{hmg+12} simulated image.
Matching radius of 15\arcsec ~was used to find counterparts.
Input parameters to the source-finding algorithms are same as those given in Figure~\ref{Fig:Reliability_algorithms}.
Here, the completeness and reliability within each SNR bin are plotted (see \S \ref{Sec:Algorithms}), which makes these diagrams 
different from the ones given in \cite{hmg+12}.
For reference, the histogram of the simulated sources is also shown (top). The completeness at SNR$\sim$70 dropping below 90\% 
is a result of highly-blended sources.}.
\label{Fig:HancockTests}
\end{figure}

There is a possibility that the superior completeness and reliability of {\tt SAD} may be due simply because the 
DR2 catalog, against which we are doing all of our comparisons, was constructed from {\tt SAD}.
However, the construction of the DR2 catalog involved much more than running {\tt SAD} (see \S\ref{sec:variables} for more details).
In any case, an independent check using simulated sources is warranted.
In Figure~\ref{Fig:HancockTests}, we compare the completeness and reliability of {\tt SAD} with other source-finding algorithms 
on the \cite{hmg+12} simulated image.
We find a similar superior behavior of {\tt SAD}.

\subsubsection{{\tt IMSAD} ({\it MIRIAD})}\label{Sec:IMSAD_efficacy}
{\tt IMSAD}\footnote{www.atnf.csiro.au/computing/software/miriad/doc/imsad.html} 
computes the image rms noise by fitting a Gaussian to the image histogram, then 
searches for contiguous pixels (islands) which are above some cutoff and fits the 
islands with Gaussian components. For fitting, the routine from the {\it MIRIAD} task {\tt imfit} 
is used, whereas the island detection is based on the {\it AIPS} task {\tt SAD}.
{\tt IMSAD} can also be used in a mode where the threshold background noise is user-specified.
The run-time for this source-finding algorithm searching for sources down to 5$\sigma$ in the DR2 image is about 2 seconds.
We used {\tt IMSAD} version 8-May-98 from MIRIAD 4.2.3 (optimized for CARMA) to carry out these tests.

As before, we searched for sources down to N$\sigma$ (specified via the {\it clip} parameter), and then selected $\geqslant$5$\sigma$ sources.
The completeness and reliability were tested using the histogram mode ($noplt,hist$ options set) as well 
as the user-specified mode (only the $noplt$ option set; rms of $8~\mu$Jy used) for background-rms determination.

The results of these tests on {\tt IMSAD} are given in Table~\ref{Tab:IMSAD_completeness}.
The $noplt,hist$ mode usually performs better than the $noplt$ mode at least by a few percent in terms of completeness, whereas the latter mode gives better completeness, again by a few percent.
For achieving completeness and reliability of 80\% or more, the detection threshold needs to be $\geqslant$6$\sigma$.
Hence, the optimum use of {\tt IMSAD} would be in this SNR cutoff regime, and the mode in which it is used will depend upon whether completeness or reliability is preferred.
Searching down to 3$\sigma$ followed by rejection of sources below the desired threshold might give better completeness.
The reason for the low completeness at a detection threshold of 5$\sigma$ is that several DR2 sources (which are detected by {\tt sfind}, {\tt SAD} and {\it Aegean}) 
close to this threshold are not detected by {\tt IMSAD}, not even at a reduced SNR.
The reliability at this threshold is also rather low. This can be attributed to several false point sources detected 
near the edges of the DR2 image, which are mainly low-SNR features mistaken to be high-SNR due to incorrect assumption of the local rms.
Since {\tt IMSAD} does not give the rms for each detected source, one needs to assume this quantity 
(a constant equal to $8~\mu$Jy in our case), and thus, the SNR calculated in regions of increased rms, 
e.g. regions close to the edge of the DR2 image, is expected to be erroneous.
By using an rms image to find the local rms, this false-detection problem can be tackled.
Another issue related to false sources is that {\tt IMSAD} does not reject a single sidelobe. 
All the sidelobes are reported as (genuine) sources.
Thus, with respect to false sources detected in noisy regions, missed sources, and sidelobe recognition, {\tt IMSAD} behaves like {\it SExtractor}.
The sources detected by {\tt IMSAD} in some of these scenarios are shown in the image cutouts in Figure~\ref{Fig:source_cutouts}.
\cite{hmg+12} tested {\tt IMSAD} on a simulated image and found that its completeness and reliability is lower than other source-finding algorithms.
However, through our tests on the DR2 image, it appears that {\tt IMSAD} has a competitive performance if used with detection thresholds higher than 5$\sigma$.

The completeness of the {\tt IMSAD} catalog as a function of the detection SNR is shown in the lower left panel of 
Figure~\ref{Fig:Completeness_algorithms}.
The upper panel of Figure~\ref{Fig:Reliability_algorithms} shows the {\tt IMSAD} completeness for optimum input parameters.
Several sources having 10$<$SNR$<$20 as well as SNR$\gtrsim$100, which are detected by other source-finding algorithms, are not detected by {\tt IMSAD}.
This is due to the inability of this algorithm to individually fit blended sources and components of extended sources; 
{\tt IMSAD} tends to fit one elongated source for all components (see upper panel of Figure~\ref{Fig:source_cutouts}).
In Figure~\ref{Fig:Reliability_algorithms}, the reliability of this algorithm is plotted for the optimum input parameters.
{\tt IMSAD} shows reduced reliability between SNRs of 10 and 20, beyond which the reliability roughly flattens off at the 95\% level.
The reason for this reduced reliability is two-fold.
First, several point sources detected close to the edge of the DR2 image have integrated flux densities less than their peak by a factor of a few, 
clearly indicating unphysical fitting of sources. These false sources can easily be rejected by using a peak-to-integrated flux density ratio criterion.
Second, {\tt IMSAD} does not break up islands into components as mentioned above, resulting in extended 
source positions to be substantially different from those listed in the DR2 catalog.
This was also found in the tests carried out by \cite{hmg+12}, which resulted in several false detections.
Figure~\ref{Fig:SNR_offset_algorithms} shows the SNR of {\tt IMSAD} sources compared with their counterparts in the DR2 catalog.
The assumed constant rms as well as the somewhat lower peak flux density reported by {\tt IMSAD} (for extended sources) are responsible for the lower SNR with respect to DR2 sources.

\subsubsection{{\it Aegean}}\label{Sec:Aegean_efficacy}
{\it Aegean}\footnote{www.physics.usyd.edu.au/\%7Ehancock/index.php/Programs/Aegean} 
uses the {\it FloodFill} algorithm, which separates the foreground pixels from the background and groups them into islands.
These ``islands'' are then passed on to the source characterization stage. 
Each island of pixels is fit with multiple Gaussian components.
The number of components to be fit is determined from a surface-curvature map (cmap), derived from the input image with a Laplacian transform.
{\it Aegean} thus performs a well-constrained multiple Gaussian fitting.
A detailed description of {\it Aegean} and its implementation can be found in \cite{hmg+12}, where the authors demonstrate 
this source-finding algorithm on a simulated image to produce catalogs for better reliability and completeness than other source finding algorithms.
The run-time for {\it Aegean} searching for sources down to 5$\sigma$ in the DR2 image is about 4 minutes (using two CPU cores).
We used $Aegean$ r706 (25-Jul-2012 release) for testing this source-finding algorithm.

We tested the effect of the parameters: (i) $innerclip$ ($seedclip$) and $outerclip$ ($floodclip$), and (ii) $csigma$ on the 
completeness and reliability of the {\it Aegean} catalog.
innerclip defines the sigma clipping (lower limit) which is used for the 
detection (``seeding'') of islands, whereas outerclip defines that used for reporting (``flooding'') the islands in the catalog.
$csigma$ is the sigma clipping parameter for the curvature map.
Note that the r706 version of {\it Aegean} uses an immutable mesh-size of 20 beamwidths.
We used $innerclip = outerclip$ for our tests.
Setting the innerclip to 1$\sigma$ lower than outerclip gives results alike equating these two clipping parameters, while
using a 1$\sigma$ lower outerclip than innerclip seems to give a few percent better completeness and a slightly reduced reliability.

Our results for {\it Aegean} are given in Table~\ref{Tab:Aegean_completeness}.
$csigma$ does not have a significant effect on either the completeness or reliability.
As with our tests on other source-finding algorithms, we searched for sources down to N$\sigma$.
The best completeness and reliability require a 7$\sigma$ or higher detection threshold.
Using a 3$\sigma$ detection threshold followed by the selection of greater than 7$\sigma$ might give better completeness at the cost of reliability.
Surprisingly, the completeness and reliability are not $>$95\%, as seen in the tests with the \cite{hmg+12} simulated image.
This is because {\it Aegean} performs well with isolated point sources, but not quite with extended or blended ones.
For extended sources, {\it Aegean} tends to fit some of the components at unexpected locations and with quite elongated Gaussians 
(the fitting-error flag is also set for these components). Numerous small (but unphysical; size equal to the synthesized beam) components are also 
reported for several extended sources.
Since \cite{hmg+12} have demonstrated that {\it Aegean} successfully fits all the components of blended simulated sources, we think that 
diffuse components and imaging artifacts such as negative sidelobes adversely affect the curvature map produced by this source-finding algorithm, 
which in turn determines the sources reported. An example of such a case is given in the upper panel of Figure~\ref{Fig:source_cutouts}.
Note that in the latest release of Aegean (r808; 13-Feb-2013) this situation is somewhat improved, since some of the elongated components from r706 
are reported with more realistic source sizes (and also without any fitting-error flags), and relatively fewer small-size components are fit.

\begin{table}[htp]
\centering
\caption{Completeness and Reliability of {\tt sfind} catalog}
\label{Tab:sfind_completeness}
\begin{tabular}{c ccccc} 
\hline \hline
rmsbox     & \multicolumn{5}{c}{$\alpha$}\\
\cline{2-6}
(beamwidths) & 10 & 5 & 2 & 1 & 0.1\\
\hline
\multicolumn{6}{c}{Completeness}\\
\hline
  5 &   80.9\% &  77.8\% &  74.3\% &  72.4\% &  61.9\% \\ 
 10 &   85.7\% &  83.7\% &  79.1\% &  75.9\% &  65.5\% \\
 20 &   88.4\% &  84.5\% &  81.2\% &  77.8\% &  67.8\% \\
 50 &   89.1\% &  85.7\% &  82.1\% &  78.4\% &  68.0\% \\
\hline
\multicolumn{6}{c}{Reliability}\\
\hline
  5 &   78.8\% &  84.9\% &  90.6\% &  93.1\% &  96.4\% \\
 10 &   86.8\% &  91.6\% &  95.2\% &  95.6\% &  97.4\% \\
 20 &   88.7\% &  92.5\% &  95.3\% &  96.0\% &  96.4\% \\
 50 &   87.5\% &  91.4\% &  93.8\% &  94.8\% &  95.3\% \\
\hline
\multicolumn{6}{p{2.9in}}{Notes$-$ Only $\geqslant$5$\sigma$ sources have 
been selected from the {\tt SFIND} catalogs prior to comparison with 
the DR2 catalog.}
\end{tabular}
\end{table}

\begin{table}[htp]
\centering
\caption{Completeness and Reliability of {\it SExtractor} catalog}
\label{Tab:SExtractor_completeness}
\begin{tabular}{l ccccc} 
\hline \hline
BACK\_SIZE               & \multicolumn{5}{c}{DETECT\_THRESH}\\
\cline{2-6}
(beamwidths)             &3($\geqslant$5) & 5 & 6       & 7       & 10\\
\hline
\multicolumn{6}{c}{Completeness*}\\
\hline
 5                       &   82.2\% &  72.1\% &  76.8\% &  79.1\% &  77.9\% \\
 5 + $ 3\times$smoothing &   83.4\% &  73.7\% &  78.7\% &  80.0\% &  79.0\% \\
 5 + $10\times$smoothing &   83.3\% &  73.0\% &  78.4\% &  80.0\% &  79.0\% \\
10                       &   81.7\% &  71.4\% &  76.9\% &  78.7\% &  78.2\% \\
10 + $ 3\times$smoothing &   82.1\% &  71.1\% &  76.7\% &  79.1\% &  77.6\% \\
10 + $10\times$smoothing &   81.7\% &  71.2\% &  76.4\% &  78.6\% &  77.6\% \\
20                       &   81.6\% &  70.6\% &  75.9\% &  78.6\% &  77.6\% \\
50                       &   81.3\% &  70.2\% &  75.3\% &  78.0\% &  77.3\% \\
\hline
\multicolumn{6}{c}{Reliability}\\
\hline
 5                       &   70.4\% &  87.5\% &  95.1\% &  96.3\% &  97.0\% \\
 5 + $ 3\times$smoothing &   72.4\% &  89.0\% &  96.1\% &  97.2\% &  97.4\% \\ 
 5 + $10\times$smoothing &   72.2\% &  89.5\% &  96.1\% &  97.2\% &  97.4\% \\
10                       &   75.2\% &  90.8\% &  96.2\% &  96.5\% &  97.4\% \\
10 + $ 3\times$smoothing &   75.8\% &  90.1\% &  96.2\% &  97.0\% &  97.4\% \\
10 + $10\times$smoothing &   75.6\% &  90.2\% &  96.2\% &  96.9\% &  97.4\% \\
20                       &   77.0\% &  90.3\% &  96.4\% &  96.9\% &  97.4\% \\
50                       &   78.1\% &  91.1\% &  96.5\% &  96.9\% &  97.4\% \\
\hline
\end{tabular}
\end{table}

\begin{table}[htp]
\centering
\caption{Completeness and Reliability of {\tt SAD} catalog}
\label{Tab:SAD_completeness}
\begin{tabular}{c ccccc} 
\hline \hline
IMSIZE        & \multicolumn{5}{c}{CPARM}\\
\cline{2-6}
(beamwidths)  & 3($\geqslant$5)& 5 & 6       & 7        & 10\\
\hline
\multicolumn{6}{c}{Completeness*}\\
\hline
 5            &   77.8\% &  75.5\% &  80.0\% &  83.7\% &  78.4\% \\
10            &   89.9\% &  88.7\% &  93.1\% &  92.1\% &  93.4\% \\
20            &   93.4\% &  92.2\% &  95.9\% &  95.4\% &  97.1\% \\
50            &   95.9\% &  94.8\% &  96.5\% &  95.8\% &  97.1\% \\
actnoise\ddag &   87.4\% &  86.7\% &  95.2\% &  96.7\% &  95.7\% \\
\hline
\multicolumn{6}{c}{Reliability}\\
\hline
 5            &   88.1\% &  89.0\% &  97.3\% &  99.1\% &  99.3\% \\ 
10            &   91.2\% &  93.0\% &  97.4\% &  97.4\% &  96.6\% \\
20            &   92.5\% &  93.9\% &  96.7\% &  96.9\% &  97.0\% \\
50            &   85.1\% &  87.3\% &  88.1\% &  88.6\% &  92.1\% \\
actnoise\ddag &   58.2\% &  60.2\% &  86.6\% &  94.6\% &  96.4\% \\
\hline
\multicolumn{6}{p{2.9in}}{Notes$-$ 
\ddag Search using the {\tt actnoise} keyword in the FITS header ($=$7.465E-06 JY/BM).}
\end{tabular}
\end{table}

\begin{table}[htp]
\centering
\caption{Completeness and Reliability of {\tt IMSAD} catalog}
\label{Tab:IMSAD_completeness}
\begin{tabular}{l ccccc} 
\hline \hline
options     & \multicolumn{5}{c}{clip}\\
\cline{2-6}
            & 3($\geqslant$5)& 5 & 6       & 7        & 10\\
\hline
\multicolumn{6}{c}{Completeness*}\\
\hline
noplt\ddag  &   75.7\% &  73.9\% &  78.1\% &  84.2\% &  79.3\% \\
noplt,hist  &   75.4\% &  76.5\% &  91.1\% &  90.5\% &  90.6\% \\
\hline
\multicolumn{6}{c}{Reliability}\\
\hline
noplt\ddag  &   77.4\% &  83.7\% &  95.3\% &  96.5\% &  96.8\% \\
noplt,hist  &   78.2\% &  79.2\% &  91.7\% &  95.8\% &  96.2\% \\
\hline
\multicolumn{6}{p{2.7in}}{Notes$-$ 
\ddag Clipping level is manually entered 
as the appropriate multiple of the background rms chosen to be 8$~\mu$Jy. }
\end{tabular}
\end{table}

\begin{table}[htp]
\centering
\caption{Completeness and Reliability of {\it Aegean} catalog}
\label{Tab:Aegean_completeness}
\begin{tabular}{c ccccc} 
\hline \hline
csigma      & \multicolumn{5}{c}{innerclip = outerclip}\\
\cline{2-6}
(cmap rms)  & 3($\geqslant$5)& 5 & 6       & 7        & 10\\
\hline
\multicolumn{6}{c}{Completeness*}\\
\hline
0.5         &   86.1\% &  78.7\% &  83.8\% &  86.1\% &  85.8\% \\
1.0         &   86.3\% &  78.5\% &  84.2\% &  86.4\% &  85.8\% \\
2.0         &   88.2\% &  78.5\% &  84.5\% &  86.8\% &  85.5\% \\
\hline
\multicolumn{6}{c}{Reliability}\\
\hline
0.5         &   86.2\% &  91.6\% &  91.6\% &  92.1\% &  91.7\% \\
1.0         &   84.9\% &  90.6\% &  90.5\% &  90.5\% &  91.1\% \\
2.0         &   83.4\% &  91.0\% &  91.2\% &  91.4\% &  89.9\% \\
\hline
\multicolumn{6}{p{3in}}{Notes$-$ *The fraction of sources in the DR2 catalog which 
are $\geqslant$6$\sigma$, 7$\sigma$ and 10$\sigma$ are 73.0\%, 58.1\% and 37.7\% respectively.
Completeness for these detection thresholds has been normalized accordingly. 
A detection threshold of 3($\geqslant$5) implies a search down to 3$\sigma$ followed by the selection 
of only those source that are greater than 5$\sigma$.}
\end{tabular}
\end{table}

Figure~\ref{Fig:Completeness_algorithms} shows the completeness of the {\it Aegean} catalog as a function of SNR of sources in the DR2 catalog, and the 
upper panel of Figure~\ref{Fig:Reliability_algorithms} shows the completeness for $csigma$=1$\sigma_{\rm cmap}$, and searching for sources down to $3\sigma$.
{\it Aegean} completeness appears to be quite good except for sources below 7$\sigma$ in the DR2 catalog, which are detected at a decreased SNR (below 5) and hence rejected 
from the catalog, and for a few components of extended sources above $\sim$100$\sigma$, which are reported to have positions not matching those in the DR2 catalog.
The decreased SNR reported for 5--7$\sigma$ sources is due to a slightly decreased peak flux density and a slightly increased rms with respect to the DR2 catalog.
Figure~\ref{Fig:Reliability_algorithms} plots the reliability using $csigma$=1$\sigma_{\rm cmap}$, and searching for sources down to $3\sigma$.
The significant deviation of reliability from unity for SNR$>$10 sources, results from the several (false) small-size, and elongated components reported for extended sources, as mentioned above.
Figure~\ref{Fig:SNR_offset_algorithms} shows the SNR of sources detected by {\it Aegean}, in comparison with the corresponding sources in the DR2 catalog.
The disagreement in SNR for 5--7$\sigma$ sources in the DR2 catalog has been explained above.
For SNR$>$200 sources, the peak flux densities agree quite well between the Aegean and DR2 catalogs, but the rms reported by Aegean is consistently higher, which causes the 
disagreement at the high-SNR end of the diagram.

\quad \newline
\subsubsection{Summary of results from the efficiency tests}\label{Sec:sum_eff}
We find remarkable differences between  
algorithms in terms of components fitted for extended sources, sidelobe 
rejection, and point sources detected in regions where the rms is appreciably 
larger than the mean rms.
Our results for completeness and reliability are broadly similar to those 
of \citet{hhn+11} and \citet{hmg+12}. For applications that need both completeness as well 
as reliability, {\tt sfind} and {\it Aegean} are good. Additionally,
we found that the {\tt SAD} algorithm within the widely available 
$AIPS$ package had a better performance. {\tt IMSAD} also gives a 
good completeness and reliability for detection 
thresholds $\geqslant$6$\sigma$. For transient searches,
reliability takes preference over completeness, since false
positives are likely to consume follow-up resources. Most transient 
projects are likely to be searching in near real-time. However, in 
this particular project we were fortunate to have a deep reference
image that was more than three times deeper than the single-epoch images.
This allowed us to study reliability with real (rather than synthetic)
datasets. From Figure~\ref{Fig:Reliability_algorithms}, we see that the reliability of 
{\tt sfind} is better than that of {\it SExtractor}, {\tt IMSAD} and 
{\it Aegean} except for SNR near 8.
However, the best reliability is provided by {\tt SAD}.

\subsection{Transient Candidate Search}\label{Sec:AlgorithmsTransients}

Using the best-performing source finding algorithms from 
\S\ref{Sec:sum_eff} and their optimum parameter values, we carried
out a search for transient radio sources over all epochs. We ran 
{\tt sfind}, {\tt SAD}, {\tt IMSAD} and {\it Aegean} on single-epoch images
and obtained 49 single-epoch catalogs for each algorithm. We required that any potential 
transient candidate identified in the single-epoch catalog obey the 
following constraints.

\begin{enumerate}\itemsep0pt

\item The source is not found in the reference catalog (within 2\arcsec) of persistent
  sources. The reference catalog was constructed similar to DR2 but the 
  sources were selected up to the 20\% power point of the beam 
  (i.e. $\theta < 21.5\arcmin$ radius) of each pointing rather than the 
  34$\arcmin$ interior region shown in Figure~\ref{Fig:obsCoverage}.

 \item The source is at least a $7\sigma$ detection.
 
 \item Is a genuine point-like source, i.e. it has
  \begin{enumerate}\itemsep0pt
   \item $0.9 < S_{int}$/$S_{peak} < 1.5$
   \item $a<$ 2.8\arcsec $\times$ 1.5, $b<$ 1.6\arcsec $\times$ 1.5 \newline
($a=$major axis, $b=$minor axis; recall that the synthesized beam is $2.8\arcsec \times 1.6\arcsec$)
  \end{enumerate}

 \item The source is at least 20 synthesized beams (20 $\times$ geometric mean of FWHMs; 42\arcsec) away from the nearest:
   \begin{enumerate}\itemsep0pt
    \item bright source ($>500~\mu$Jy), so that any sidelobe emission is rejected
    \item extended source
   \end{enumerate}

\end{enumerate}

The multiplicative factor of three-halves used in the selection of
point sources, as well as the distance of 20 beamwidths used for 
constraining the proximity from bright and extended sources, is
somewhat arbitrary, but is based on several iterations of our
transient search code and inspection of the cutouts of the resulting
transient candidates. By investigating how the major axes of sources 
increase with their distance from the pointing center, we found that 
$\theta \simeq 21.5\arcmin$ appears also to be the threshold 
beyond which bandwidth smearing coupled with our constraints on the major 
(and minor) axes start rejecting genuine point sources.

Due to the large number of synthesized beams searched ($n=1.8 \times
10^7$) in this dataset, there is a modest probability that a transient
candidate is due to noise\footnote{From theory we know that the 
statistics of beam values of 
interferometric maps should follow a Gaussian distribution.}. We
thus carried out an analysis similar to \cite{fko+12} (see Appendix A
of that paper) to determine the SNR above which the probability of
having the highest value of $n$ Gaussian random numbers is $\leqslant$1\%.
This corresponds to an SNR of 6.1. However, following the
recommendation of \cite{fko+12} to have a higher SNR cutoff when the 
noise was not strictly Gaussian, we chose $7\sigma$ as the lower limit
for finding transients.

The search method outlined above may miss transients which are bright enough 
to be present in the reference catalog. Therefore, we also searched (with 
similar constraints as above) for sources which are detected in the reference 
catalog of persistent sources and detected in only one of the single-epoch catalogs.

For {\tt sfind}, we adopted the parameters $\alpha$=10 and $rmsbox$=20 
for transient search. We found five candidates which are present 
only in a single epoch above the $7\sigma$ detection threshold. They also show up 
in the reference catalog. However, all of these candidates seem to 
be variable but persistent, and are detected at a low SNR ($\sim$3--5$\sigma$) 
in other epochs. Thus we do not find any transient with {\tt sfind}.

For {\it Aegean}, we used a clipping level of $7\sigma$ and a curvature-map 
cutoff of $1\sigma$. We discarded all the transients corresponding to 
islands which were too small to give a 6-parameter Gaussian fit 
(sources with flag 10000 or 00100 set) since otherwise we were dealing with 
a large number of transients. Using these constraints, we found one 
candidate which is present in the reference catalog 
and is a persistent source detected at a largely reduced SNR in other epochs.
Accordingly, {\it Aegean} does not yield any transients.

For {\tt SAD}, we searched for sources down to $3\sigma$ with an rmsbox of 
20 beamwidths, and used the input parameters {\small $DPARM(3),DPARM(7)$=(1000,1)}
to ensure optimum completeness and reliability based on our efficiency test. 
We do not find any transients with {\tt SAD}.

With {\tt IMSAD}, we searched for sources down to $7\sigma$ with the histogram option 
set. The several sources found to be transient candidates, are only variables at our detection 
limit, as evident through visual inspection of the images. Thus, there are no transients 
reported by {\tt IMSAD}.

\section{Discussion and Conclusions}\label{Discussion}

In this paper we have explored the time-domain properties of a 1.4 GHz
survey made toward the Extended Chandra Deep Field South (E-CDFS)
region. Six mosaic pointings toward E-CDFS were taken in 49 separate epochs
over a period of three months. Single-epoch images allow us to explore
the transient and variable radio sky at sub-milliJansky levels on
timescales of days, weeks and months. We will now use these data to
assess the degree of variability (\S\ref{Sec:vary}) and the transient
rate (\S\ref{Sec:rate}) of the radio sky, and predict what will be seen 
by future wide-field surveys (\S\ref{Sec:future}).


\subsection{Comparison of variability with previous surveys}\label{Sec:vary}

We found in our study that only a small fraction ($7/599 = 1.2^{+1.2}_{-0.7}\%
$) of the point sources in the E-CDFS showed any significant
variability on day-week-month timescales.  Evidently, the
sub-milliJansky radio sky at 1.4 GHz is not highly variable. The only
previous sub-mJy study at 1.4 GHz was from a single deep pointing
toward the Lockman Hole. With sampling timescales of 19 days,
\citet{cif03} found less than 2\% of sources above 0.1 mJy to be highly
variable.

Our findings at sub-mJy levels are consistent with the several
previous 1.4 GHz studies at higher flux density thresholds.
\citet{thwb+11} analyzed the 8444 deg$^{2}$ of the FIRST survey and
found only 0.5\% (1627/279407) of sources above 1 mJy varied
significantly on timescales of minutes to years. \citet{fkh+94} imaged
a 2$^\circ$ region toward a gamma-ray burst on timescales of 1-96 days
and found that fewer than $\sim$1\% of the sources above a flux density of
3.5 mJy were strongly variable. \citet{vbwh04} imaged a 120.2 deg$^2$
area of Sloan Digital Sky Survey Stripe 82 finding $\leqslant$1.4\% (123/9086) of the radio sources
to be strongly variable (i.e. $>$4$\sigma$) above flux densities of 2 mJy
on a 7 year timescale. \citet{of11} do a two-epoch comparison of
FIRST and NVSS point sources brighter than 5 mJy and find that only
0.1\% (43/4367) vary by more than 4$\sigma$ over timescales ranging 
from about 300 to 1700 days. \citet{bmg+11a,bmg+11b} analyzed 22 years and
2775 deg$^{2}$ of MOST observations at 0.84 GHz and found only 0.17\%
(53/29730) strong variables above 14 mJy on timescales of days to
minutes to years.  Finally, \citet{cbk+11} used the Allen Telescope
Array to survey a 690 deg$^{2}$ area at 1.4 GHz.  They compared their
catalog to the NVSS, finding that 0.1\% (6/4408) of the sources were highly
variable on a timescale of 15 years.

\begin{figure*}[htp]
\centering
\includegraphics[width=5.6in,viewport=10 7 590 440,clip]{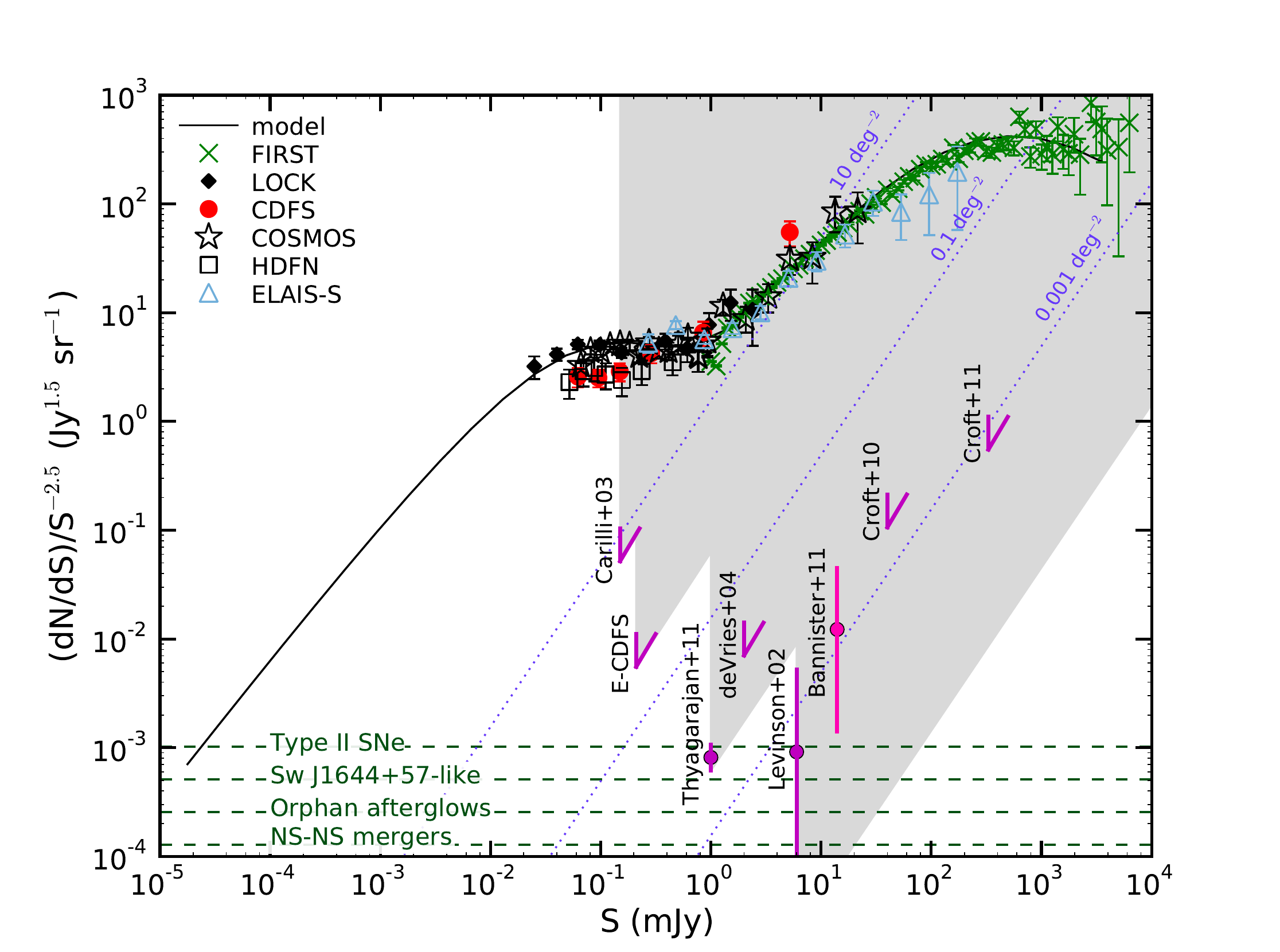}
\caption{Normalized 1.4 GHz differential radio source counts for 
  persistent sources from \cite{zmnw10} and the normalized areal density of transients 
  (or limits) as a function of the flux density for various surveys at this frequency.
  \cite{bmg+11a} survey at 0.84 GHz is colored differently than the other surveys.
  Most of the surveys are upper limits (wedge symbols) and the sampled phase 
  space is shown by the grey shaded area. Upper limits from \cite{fkh+94} and 
  \cite{bower2011} do not explore any new part of the phase space (non-grey area), 
  and hence have been left out of this diagram. Our upper-limit is labeled as ``E-CDFS''.
  Three surveys have transient detections so far, $2\sigma$ error bars for 
  which are shown according to \cite{geh86}.
  Note that \cite{thwb+11} and \cite{bmg+11a} may have identified a few strong 
  variables as transients (see \S\ref{Sec:rate}), which would make their detections move 
  downward on this plot. The black solid line is the model for AGNs and 
  star-forming galaxies from \cite{condon1984}. Lines of constant areal density 
  are shown as blue dotted lines. The horizontal dashed lines are estimates for 
  the areal density for known and expected classes of long-duration radio 
  transients taken directly from \citet{fko+12}. The areal density for Swift J1644+57-like
  tidal disruption events has been modified according to \cite{berger2012} to reflect their 
  true rate at 1.4 GHz. Upper limits from the ASKAP-VAST surveys are estimated to be an 
  order of magnitude or more below the rate of orphan gamma-ray burst afterglows, and having a rms 
  sensitivity ranging between 10 $\mu$Jy and 0.5 mJy.}
\label{Fig:logNlogS}
\end{figure*}

In Figure~\ref{Fig:logNlogS} we plot the differential source counts for 
the persistent radio sky at 1.4 GHz, normalized in the usual way by 
the Euclidean rate \citep{hjnp05}. The steep evolution of the AGN 
with decreasing flux density is apparent, as is the flattening of the 
source counts near 1 mJy. The fractional variability appears to be  
low, at a level of one percent, among the sources greater than 100 $\mu$Jy.

Comparing the variability properties at 1.4 GHz of our sub-mJy
population with those at higher flux densities, we find no obvious
change in the {\it fraction} of strong variables. 
This is despite the fact that radio-loud AGN (which dominate at milliJansky levels)
diminish in importance as radio-quiet AGN and star forming galaxies
begin to populate the radio sky at lower flux densities,
the latter constituting about 50\% of the sub-milliJansky radio
sky \citep{sss+08,pad11}.
On the other hand, in a sample of brightness
temperature-limited radio sources, it would be expected that the
variability would increase with decreasing flux density.

Our optical identifications for the seven radio variables show a mix
of AGN and star-forming galaxies. Six of these are consistent with a
nuclear source based on the carefully matched astrometry. There is no
evidence from this sub-mJy sample that we are seeing a new source of
variable radio emission such as supernovae, gamma-ray bursts,
etc. that would be expected to be offset from the nucleus.

Radio variability appears to be a strong function of frequency.  The
best study to compare to this one is the 5 GHz survey of
\citet{ofb+11}, since it used a similar observing cadence and the
identical statistical measures of variability. In this case it was
found that $\sim$30\% of point sources brighter than 1.5 mJy at 5 GHz
were variable.

Some caution is warranted in comparing fractional variability between
different experiments. Differences in the cadence, integration time,
duration, and angular resolution will have a tendency
of reducing the variability amplitude. For example, the low angular
resolution of some surveys may have the effect of reducing the strong
variability from compact radio sources embedded in diffuse
emission. This current survey with a synthesized beam of
2.8$^{\prime\prime}\times 1.6^{\prime\prime}$ has the highest angular
resolution of any previous variability survey
\citep[see][]{ofb+11}. 

Another factor to consider is that the degree of variability which can
be detected in a given experiment, as measured by the modulation index
$m$, is a function of the signal-to-noise ratio. A source in our present survey
would have to have a mean flux density of 1 mJy in order to detect a
10\% modulation, while at the 5$\sigma$ limit of the DR2 catalog
($\sim$40 $\mu$Jy) a source would have to have $m\geqslant 1.5$ in order to
be identified as a significant variable. No strong variables
(i.e. $m>0.5$) were identified in our survey, but only the about 90
sources in the DR2 catalog are bright enough ($\geqslant$126 $\mu$Jy) to
have been identified as a strong variable. In either case the fraction
of significant or strong variables is less than a few
percent of the sample.

The most robust conclusion that can be drawn is that the variable
radio sky at 1.4 GHz appears to be relatively quiet with only a
fraction of a percent of sources varying substantially over a wide
range of flux densities and timescales.

\subsection{Limits on transient areal density and rate}\label{Sec:rate}

We searched our multi-epoch data for transients but found none. The
search was conducted on each image out to a radius of 21.5$^\prime$ from
the pointing center. The single-epoch area out to that radius is 0.40
deg$^2$, or a total area of 20 deg$^2$ for all 49 epochs. However, the
sensitivity of the VLA antennas is not uniform across this area. The
primary beam response is well-described by a Gaussian with a
half-width to half-maximum of 15$^\prime$, falling to 20\% response at
our search radius of 21.5$^\prime$. At the pointing center the 7$\sigma$
flux density limit was approximately 210 $\mu$Jy for each epoch.

In order to calculate a limit on the areal density of any putative
transient population we follow \citet{ofb+11} and parameterize the
source number-count function as a power law of the form
\begin{equation}
\kappa(>S)=\kappa_{0}(S/S_{0})^{-\gamma},
\label{Eq:CountFun}
\end{equation}
where $S$ is the peak flux density, $\kappa(>S)$ is the sky surface density of
sources brighter than $S$, $\kappa_{0}$ is the sky surface density of
sources brighter than $S_{0}$, and $\gamma$ is the power law index of
the source number-count function. We assume, for simplicity, a
homogeneous source distribution in Euclidean Universe so $\gamma=3/2$.
The one-sided 2-$\sigma$ upper limit on the areal density is three
events \citep{geh86}. Therefore, using Equation C5 in \citet{ofb+11}
we find that the 2$\sigma$ upper limit on areal density to a flux limit of
210\,$\mu$Jy is 18.0\,deg$^{-2}$ per epoch.  Given that we have 49
epochs the 2$\sigma$ upper limit on the areal density is
$\kappa(>0.21\,{\rm mJy})<0.37$\,deg$^{-2}$. We can further estimate an
upper limit on the transient rate assuming a duration $t_{{\rm dur}}$
less than the shortest time between epochs of $\Re(>0.21\,{\rm
  mJy})<268 (t_{{\rm dur}}/{0.5\,{\rm day}})^{-1}$ $\, {\rm
  deg}^{-2}\,{\rm yr}^{-1}$.

Our upper limit on the areal density of transient sources at
sub-milliJansky levels can be compared with the predictions based on
previous surveys. The \citet{bsb+07} survey is a useful benchmark
since their areal density dominates all known classes of transients.
Adopting their measured two epoch rate of $\kappa(>0.37\,{\rm
  mJy})=1.5$\,deg$^{-2}$ and assuming a Euclidean source distribution
(i.e.  $\gamma=3/2$) we predict $\kappa(>0.21\,{\rm
  mJy})=3.5$\,deg$^{-2}$ at the flux density limit of our current
survey.

An alternative way to look at our results is to compare our null
detection to the expected number of Bower et al. transients expected
in our dataset. We use the parameterization of \cite{fb11} for the
predicted Bower et al. transient rate as a function of flux density, 

\begin{equation}
\mbox{log} \left(\frac{\kappa}{\mbox{deg}^{-2}}\right) = -1.5 ~\mbox{log} \left(\frac{S_{\nu}}{\mbox{Jy}}\right) - 5.13 
\label{Eq:transientRate}
\end{equation}

where $\kappa$ is the snapshot rate, and $S_\nu$ denotes the detection
threshold of the observations at the pointing center (i.e.
7$\sigma=210~\mu$Jy). Integrating both sides of equation
\ref{Eq:transientRate} over the azimuthal angle and in $\theta$ out to
$21.5\arcmin$ we get about 0.42 transients per epoch, if the
\cite{bsb+07} transients are real. Since we have 49 epochs, we expect
to have about 21 Bower et al. transients in our E-CDFS dataset.

Our search on the E-CDFS field suggests that the areal density of
radio transients is an order of magnitude or more below the rate
measured by \citet{bsb+07} (i.e. $<$0.37 \,deg$^{-2}$ vs.
3.5\,deg$^{-2}$). Alternatively, we find a 2$\sigma$ upper limit of
$<$3 transients while the predicted number is $\sim$21 transients.
Our work therefore appears to support that of \cite{fko+12}, which found from
a re-analysis of the Bower et al. data, that the transient rate was as
much as an order of magnitude smaller than previously reported. This
conclusion would be more robust if the spectral index of the putative
transient population was better known. \citet{obg+10} was able to use
other surveys to constrain the spectra index $\alpha>0$ (where
S$_\nu\propto\nu^\alpha$). Since the \citet{bsb+07} rates were derived
based on observations made mostly at 5 GHz, our only data provide
strong constraints for $0\leqslant\alpha\leqslant 1.1$.  A population of
optically thick $\alpha\simeq5/2$ sources with a rate similar to that
of \citet{bsb+07} would be undetected in our 1.4 GHz E-CDFS fields.

Many of the same variability surveys discussed in \S\ref{Sec:vary}
were also sensitive to transients. Superposed on the radio source-count 
plot of Figure~\ref{Fig:logNlogS} are the results of several of
these transients surveys. Light grey shaded areas represent the
transient phase space covered by each of the surveys, white space
represents open phase space for future narrow-deep or wide-shallow
surveys. With few exceptions most of these transient surveys result in
upper limits. Also shown are the normalized areal density of several
known and expected classes of long-duration radio transients, based on \citet{fko+12}. 
The nominal rates for the putative \citet{bsb+07}
sources are about two orders of magnitude above the tidal disruption or
Sw\,J1644+57-like objects. \cite{thwb+11} define a transient as having either 
a single detection in the analyzed epochs, or the highest 
flux density 5 times greater than the next highest one (detection/upper-limit).
Thus, \cite{thwb+11} and 2-epoch surveys like \cite{bmg+11a} may identify a strong variable as a 
true transient, which will move the source-count of detected transients lower in 
Figure~\ref{Fig:logNlogS}.

\subsection{Future Radio Surveys}\label{Sec:future}

There are several facilities built or under construction that will be
capable of synoptic imaging at 1.4 GHz. All of these facilities have
the exploration of the time domain as part of their core science
programs. The Australian Square Kilometer Array Pathfinder (ASKAP) and
the Apertif instrument on the Westerbork Synthesis radio Telescope
(WSRT) will be using focal plane array technology to instantaneously
image an instantaneous field of view (FoV) of 30 deg$^2$ and 8
deg$^2$, respectively \citep{mck12,ovv10}. South Africa is building
MeerKAT, an array of 64 13.5-m diameter dishes, with a FoV of 1
deg$^2$ at 1.4 GHz \citep{bj12}.  Finally, there is the newly
refurbished Karl G.  Jansky Array (VLA) has 27 25-m dishes with a FoV
of 0.25 deg$^2$ at 1.4 GHz \citep{pcbw11}.

We list the capabilities of each of these telescopes in Table
\ref{Tab:Experiments}. Survey speed (SS), normalized here to the VLA, is a
useful figure of merit for inter-comparison of survey capabilities of
long duration transients and is expressed as

\begin{equation}
{\rm SS} \propto {\rm BW} \times \Omega ({\rm A}_e/{\rm T}_{sys})^2
\label{Eq:SurveySpeed}
\end{equation}

\noindent where BW is the bandwidth, $\Omega$ the FoV, A$_e$ is the total
collecting areas times the aperture efficiency $\epsilon_e$, and
T$_{sys}$ is the antenna system temperature \citep{cordes08}. The
relative survey speeds are only approximate since some of the system
parameters for MeerKAT and ASKAP have not been confirmed with on-the-sky 
testing.  Likewise, we have assumed that radio frequency
interference (RFI) limits the VLA bandwidth to only 50\% of its
maximum BW. Larger fractions are achievable in the more extended array
configurations and with better RFI excision of the data. We have
calculated the ASKAP survey speed with both the 18 phased array feeds
which are currently funded and with the full 36 as originally
specified. Despite these uncertainties, it is clear from Table
\ref{Tab:Experiments} that to within factors of a few, these are all
powerful wide-field imaging facilities.


\begin{table}[htp]
\centering
\footnotesize{
\caption{Telescope Specifications}
\label{Tab:Experiments}
\begin{tabular}{l|*{6}{c}l} 
\hline \hline
Telescope     & BW    & $\Omega$    & D   & N & $\epsilon_e$  & T$_{sys}$ & SS \\
              & (MHz) & (deg$^{2}$) & (m) &   &               & (K)       &  \\
\hline
VLA      & 512 & 0.25 & 25   & 27     & 0.5  & 26 & 1.0  \\
ASKAP    & 300 & 30   & 12   & 18(36) & 0.8  & 50 & 1.1(4.6) \\
Apertif  & 300 & 8    & 25   & 13     & 0.75 & 70 & 1.3  \\
MeerKAT  & 750 & 1    & 13.5 & 64     & 0.7  & 30 & 4.1  \\
\hline
\multicolumn{8}{p{3.3in}}{Notes$-$ Here BW is bandwidth in MHz, $\Omega$ is the field of
  view in deg$^2$, D is the antenna diameter in meters, N is the
  number of antennas in the array, $\epsilon_e$ is the aperture
  efficiency, T$_{sys}$ is the system temperature in Kelvin, and SS is the
  survey speed normalized to the VLA.}
\end{tabular}
}
\end{table}        

To illustrate these survey capabilities and compare them to what we
currently know about the transient and variable radio sky, we will use
the example of a electromagnetic counterpart search for gravitational
waves. For a good overview of the topic of EM-GW searches, and the
main issues, we refer the reader to \citet{mb12} and \citet{nkg13}.

Long-duration radio emission has been predicted to originate from the
merger of a neutron star binary from several sources including the
merger shock \citep{kis12}, afterglow emission from the beamed outflow
in the relativistic and non-relativistic phases \citep{mb12}, and from
quasi-isotropic, mildly relativistic outflows ejected during the
merger \citep{np11,pnr12}. All of these mechanisms depend on the
amount of energy put into shocked material and the density of the
ambient medium.  Predicted flux densities and timescales therefore
vary over a wide range. We take as an example the detection of a
signal with a peak flux density of 100 $\mu$Jy. Such a signal might be
expected to occur for an ambient medium with a density of 0.1
cm$^{-3}$ on day timescales for a merger shock, or on year-long
timescales for mildly relativistic ejecta. We note that if short
duration gamma-ray bursts are the dominant population of neutron star
mergers, then the predicted radio signal would be difficult to detect
with the telescopes in Table \ref{Tab:Experiments}, given the the
canonical energy and ambient density inferred for this population
\citep{fbm12,mb12}.

The median sky localization of a gravitational-wave source will be 60
deg$^2$ with a three-element GW network, and 7 deg$^2$ with a five-
element GW network \citep{nkg13}. With such large error boxes, the
main challenge for the identification of a EM-GW counterpart will be to
distinguish it from the foreground of false positives. Optical-only
searches for EM-GW counterparts are expected to be overwhelmed by
false positives at the required depth of 22-23 mag and special
strategies are required \citep{nkg13}. 
However, as we have shown from this paper, the radio sky at 1.4 GHz is 
relatively quiet. Integrating the differential source counts in 
Figure~\ref{Fig:logNlogS} using the fit from \citet{hjnp05}, we estimate 
that the number of {\it persistent} radio sources above 100 $\mu$Jy 
to be 910 deg$^{-2}$. We have estimated that the fraction of strong 
variables, on a wide range of timescales, is likely to be one percent 
or less, or 9 strong variables per square degree. Further, the radio variables 
that we have seen to date have all been nuclear 
sources (\S\ref{sec:VariablesNotes}). Such variable or transient
sources could be easily rejected as EM-GW counterparts since 
significant offsets are predicted from the host galaxy based on binary 
neutron-star population synthesis models and measurements from 
short-hard gamma-ray bursts \citep{bpb+06,fbf10}. 
Lastly, the number of unrelated transients is also expected 
to be much less. Our derived limit on the transient areal rate of
$\kappa(>0.21\,{\rm mJy})<0.37$\,deg$^{-2}$ translates to a $<$1.1 deg$^{-2}$ at
100 $\mu$Jy (\S\ref{Sec:rate}) for a Euclidean distribution. The limit on the {\it known}
transient populations is even smaller \citep{fko+12}.


We believe that the multi-wavelength approach which we have taken here
should inform future searches. For the radio variables that we
found in the E-CDFS, we were able to identify the source of 
the emission using optical images and spectra 
(\S\ref{sec:VariablesNotes}). A similar strategy could be employed to 
identify false-positives for the small number of radio variables or 
transient sources identified in EM-GW counterpart searches. Whether this
approach will ultimately lead to a robust EM-GW counterpart detection
is uncertain, but in terms of characterizing the variables, minimizing 
false-positives and getting an early sense on the nature of transients, 
we suggest that joint radio-optical searches will be fruitful for 
exploring the dynamic sky.

\acknowledgments

M. Kunal wishes to thank Rick Perley, Eric Greisen, Sanjay Bhatnagar, Andrea Petric, 
Margherita Bonzini, Bill Cotton, and Paul Hancock for useful discussions. 
We thank the anonymous referee for useful comments. 
The National Radio Astronomy Observatory is a facility of the National 
Science Foundation operated under cooperative agreement by Associated 
Universities, Inc. SRK's research in part is supported by NASA and 
NSF.  This research has made use of NASA's Astrophysics Data System, 
Vizier and NED.

\end{document}